\documentclass[iop]{emulateapj}
\usepackage{apjfonts}
\usepackage{rotating}
\usepackage{color}

\shorttitle{A MegaCam Survey of Halo Satellites}
\shortauthors{Mu\~noz et al.}

\begin{document}

\title{A MegaCam Survey of Outer Halo Satellites. III. 
Photometric and Structural Parameters\altaffilmark{1,2}}

\author{
Ricardo\ R.\ Mu\~noz\altaffilmark{3,4},
Patrick\ C\^ot\'e\altaffilmark{5},
Felipe\ A.\ Santana\altaffilmark{3},
Marla\ Geha\altaffilmark{4},
Joshua\ D.\ Simon\altaffilmark{6},
Grecco\ A.\ Oyarz\'un\altaffilmark{3},
Peter\ B.\ Stetson\altaffilmark{5} \&
S.\ G.\ Djorgovski\altaffilmark{7}
}

\altaffiltext{1}{Based on observations obtained at the Canada-France-Hawaii 
Telescope (CFHT) which is operated by the National Research Council of 
Canada, the Institut National des Sciences de l'Univers of the Centre 
National de la Recherche Scientifique of France,  and the University of Hawaii.}

\altaffiltext{2}{This paper includes data gathered with the 6.5 meter Magellan 
Telescopes located at Las Campanas Observatory, Chile.}

\altaffiltext{3}{Departamento de Astronom\'ia, Universidad de Chile, 
Camino del Observatorio 1515, Las Condes, Santiago, Chile ({\tt rmunoz@das.uchile.cl})}

\altaffiltext{4}{Astronomy Department, Yale University, New Haven, CT 06520, USA}

\altaffiltext{5}{National Research Council of Canada, Herzberg Astronomy \& Astrophysics Program, 5071 W. Saanich Road, Victoria, BC, V9E 2E7, Canada}

\altaffiltext{6}{Observatories of the Carnegie Institution of Washington, 
813 Santa Barbara St., Pasadena, CA 91101, USA}

\altaffiltext{7}{Astronomy Department, California Institute of Technology, 
Pasadena, CA, 91125, USA}

\begin{abstract}
We present structural parameters from a wide-field
homogeneous imaging survey of Milky Way satellites carried out with the
MegaCam imagers on the 3.6m Canada-France-Hawaii Telescope (CFHT) 
and 6.5m Magellan-Clay telescope.
Our survey targets an unbiased sample of ``outer halo" 
satellites (i.e., substructures having Galactocentric distances greater than 25 kpc) and includes 
classical dSph galaxies, ultra-faint dwarfs, and remote globular clusters. We combine deep, panoramic $gr$ 
imaging for 44 satellites and archival $gr$ imaging for 14 additional objects (primarily 
obtained with the DECam instrument as part of the Dark Energy Survey), to measure
photometric and structural parameters for 58 outer halo satellites. This is the largest and most uniform analysis 
of Milky Way satellites undertaken to date and represents roughly three quarters 
($58/81 \simeq$~72\%) of all known outer halo satellites. We use a maximum-likelihood method 
to fit four density laws to each object in our survey: exponential,  Plummer, King and S\'ersic models. 
We examine systematically the 
isodensity contour maps and color magnitude diagrams for each of our program objects, present a 
comparison with previous results, and tabulate our best-fit photometric and structural parameters, including ellipticities, position 
angles, effective radii, S\'ersic indices, absolute magnitudes, and surface brightness measurements. We investigate the
distribution of outer halo satellites in the size-magnitude diagram, and show that the current sample of outer halo substructures 
spans a wide range in effective radius, luminosity and surface brightness, with little evidence for a clean
separation into star cluster and galaxy populations at the faintest luminosities and surface brightnesses.
\end{abstract}


\section{Introduction}
\label{sec:introduction}

The halo of the Milky Way contains substructures that hold important clues to
formation and evolution of the halo itself. Historically, these substructures (i.e., satellites) were divided into two distinct populations --- i.e.,
{\it globular clusters} and {\it dwarf galaxies} --- presumed to have fundamentally different formation channels.
At the same time, history has also shown that the census of satellites at any time depends sensitively on observational selection 
effects, with surface brightness being a critical factor in our ability to identify and characterize halo substructures.

Over the past two decades, a number of wide-field optical surveys (having point-source detection limits that allow extremely
faint surface brightness thresholds to be reached) have revolutionized our view of the halo and its embedded substructures.
In addition to revealing numerous stellar streams and large-scale density fluctuations, these surveys (most notably the Sloan Digital 
Sky Survey, Pan-STARRS and the Dark Energy Survey; \citealt{york00a,chambers16a,diehl14a}) have led to the discovery of more 
than 50  satellites since 2000: i.e., roughly two thirds of all outer halo substructures known at this time.

In a number of cases, the newly discovered satellites cannot easily be identified as globular clusters or dwarf galaxies ---
the two-category scheme historically used to classify halo substructures. In these instances, spectroscopic data
is essential for measuring dynamical mass-to-light ratios and/or element abundances of individual stars.
Still, it is worth bearing in mind that classifying as a star cluster or dwarf galaxy on the basis of 
photometric and structural parameters can itself be problematic.
First, published catalogs for 
satellites tend to focus on either globular clusters or dwarf galaxies, rather than
taking a holistic approach to halo substructures in general. Second, existing compilations often rely on shallow and 
heterogenous  data, some of  it dating back to the 1960s 
(see, e.g., \citealt{djorgovski93a,pryor93a,trager95a,irwin95a,harris96a,mateo98a,mclaughlin05a,mcconnachie12a}
and references therein). Finally, photometric and structural parameters are usually derived by fitting parametric
models to the observed one- or two-dimensional density profiles, with different choices of the density law commonly
made for globular clusters and dwarf galaxies. 

Between 2009 and 2011, we carried out an extensive imaging survey that aimed to address these issues by using  
the 3.6m Canada France Hawaii Telescope (CFHT) and the 6.5m Magellan-Clay telescope to acquire panoramic $gr$
images for a nearly complete sample of substructures in the outer halo of the Milky Way (i.e., at Galactocentric
radii of $r_{\rm GC}$ = 25~kpc or more). In this paper, we use the point-source photometric catalogs from this program 
to derive homogeneous photometric and structural parameters for each of our program objects.

This paper is the latest in a series that explores the properties of outer halo substructures based on these 
CFHT and Magellan data. \citet[][hereafter {Paper~I}]{munoz18a} have presented an overview of the survey, 
including observational material, target selection, reduction 
procedures and data products. In an upcoming paper, we will use the structural and photometric parameters from this paper 
to explore the scaling relations of outer halo satellites.
\citet{bradford11a} used imaging from this survey to carry out a dynamical analysis
of the globular cluster Palomar~13, while \citet{munoz12b} reported the discovery of an ultra-faint star cluster (Mu\~noz~1)
in the direction of the Ursa Minor dwarf galaxy.  \cite{santana13a} presented a study of blue straggler stars 
across satellites of all types, while \cite{carballo15a} reported the detection of possible foreground populations associated with
Monoceros substructure in the direction of NGC2419 and Koposov 2. Recently, \cite{santana16a} used imaging for the 
Carina dwarf galaxy to investigate its spatially revolved star formation history, and finally \cite{carballo17a} studied the 
leading and trailing arms of the Sagittarius tidal stream around the globular cluster Whiting~1. 

This paper is structured as follows. In \S\ref{sec:obs}, we briefly review our observations and sample selection. Our maximum
likelihood method for measuring structural parameters and density profiles is described in \S\ref{sec:method}. \S\ref{sec:results}
presents a comparison to previous results, including a case-by-case discussion of our survey targets. 
A discussion of our results is presented in \S\ref{sec:discussion} and we conclude in \S\ref{sec:summary}.

\section{Observations and Target Selection}
\label{sec:obs}

The scientific justification for our survey, including target selection, observing strategy,
data reduction methods and photometric calibration, are described in detail 
in a companion paper (\citeauthor*{munoz18a}). Briefly, our sample consists of 44 {\it primary targets}
and 14 {\it secondary targets} located in the outer halo of the Milky Way. 
Here, we consider the ``outer halo" to begin at a Galactocentric distance of $R_{\rm GC}=25$\,kpc.
Data acquisition for targets belonging to our primary sample was
completed in 2010 and, after including our 14 secondary targets, the combined 
sample of satellites analyzed in this paper represented --- {\bf at the time of writing } --- 58 of the 81 known Galactic satellites 
beyond $R_{\rm GC}$ = 25~kpc (for an overall completeness level of $\approx$ 72\%). Two massive 
satellites --- the Large and Small Magellanic Clouds --- were omitted from our study due to their large sizes, 
which render them impractical for a program of this scope. Although the same can be said of the Sagittarius dwarf 
spheroidal galaxy, this system lies at a Galactocentric distance of 18~kpc and thus does not strictly satisfy our 
criterion for membership in the outer halo.

\subsection{Primary Sample}
\label{sec:primary}

Our 44 primary targets were observed in $g$- and $r$-band filters with the wide-field imagers on the 
Canada-France-Hawaii Telescope (CFHT) and Magellan-Clay telescopes. This sample thus includes
objects in both the northern and southern hemispheres. In all, images for 30 and 14 satellites were
collected using CFHT and Clay, respectively. Three satellites --- the faint globular clusters Palomar~3,
NGC7492 and the ultra-faint dwarf Segue~1 --- were observed with both facilities, with the intention
of using them as cross checks on our photometry and astrometry. Twenty two of the northern 
satellites were covered by a single pointing as this provided full spatial coverage. The 
remaining eight objects were observed using either a 2$\times$2 or 2$\times$1 grid to ensure 
adequate coverage.

The Megacam imagers are not identical instruments. CFHT-MegaCam 
is a wide-field imager composed of
$36$ CCD chips that cover roughly a 1$\times$1\,deg$^{2}$ on the sky \citep{boulade03a}. Clay-Megacam  
also consists of $36$ chips but covers a smaller field of 0.4$\times$0.4\,deg$^{2}$ \citep{mcleod15a}.
In both cases, the images delivered by the observatory were pre-processed to correct for 
instrumental signatures across the mosaic.
Image processing was then carried out using the DAOPHOT, ALLSTAR and ALLFRAME packages 
\citep{stetson94a} and astrometric solutions were refined using the SCAMP 
package\footnote{http://astromatic.net/software/scamp}.  The typical $5\sigma$
point-source limits are $g_{\rm lim} \simeq 25.6$ and $r_{\rm lim}\simeq 25.3$ AB magnitudes,
with typical seeing of $0\farcs7$--$0\farcs9$ for CFHT and $0\farcs7$--$1\farcs1$ for Clay.

\subsection{Secondary Sample}
\label{sec:secondary}

As discussed in \citeauthor*{munoz18a}, data collection for our primary sample was completed in mid-2010.
Since that time, an impressive number of new Galactic satellites have been discovered, most by the Dark Energy Survey
team \citep{bechtol15a} and independently by \citet{koposov15a}. In 2015, we therefore
retrieved the publicly available DECam images for a number of these satellites and performed photometry in a
manner similar to that used for our CFHT and Clay data. Photometry for a few other systems whose images
were not publicly available was kindly provided by their respective discovery teams. 

Table~\ref{t:cat1} lists the 44 objects belonging to our primary sample, along with their center
equatorial coordinates (from this work). This table also includes estimates for their heliocentric distances, 
metallicities, metallicity dispersions, heliocentric systemic velocities, radial velocity dispersions and 
mass-to-light ratios, when available, from the literature. At the bottom of Table~\ref{t:cat1} we present the same
information for our secondary sample of 14 satellites.

\subsection{Satellites Not Included in our Survey: Tertiary Sample}

The discovery of Galactic satellites has proceeded apace, and many new faint stellar stellar systems have 
been reported during the past two years. In all, 21 newly discovered outer halo satellites are absent from our 
sample defined in 2015:  

\begin{enumerate}
  \item Fourteen objects discovered in DES imaging: 
  Kim~2 \citep{kim15a},
  Peg~III \citep{kim15d}, 
  Tuc~II \citep{bechtol15a,koposov15a}, 
  Tuc~IV, Cet~II, Ret~III, Col~I, Ind~II, Gru~II, Tuc~V and Tuc~III \citep{drlica15a}, 
  DESJ0034-4902 \citep{luque16a},
  DESJ0111-1341 and DESJ0225+0304 \citep{luque17a}.
  \item Two objects discovered in Pan-STARRS imaging: Sgr~II \citep{laevens15b} and Laevens~3 \citep{laevens15b}.
  \item Two objects discovered in VST/ATLAS imaging: Crt~II \citep{torrealba16a} and Aqu~II \citep{torrealba16b}.
  \item Two objects discovered in Subaru/HSC imaging: Vir~I \citep{homma16a} and Cetus~III \citep{homma17a}.
  \item {\bf Two new objects have been discovered in the constellation of Carina during the revision phase of this article: Carina~II and III \citep{torrealba18a}}.
\end{enumerate}

To summarize, 81 cataloged stellar systems meet our definition of outer halo members: i.e., our 58 primary
and secondary targets, the two Magellanic Clouds, and the 21 recently discovered stellar systems listed above. Thus,
the sample analyzed here represents 72\% of all known outer halo satellites. If one restricts the sample to  
systems fainter than $M_V = -13.45$ (the absolute magnitude of Fornax dSph galaxy, the brightest object in our sample), 
then the overall completeness is 58/77 $\simeq$ 75\%.

We note in passing that three other newly discovered satellites --- Dra~II \citep{laevens15b} and Gaia~1 and 
2 \citep{koposov17a} --- are located at $R_{\rm GCC} < 25$~kpc and so do not satisfy our criterion 
for membership in the outer halo.

\section{Structural Parameters and Density Profiles}
\label{sec:method}

\subsection{Two-Dimensional Analysis: Parameter Estimation with a Maximum Likelihood Approach}
\label{sec:2d}

The extreme sparseness of stellar systems at the faint end of the galaxy luminosity function
challenges our ability to derive reliable photometric and structural parameters  
\citep[e.g.,][]{martin08a, sand09a, munoz12a}. Such parameters --- including 
integrated luminosities, effective radii, central and
mean effective surface brightness, to name a few --- are key ingredients 
in the measurement of dynamical masses, and ultimately, in using faint
stellar systems in near-field cosmological studies of galaxy formation.

\citet{martin08a} presented a comprehensive analysis of SDSS photometric
data available for the ultra-faint galaxies and derived
structural parameters for them using a technique that: (1) relies on all stars
observed in a given field; and (2) does not require binning of the
photometric data. \citet{sand09a} and \citet{munoz10a},  showed that
the structural parameters derived via this method often suffer from significant 
associated uncertainties ---
in some cases as large as 80\%. This is largely a consequence of the 
relatively shallow depth of the SDSS, which has 5$\sigma$ point-source limiting
magnitudes in the $g$ and $r$ bands of $\sim$23.3 and $\sim$23.1, 
respectively \citep[][]{york00a}, although \citet{martin08a} only used stars
with $g<22.5$ and $r<22$for most objects. Indeed,
\citet{munoz12a} quantified the impact of stellar sample size
on the measurement of structural parameters, showing
that photometry significantly deeper than SDSS would be typically needed to measure 
parameters to a precision of $\sim$20\% or better.

Our survey has been specifically designed to produce improved estimates for the photometric
and structural parameters of stellar systems in the outer halo.
As shown in Figure~15 of \citeauthor*{munoz18a}, our imaging reaches typical $5\sigma$
point-source depths of $g \simeq 25.6$ and $r \simeq 25.3$, or about 2.2 mag deeper than
SDSS. On average, our photometry reaches $\sim$2 mag below the
main sequence turnoff (MSTO) in our program objects (with some variation between
objects, depending mainly on their distance: i.e., in the case of our most distant target, 
Leo~T, we fall $\sim2$ mag short of the expected location of the MSTO). In addition, our analysis 
benefits from improved star-galaxy separation compared to SDSS, thanks to the superior image
quality of CFHT and Clay. The median FWHMs for these two datasets are 0\farcs8 and 0\farcs9, 
respectively, which are significant improvements to what is available from SDSS (1\farcs4).

We employ the methodology described in \citet{martin08a} and adopted by several
other authors. In essence, the technique determines six structural parameters simultaneously.
These parameters are:  the equatorial coordinates, $\alpha_{\rm 0}$ and $\delta_{\rm 0}$ of the 
center of the satellite; the radius, $r_{\rm h}$, containing half the luminosity;  the ellipticity, 
$\epsilon$, of the projected two-dimensional isophotes; the position angle, $\theta$, measured 
north through east, of the major axis; and the background density of stellar sources, $\Sigma_{b}$,
not belonging to the satellite. This last parameter reflects the unavoidable presence of Galactic 
stars and unresolved galaxies in any field.

\citet{munoz12a} showed that the reliability of the measured structural
parameters depends not just on the total number of satellite stars, but also on the
stellar density contrast between the satellite and the background density. Thus, it is
critically important to maximize the ratio $\Sigma_{0}/ \Sigma_{b}$, where $\Sigma_0$
is the central surface density of the satellite. For each program object, we therefore 
select point sources having DAOPHOT morphological
classification indices --- $\chi$ and $sharp$ --- that are consistent with stellar 
detections \citep{stetson94a}. In our specific case, we adopt $-0.4<sharp<+0.4$ and $\chi<3$.
Additionally, we select candidate stars only in regions of the color magnitude diagram (CMD) that 
are close to the primary sequences of each satellite: i.e., the main sequence, MSTO, 
red giant branch (RGB), horizontal branch (HB) or red clump, when present.

As in our previous studies \citep{munoz10a,munoz12a}, we initially fitted three density models
that have been commonly used to fit the surface density or surface brightness profiles of
Local Group galaxies: (1) an {\it exponential} profile; (2) a {\it Plummer} profile \citep{plummer11a};
and (3) a {\it King} profile \citep{king62a}. The first two models are known to provide
adequate descriptions for dwarf galaxies, while the third model, although most commonly used
to fit the profiles of globular clusters, has been used successfully to fit the density profiles
of dwarf galaxies as well \citep[e.g.,][]{irwin95a}. We note that, as has been argued in the past,
King model parameters do not have obvious physical interpretation for dark-matter dominated 
systems (e.g., \citealt{koch06a,gilmore07a}).

The three profiles investigated here have the following functional forms:
\begin{equation}
\begin{array}{lcl}
\Sigma_{\rm exp}(r) & = & \Sigma_{\rm 0,E}{\rm exp}\left({-{r \over r_{E}}}\right), \\
\end{array}
\end{equation}

\begin{equation}
\begin{array}{lcl}
\Sigma_{\rm Plummer}(r) & = & \Sigma_{0,\rm P}\left(1+{r^{2} \over r_{P}^{2}}\right)^{-2}, \\
\end{array}
\end{equation}

\begin{equation}
\begin{array}{lcl}
\Sigma_{\rm King}(r) & = & \Sigma_{0,\rm K} \left[\left(1+ \frac{r^{2}}{r_{c}^{2}}\right)^{-\frac{1}{2}} - \left(1+ \frac{r_{t}^{2}}{r_{c}^{2}}\right)^{-\frac{1}{2}}\right]^{2}. \\
\end{array}
\end{equation}
Here $r_{E}$ and $r_{P}$ are the exponential and Plummer scale lengths, respectively, while
and $r_{c}$ and $r_{t}$ are the King core and tidal radii. Note that the exponential scale length is 
related to the half-light radius by the relation $r_{\rm h}=1.68\times r_{E}$. In the case of the
Plummer profile, $r_{P}$ is equivalent to $r_{\rm h}$. 

We also include a fourth parameterization in our analysis. Although originally used to fit the 
luminosity profiles of early-type galaxies, we also fitted a S\'ersic model  to each of our program 
objects \citep{sersic68a}. This model takes the form:

\begin{equation}
\begin{array}{lcl}
\Sigma_{\rm Sersic}(r) & = & \Sigma_{0,S}{\rm exp}\left[{-\left({r \over r_e}\right)^{1/n}}\right] \\
\end{array}
\end{equation}
where $n$ is the S\'ersic index, a measure of concentration. The effective radius, $r_e$, the
radius which contains half the total luminosity, is defined as $r_{e}=b_n^{n} \alpha$, where 
$b_n=1.999n-0.327$ \citep{caon90a}. This parameterization is known to provide a good representation
of the brightness profiles of early-type galaxies in local clusters, including low-mass dSph-like 
galaxies similar to those in the outer halo \citep[see, e.g.,][]{jerjen97a,graham03a,graham03b,ferrarese06a,cote07a}.

The maximum likelihood technique that we use for our analysis is predicated on the assumption
that we know the shape of the satellite's light distribution beforehand (i.e., one of the four
models described above). The position of stars in our photometric catalog should then follow 
this distribution which is well represented by a set of parameters $p_{1}$,\,$p_{2}$,\,...,$p_{j}$. 
Thus, we maximize a function of the form
\begin{equation}
\begin{array}{lcl}
L(p_{1},p_{2},...,p_{j}) & = & \prod_{i}l_{i}(p_{1},p_{2},...,p_{j})
\end{array}
\end{equation}
where $l_{i}(p_{1},p_{2},...,p_{j})$ is the probability of finding the datum $i$ 
given the set of parameters $p_{1}$, $p_{2}$,...,$p_{j}$. For example, in
the case of a Plummer profile, this function takes the form
\begin{equation}
l_{i}(p_{1},p_{2},...,p_{j})=S_{\rm 0}\left(1+{r^{2}_{i} \over r_{P}^{2}}\right)^{-2}+\Sigma_{b}
\end{equation}
where $S_{0}$, $r_{i}$ and $r_{P}$ are expressed in terms of the structural
parameters we want to determine.

In practice, to identify the best-fit parameters, we look for a global maximum 
$log(L(\hat{p_{1}}, \hat{p_{2}},...,\hat{p_{j}}))$ by searching 
the $j$-dimensional parameter space. In the case of an exponential and Plummer profile, 
the parameter space is $6$-dimensional, with free parameters 
$\alpha_{0}$, $\delta_{0}$, $\epsilon$, $\theta$, $r_{h}$ and $\Sigma_{b}$.
For a King profile, the approach is slightly different. In this case, there are seven parameters
to be determined because the tidal truncation introduced by this density law
results in two characteristics radii --- the core radius, $r_c$, and tidal radius,  
$r_{t}$. Finding a set of parameters that maximize $L$ has the extra
complication that $r_{t}$ is degenerate with $\Sigma_{b}$. We therefore
fix the background density to the value found for the Plummer profile.
The case of the S\'ersic model is similar because are again seven parameters 
to fit including the background density. In this instance, the S\'ersic index is degenerate with the background
density, so we follow the same strategy adopted for the King profile: i.e., we fix the 
value of the background density to that found for the Plummer model and 
solve for the remaining six parameters.

We find a solution by searching the parameter space using the amoeba
simplex algorithm (\citealt{press88a}).
This method is somewhat sensitive to the specified region of parameter 
space to be searched (i.e., the initial guess and allowed range for the parameters) 
but it runs considerably faster than an iteratively refined grid.
To ensure convergence, we re-started the amoeba three times
using the previously derived values.
To derive uncertainties for the structural parameters, we carry out 10\,000 
bootstrap realizations of our data (i.e., a resampling with replacement). 
In most cases, the distribution of a given parameter is well described by a Gaussian distribution, except
for the King tidal radius which tends to deviate slightly from this functional form. We have therefore fitted
Gaussian functions and report their mean and standard deviation as the mean and 1-$\sigma$ 
uncertainty for a given parameter.  

Table~\ref{t:structural1} presents the best-fit structural parameters and their errors for the
exponential and Plummer profiles. In all cases, we tabulate the position angles ($\theta$), 
the overall ellipticities ($\epsilon$) and half-light radii (in both angular and
physical units). In Tables~\ref{t:structural2} and \ref{t:structural3}, we list the best-fit King 
and S\'ersic parameters. 

\subsection{Absolute Magnitude and Central Surface Brightness}

In the past, total luminosities for resolved stellar systems have usually been estimated by adding the
fluxes of individual stars down to a certain limiting magnitude, and then correcting for the  
``missing" light contributed by stars below this threshold.  Given the small number of stars 
in most ultra-faint dwarfs, and some of the outer halo globular clusters, it has been argued
\citep[e.g.][]{martin08a, walsh08a,sand09a, munoz10a, munoz12a} that this methodology is 
prone to error due to shot-noise, i.e., the inclusion, or exclusion, of even a single RGB star 
can significantly change the measured total luminosity for some systems.
Therefore, an alternative method is often used to calculate the integrated luminosity,
or absolute magnitude, in a given bandpass. In this study, we apply 
this alternative technique to all objects regardless of the number of stars, with the exception of
the brighter classical dSphs which can present complex star formation histories.

The method relies on the number of stars, $N_{*}$, that belong to the satellite
down to the adopted magnitude threshold.  $N_{*}$ is related to the background density 
$\Sigma_{b}$ obtained from the maximum likelihood method by,
\begin{equation}
\begin{array}{lcl}
N_{*} & = & N_{\rm total} - A\Sigma_{b},\\
\end{array}
\end{equation}
where $N_{\rm total}$ is the total number of stars (both satellite stars and background objects) 
used to determine the best-fit structural parameters, and $A$ is the total area of the field.

We then assume that the satellites in our survey are well described by old, single 
stellar populations. This assumption is certainly reasonable for the globular clusters,
and is appropriate for most of the dwarf galaxies as well. Indeed, the faintest dwarfs in our 
sample seem to consist exclusively of old, metal-poor populations \citep{brown12a}. This is 
also true of at least some of the brighter dSph systems like Draco, Ursa Minor and 
Sextans \citep{santana13a}. 

We then model the respective stellar populations using theoretical luminosity functions 
(LF). In particular, we use LFs from \citet{dotter08a} generated for each object 
using the adopted metallicity and distance information and assuming a Salpeter initial mass function 
\citep{salpeter55a}.
The theoretical LF then gives us the relative number of stars in different magnitude
bins, from which we can derive the integrated flux down to an arbitrary
threshold.  Comparing this flux with what one obtains by integrating the entire LF then yields the 
amount of light that is contributed by stars below the adopted magnitude limit. $N_{*}$ is 
used to normalize the values obtained from the theoretical LF to the values corresponding 
to our actual program objects.  

In the case of brighter satellites, which can have more complex star formation histories (e.g., Carina, 
Fornax, Sculptor and  Leo~I), we follow the traditional methodology. In these cases, 
rather than modeling the population using a theoretical LF, we sum the fluxes for 
all stars above the appropriate magnitude limit. Typically, this is chosen so that the completeness is higher 
than $90$\% (after removing the estimated background). We then use the theoretical LFs to 
correct for the missing flux below this limit.

The mean absolute magnitudes and associated uncertainties are derived using a bootstrap analysis. 
The procedure is as follows: we treat the theoretical LF used to calculate
the luminosities as a cumulative probability function (down to our chosen magnitude
threshold) for the number of stars expected as a function of magnitude. We then randomly draw 
a number $N_{*}$ of stars from the luminosity function and add their fluxes. We repeat this process 
10\,000 times and use the distribution of magnitudes to estimate 1-$\sigma$ errors. 
Table~\ref{t:mags} records our best-fit absolute magnitudes and luminosities for all objects
in the $g$ and $r$ bandpasses.

With structural parameters and total luminosities in hand, we can calculate $\mu_{0}$, the central surface brightness, 
for each object in the survey. For this calculation, we rely on the fitted S\'ersic model which, as
we discuss below, provides a very good representation of the density and brightness distributions 
for most of our program objects, regardless of prior classification as a star cluster or dwarf galaxy. In physical units 
($L_{{\sun}}$\,pc$^{-2}$), the central  surface brightness is given by
\begin{equation}
\begin{array}{lcl}
I_{0} & = & {L_{}} / \left[ {{2\pi \alpha^{2} n \Gamma(2n)(1-\epsilon)}} \right]. \\
\end{array}
\end{equation}
From this, we can calculate the central surface brightness, in units of mag\,arcsec$^{-2}$, as
\begin{equation}
\begin{array}{lcl}
\mu_{0} & = & M_{\sun}+21.572-2.5\log{I_{0}}. \\
\end{array}
\end{equation}
We also report the value of $\mu_e$, the effective surface brightness. For the S\'ersic profile, 
this variable is related to the central surface brightness by
\begin{equation}
\begin{array}{lcl}
\mu_{e} & = & \mu_{0} + 1.086 b_n \\
\end{array}
\end{equation}
with the value of $b_n$ given in \S\ref{sec:2d}. Our measurements for absolute magnitude,
integrated luminosity, and central and effective surface brightness are given in Table~\ref{t:mags}.
We quote absolute magnitudes in the $g$, $r$ and $V$ bandpasses, while luminosities and surface
brightness estimates are specified in the $V$ bandpass. To obtain the $V-$band values we used
the transformation  $V=g-0.569\times(g-r) -0.021$  derived by \citet{jordi06a} which is
generally useful for metal-poor Population~II stars.

\subsection{One-Dimensional Analysis: High Surface Brightness Objects}
\label{sec:1d}

The maximum likelihood procedure described above will yield reliable results as long as the stellar 
completeness function remains roughly constant across the field. While this is true for most objects
in our sample, it is not the case for Palomar~2, NGC2419, NGC5694, NGC5824, NGC6229, NGC7006
and, to a lesser extent, NGC7492. These stellar systems --- all relatively bright globular clusters --- have 
central $V$-band surface brightnesses in the range 11.15--18.33 mag arcsec$^{-2}$ 
(with $\mu_{V,0} \simeq$~21~mag~arcsec$^{-2}$ for NGC7492) and are therefore limited by crowding.

For these seven objects, we must resort to a more standard approach for measuring photometric and structural
parameters. In the inner regions of each cluster, where crowding reduces completeness in the star
counts, we performed surface photometry in the manner described by \citet{fischer92a}. The central
CCD images were divided into concentric annuli positioned on the centroid of the star count 
distribution. The annuli were then divided into eight azimuthal sections, and the median pixel value
for these sectors was adopted as the surface brightness for the annulus, at the area-weighted radius.
The standard error in the median of the eight sectors was adopted as the uncertainty in the surface
brightness at that radius. In the outer regions of the cluster, we used star counts to derive surface
density profiles (with assumed Poisson errors) that were then matched via least-squares to the
surface photometry profiles at intermediate radii. 

\begin{figure}
\plotone{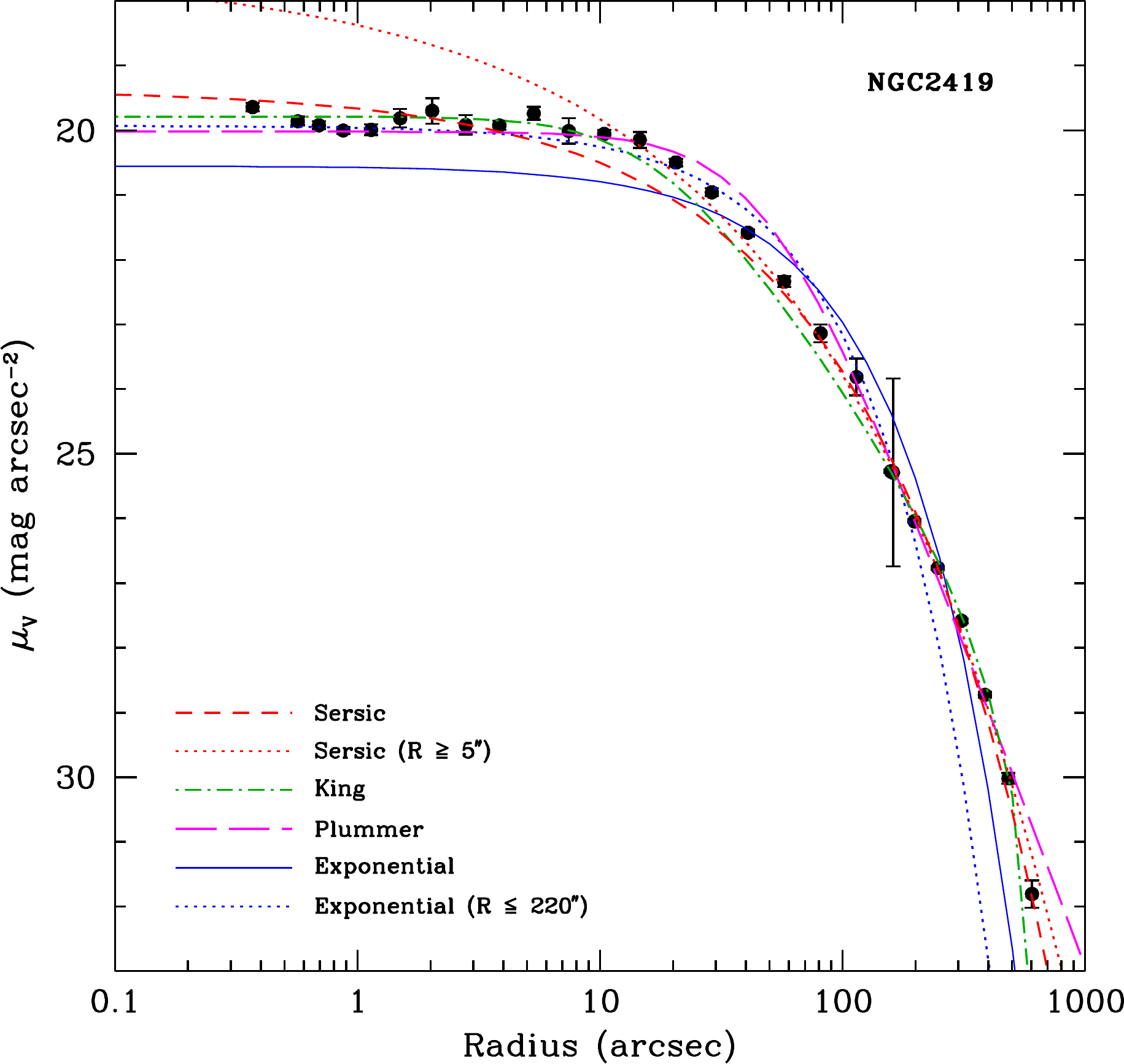}
\caption{
Surface brightness profile for NGC2419, one of seven high surface brightness satellites in our sample for which our two-dimensional maximum 
likelihood approach is not appropriate because of varying stellar completeness. This is a composite profile based on surface photometry in the 
core and star counts in the outer regions, matched via least-squares at intermediate radii. The smooth curves show the best-fit S\'ersic, King, 
Plummer and exponential models (red, green, magenta and blue curves, respectively). For the  S\'ersic  and exponential models, two models 
are shown: one that best fits the full profile (dashed or light solid curves curves) and one that best fits the observed profiles over a restricted radial range, as 
indicated in the legend (dotted curves).
} \label{ngc2419}
\end{figure}

The resulting composite surface brightness profiles were then fitted with exponential, Plummer, King
and S\'ersic surface brightness profiles using Levenberg-Marquardt minimization to determine the
best-fit parameters and their errors. The derived parameters are recorded in Tables~\ref{t:structural1}--\ref{t:structural3},
and the composite brightness profile for one representative cluster in this class, NGC2419, is shown in 
Figure~\ref{ngc2419} along with the various best-fit models.

The results for NGC2419 are typical for this sample of high surface brightness objects.
As expected, the King (1962) is found to provide an excellent representation of the surface brightness
profiles for these bright globular clusters. Though less widely appreciated, the S\'ersic model
is also able to provide reasonable fits to these cluster brightness profiles, although we show two S\'ersic models
in Figure~\ref{ngc2419} --- one that best fits the full profile, and one that excludes the central few arcseconds (see also
\citealt{baumgardt09a}). At the same time, an 
exponential profile provides a poor parameterization for these seven systems (even over a restricted radial
range) and we omit the best-fit 
parameters for this model from Table~\ref{t:structural1}. This is largely the case for the Plummer model as well.
However, for this model, we find marginally acceptable fits for Palomar~2, NGC2419, NGC6229 and NGC7492
and thus record the best-fit Plummer parameters in Table~\ref{t:structural1}. No acceptable Plummer model fit 
could be found for NGC5694, NGC5824 or NGC7006.  

\section{Results}
\label{sec:results}

\subsection{Critical Evaluation of Density Models}
\label{sec:critmodels}

It is natural to ask which family of models fitted in this paper are best able to match the surface brightness and surface
density distributions of our program objects. As discussed in \S\ref{sec:2d}, the choice of parameterization for different types 
of halo substructures has often been a matter of historical precedent, with King models being the standard choice for 
globular clusters, and exponential or Plummer models widely used for Local Group dwarf galaxies (of both classical and ultra-faint 
varieties). Meanwhile, outside of the Milky Way and M31 systems, S\'ersic models are usually the parameterization of choice for the 
surface  brightness profiles of early-type galaxies, including both high-mass (giants) and 
low-mass (dE-type) systems.  \citep[see, e.g.,][]{jerjen97a,graham03a,graham03b,ferrarese06a,cote07a}.

\begin{figure}
\plotone{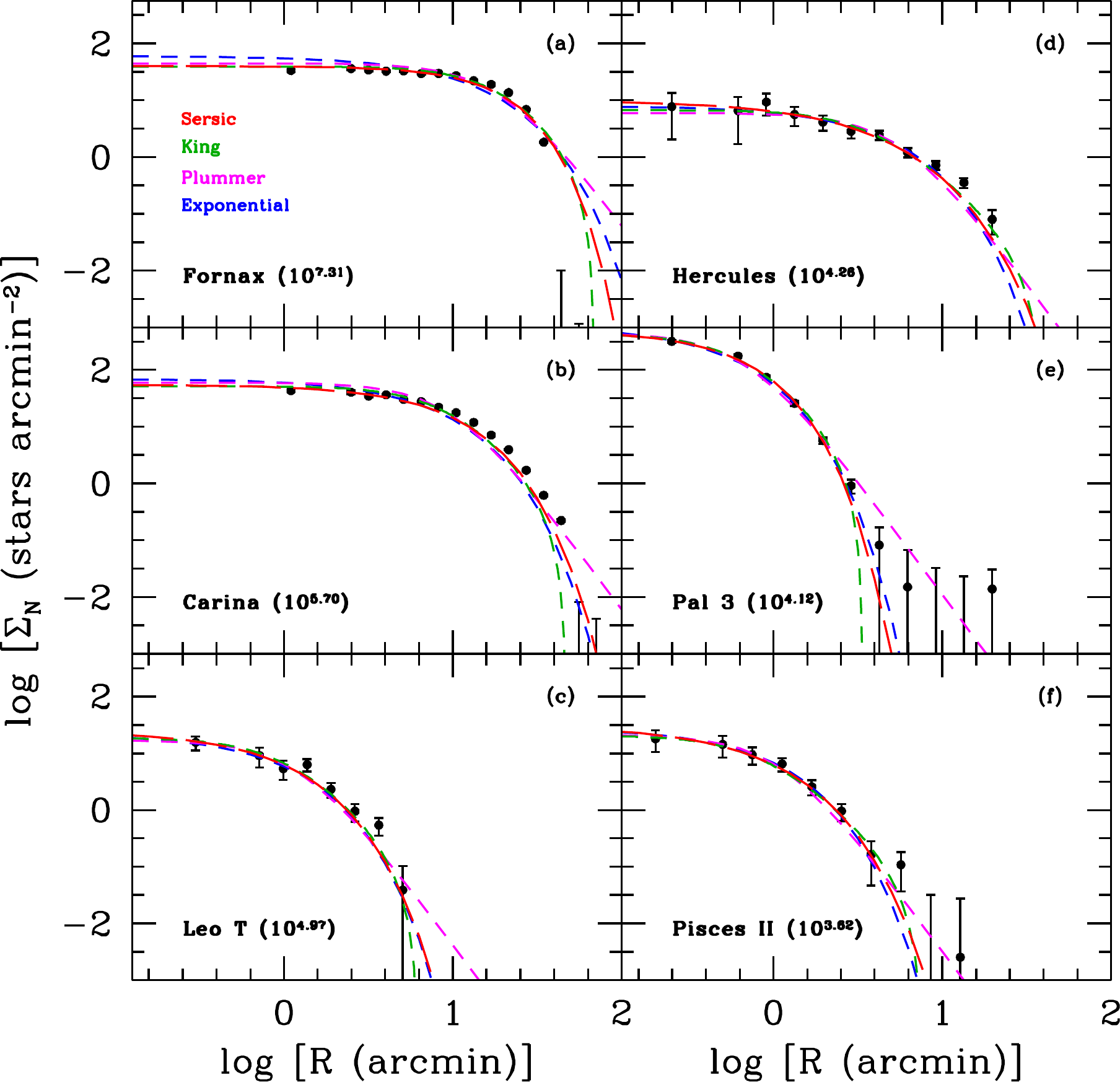}
\caption{
One-dimensional surface density profiles for six representative satellites from our survey. The subsample of objects shown here includes 
three classical dwarf galaxies (Fornax, Carina, Leo T), a globular cluster (Palomar 3) and two ultra-faint dwarfs (Hercules and  Pisces II). 
Total luminosities (given in parentheses) decrease monotonically from {\it panels~(a)} to {\it (f)}. In each panel, the best-fit (two-dimensional) 
S\'ersic, King, Plummer and exponential models are shown by the red, green, magenta and blue profiles, respectively.
} \label{demo}
\end{figure}

As described in \S\ref{sec:diag}, we can use the structural parameters derived using our maximum likelihood
method to generate a one-dimensional surface density profile for each of our program objects. In 
Figure~\ref{demo}, we show surface density profiles computed in this way for six representative 
objects from our survey. In order of decreasing luminosity, these are: Fornax, Carina, Leo~T, Hercules,
Palomar~3 and Pisces~II. These six satellites span a factor of nearly 5000 in luminosity, and include
three classical dwarf galaxies, two ultra-faint dwarf galaxies and one low-mass globular cluster. Note that this
sample excludes the high surface brightness globular clusters that, as discussed 
in \S\ref{sec:1d}, are well fitted by King or S\'ersic models (but not exponential or Plummer models).

In each panel of Figure~\ref{demo}, we show four different models (i.e., S\'ersic, King, Plummer and exponential) 
having best-fit parameters derived from our two-dimensional analysis. For each of 
the satellites fitted with this maximum likelihood technique, we have computed $\chi^2$ 
values for these four models using the observed density 
profiles and fitted models. We find median $\chi^2$ values of
0.33 (S\'ersic), 0.44 (King), 0.49 (Plummer) and 0.50 (exponential). Thus, the slightly preferred parameterization
for these systems is the S\'ersic model --- consistent with our findings for the high-luminosity globular
clusters examined in \S\ref{sec:1d}. 

This finding should perhaps not come as a surprise, for two reasons. First, unlike the Plummer or exponential 
laws, which have two free parameters, the S\'ersic model has three: a scale density or surface
brightness, a scale radius, and a concentration parameter that serves to change the shape, or
curvature, of the profile. As a result, this model has greater flexibility in reproducing the observed
density profiles of satellites, from classical dSphs to ultra-faint systems and globular clusters. 
Although the King model has three free parameters as well, including
a concentration index that governs the global shape of the profile, it features (by definition) a 
tidal truncation that limits its ability to fit the extended profiles exhibited by some of the satellites. In addition, it is 
now recognized that S\'ersic models are also flexible enough to accurately match to the surface brightness 
profiles of early-type galaxies spanning a wide range in luminosity --- from brightest cluster galaxies down to 
the level of dwarf galaxies with luminosities comparable to the brightest systems in our sample, like Fornax, 
Leo~I or Sculptor (see, e.g., \citealt{ferrarese16a}). 

\begin{figure}
\plotone{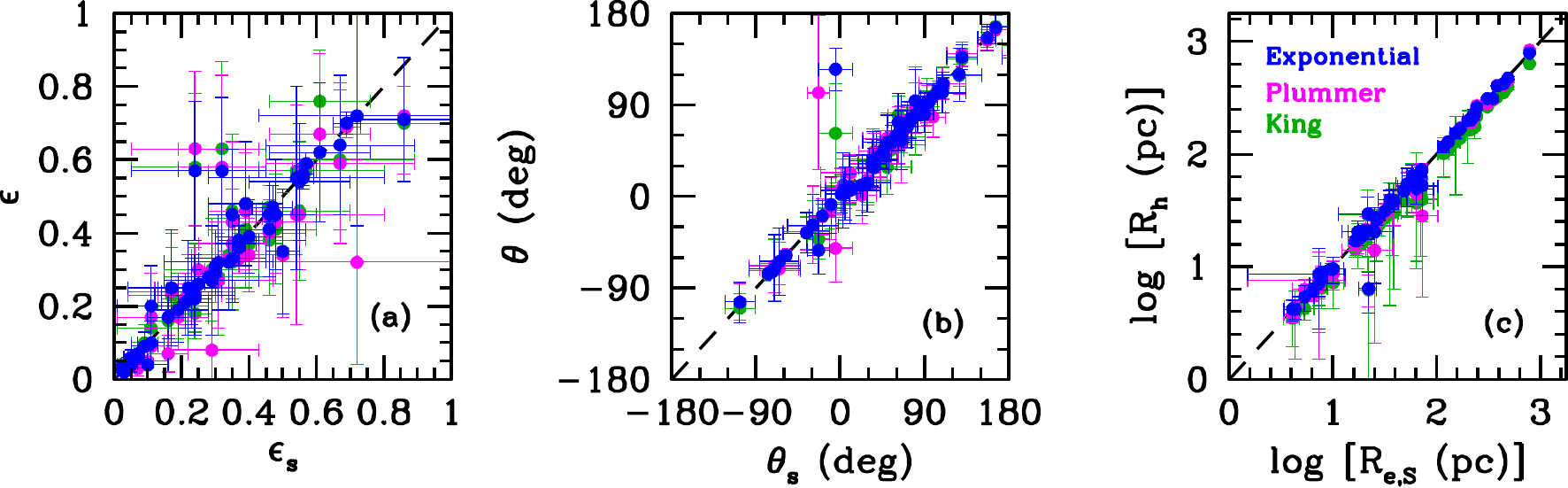}
\caption{
Comparison between the baseline S\'ersic model parameters and those found assuming King, Plummer and exponential 
models (green, magenta and blue points, respectively). {\it Panels (a-c)} show results for ellipticity, position angle and 
effective or half-light radius, respectively. The dashed line in each panel shows the one-to-one relation.
} \label{modelpars}
\end{figure}

For the remainder of this paper, we therefore adopt the  S\'ersic parameters as the default parameters
for our program objects, though structural parameters are also provided for
exponential, Plummer and King fits so that the reader is free to choose from these four options. 
In practice, the precise choice of model has no impact on our conclusions for the 
satellite population as a whole, or for individual systems. For instance, Figure~\ref{modelpars} shows
a comparison between the best-fit S\'ersic parameters (ellipticity, position angle and effective or half-light 
radius) and those found using King, Plummer or exponential models for the 37 objects fitted using
our maximum likelihood approach. There is generally good agreement between the different models.

\subsection{Comparison with Previous Results}
\label{subsec:prev}

Although \S\ref{sec:individ} presents a detailed comparison of our photometric and structural parameters to those
in the literature for each object in our survey, it is useful to begin by comparing our measurements to those 
reported in the most widely used databases for dwarf galaxies and globular clusters. 

\subsubsection{McConnachie (2012)}

The most extensive compilation of photometric and structural parameters for Local Group galaxies remains that of 
\citet{mcconnachie12a}. This catalog includes basic information for about two dozen 
Galactic dwarf galaxies. Note that while the McConnachie catalog has been updated since publication a number
of the most recently discovered Galactic satellites --- such as those uncovered in recent DES, Pan-STARRS and
Subaru imaging surveys --- are not included in the current database. 

\begin{figure}
\plotone{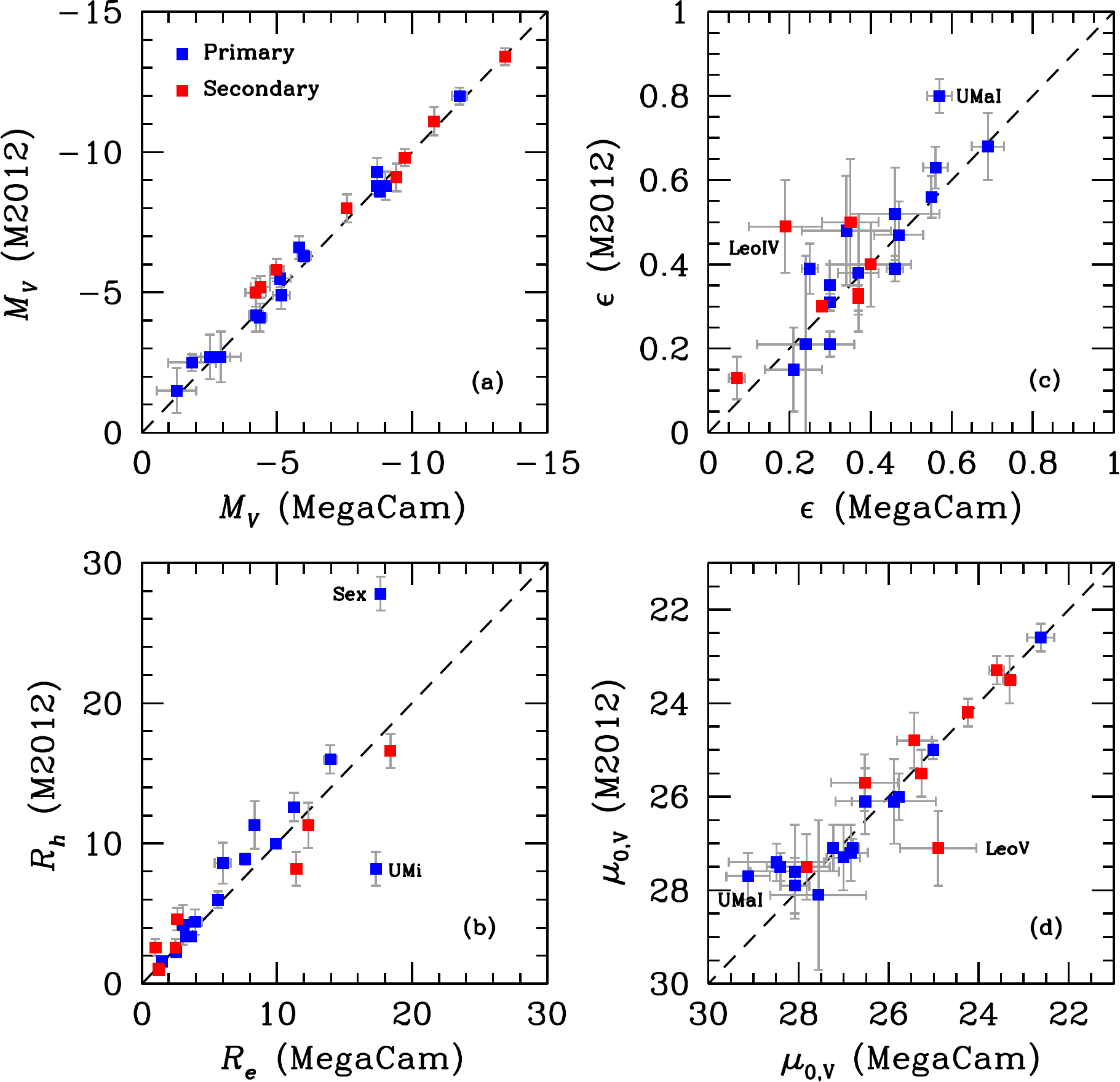}
\caption{
Comparison of the best-fit S\'ersic photometric and structural parameters for ``classic" and ultra-faint dwarf galaxies in 
our survey with parameters for Local Group dwarfs galaxies taken from the compilation of \citet{mcconnachie12a}. The four 
panels, in clockwise order beginning at the upper left, compare absolute $V$-band magnitude, ellipticity, half-light 
vs. effective  radius, and central surface brightness. A total of 23 satellites are shown in this figure --- 15 and 8
objects  belonging to our primary and secondary samples, respectively.
} \label{comp1}
\end{figure}

In Figure~\ref{comp1} we compare our best-fit parameters to those tabulated in \citet{mcconnachie12a}.
Panels $(a)$ to $(d)$ of this figure show the results for four key parameters: absolute magnitude, $M_V$, 
effective  (or half-light) radius, $R_e$, mean ellipticity, $\epsilon$, and central surface brightness, $\mu_{0,V}$.
The dashed line in each panel shows the one-to-one relation.  A total of 23 objects are shown in 
Figure~\ref{comp1}, of which 15 and eight objects belong to our primary and secondary samples (blue 
and red symbols, respectively). Note that some caution must be exercised in these comparisons because 
the \citet{mcconnachie12a} measurements were standardized assuming exponential profiles whereas 
our parameters come from the best-fit S\'ersic models (cf., \S\ref{sec:critmodels}). 

This caveat aside, the comparisons in Figure~\ref{comp1} show very good agreement apart from a few 
outliers.  Most notably, our measured effective radii for Sextans and Ursa Minor differ significantly from those 
given in \citet{mcconnachie12a}. For Sextans, we measure $R_e = 17\farcm67\pm0\farcm17$ --- much smaller 
than the value of $27\farcm8\pm1\farcm2$ in \citet{mcconnachie12a}. For Ursa Minor, we find $R_e = 17\farcm32\pm0\farcm11$, 
which is more than double the value of $8\farcm2\pm1\farcm2$  given in \citet{mcconnachie12a}. 
In both cases, though, the \citet{mcconnachie12a} values were taken from the photographic survey of \citet{irwin95a}.

Ursa Major~I is a slight outlier in panel $(c)$, where we measure an ellipticity of $\epsilon = 0.57\pm0.03$.
This is significantly lower than the value of $0.80\pm0.04$ from \citet{mcconnachie12a} which is, in turn, 
based on relatively shallow SDSS star counts analyzed by \citet{martin08a}. Using much deeper 
Suprime-Cam imaging, \citet{okamoto08a}
found an ellipticity of $0.54$, which is in excellent agreement with our value. 
Finally, in panel $(d)$, we see a
slight (1.2$\sigma$) discrepancy between our central surface brightness measurement for Leo~V, $\mu_{0,V} \simeq 24.9^{+0.90}_{-0.79}$
mag~arcsec$^{-2}$, and that given in \citet{mcconnachie12a}, $\mu_{0,V} \simeq 27.1\pm0.8$ mag~arcsec$^{-2}$, which 
is based on the the analysis of \citet{dejong10a}. The origin and significance of this disagreement is unclear,
since the photometric catalogs used in our analysis and in \citet{dejong10a} have comparable depth 
and areal coverage, although we note that Leo~V is located at the field edge in the latter study.
We note that the analysis of Leo~V by \citet{sand12a} using deeper data yields a central surface brightness
of $\mu_{0,V} \sim 26.3$.

\subsubsection{Harris (1996, 2010)}

\begin{figure}
\plotone{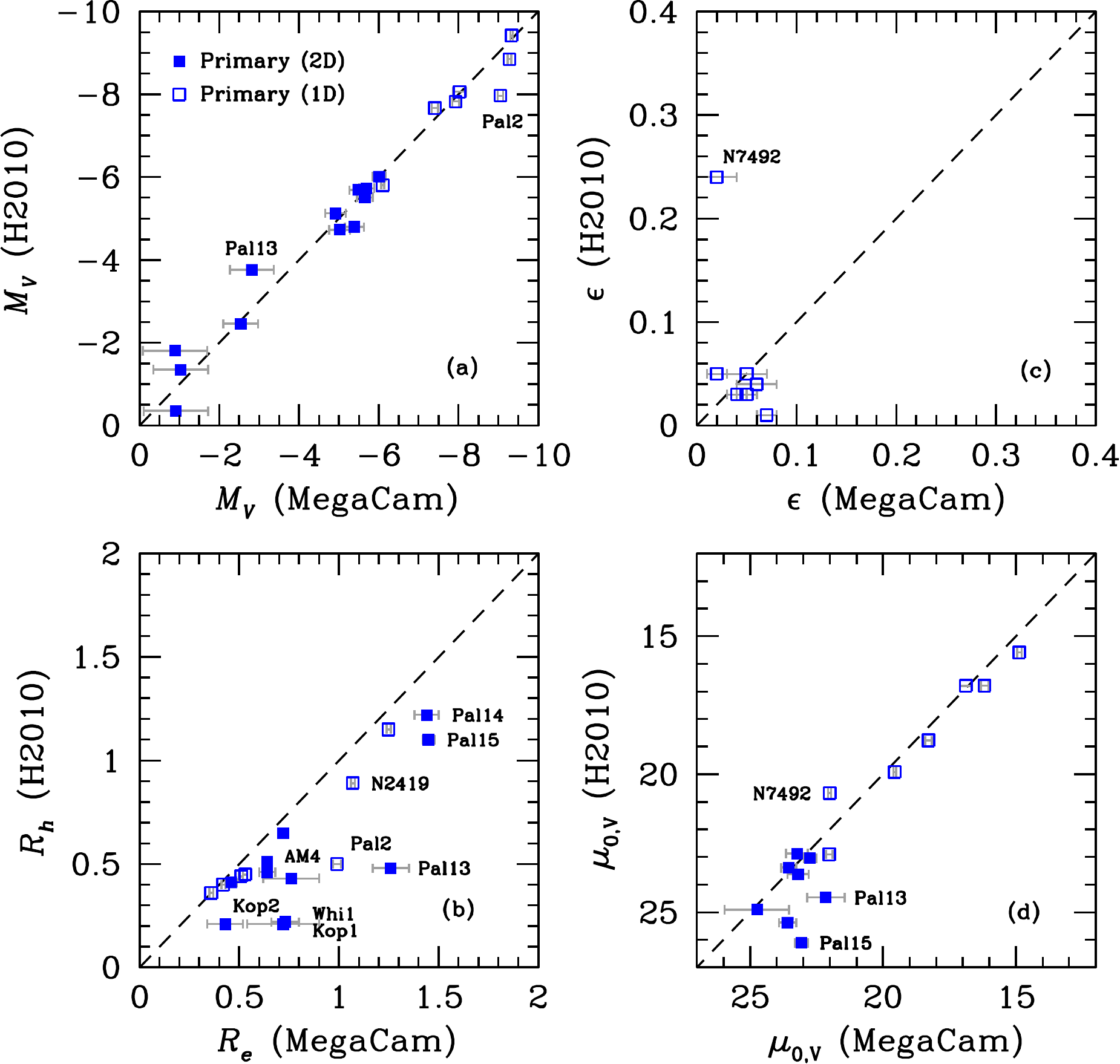}
\caption{
Comparison of the best-fit S\'ersic photometric and structural parameters for halo clusters in our survey with those 
in the 2010 version of the \citet{harris96a} catalog of Galactic globular clusters. The four panels, in clockwise
order beginning at the upper left, compare absolute $V$-band magnitudes, ellipticities, King half-light vs. S\'ersic
effective radius, and central surface brightness. A total of 19 globular clusters are shown in this figure.
} \label{comp2}
\end{figure}

The most widely used catalog of Galactic globular clusters is that of \citet{harris96a}.
Figure~\ref{comp2} compares our photometric and structural parameters for the
19 objects that appear in the 2010 version of this catalog. 
This comparison sample consists of 12 low surface brightness objects with parameters derived from our two-dimensional
maximum likelihood method (\S\ref{sec:2d}), and seven high surface brightness objects whose parameters 
were derived from a one-dimensional analysis as described in \S\ref{sec:1d}. Filled and
open squares indicate the low and high surface brightness clusters, respectively.

As was the case in Figure~\ref{comp1}, the four panels of this figure show absolute magnitude, $M_V$, effective
(or half-light) radius, $R_e$, mean ellipticity, $\epsilon$, and central surface brightness, $\mu_{0,V}$. Note that 
most of the Harris magnitude, size and surface brightness measurements are, in fact, taken from 
the study of \citet{mclaughlin05a} who fitted King models to the (rather hetereogeneous) star count
and surface brightness data of \citet{trager95a}. For the remote halo clusters that are the focus
of this study, the \citet{trager95a} profiles often come from a variety of photographic, photoelectric and
CCD sources, sometimes taken in different filters, and having differing depths and resolution. 

For the most part, we find reasonable agreement between our S\'ersic measurements and those
tabulated in the \citet{harris96a} catalog. In terms of absolute magnitude, there is quite good 
agreement, although our magnitude for Palomar~2 ($M_V \simeq -9$) is significantly brighter 
than the \citealt{harris96a} value ($M_V \simeq -8$). However, integrated light
measurements for this object are problematic given its low Galactic latitude ($b \simeq -9^\circ$) and 
large and variable reddening. The two sets of ellipticity measurements are also in good agreement, 
although \cite{harris96a} reports a rather high ellipticity, $\epsilon = 0.24$, for NGC7492. We
find a much smaller value, $\epsilon =  0.02\pm0.02$, as would be expected from the 
round isopleths shown in Figure~\ref{dens11}. 

Care must be taken when comparing central surface brightness measurements. The values given in 
\citet{harris96a} are based on King model fits, which have isothermal cores, so these may not be directly
comparable to the inward extrapolations of S\'ersic models (i.e., the values reported in Table~\ref{t:mags}). 
For the seven high surface brightness clusters, we therefore rely on median (non-parametric) values within a 
radius of 5\arcsec. All in all, there is reasonable agreement between the surface brightness measurements, 
although our values for Palomar 13, 14 and 15 are  somewhat brighter than suggested by the earlier 
\citet{trager95a} data. We note that, in the case of Palomar 13, our estimate of $\mu_{0,V} \simeq 22.15\pm0.70$ 
mag arcsec$^{-2}$ is in agreement with the value of $\mu_{0,V} = 22.54^{+0.20}_{-0.17}$ mag arcsec$^{-2}$ given
in \citet{cote02a}.

Finally, Figure~\ref{comp2} shows that our effective radii are often 
larger than the half-light values reported in \cite{harris96a}. 
The most discrepant objects are
Whiting~1 ($0\farcm73$ vs. $0\farcm22$),
 Koposov~1 ($0\farcm72$ vs. $0\farcm21$),
 Palomar~13 ($1\farcm26$ vs. $0\farcm48$),
Palomar~2 ($0\farcm99$ vs. $0\farcm5$) and 
 AM4 ($0\farcm76$ vs $0\farcm43$).
 We have carefully inspected our fits and found that the larger
values are in fact much more consistent with the deeper Megacam data than
the Harris ones.
In the case of Whiting~1 and AM4, the Harris value was taken from \citet{carraro07a}
and \citet{carraro09a}, respectively. In both cases, the light distribution appeared
to extend out to several arcmin following a shallow slope in the outer region
and was thus fitted using a King plus power law profile to account for the extended light 
distribution.  Our photometry is at least two magnitudes deeper in both cases. We do 
not see an obvious break in the light distribution and consequently our best fit S\'ersic 
profiles yield larger half-light radii.  In Palomar~2, the central stellar density is 
severely affected by a dust lane crossing the cluster, which significantly modifies
the light distribution, especially that of the brighter stars.  To derive its structural 
parameters we excluded the brighter giants, which results in a  larger half-light
radius.  

The case of Palomar~13 is interesting: our effective radius is
roughly three times larger than the Harris value, though we note that Harris
half-light radius was not measured directly  but instead estimated based on 
the cluster's core radius following the expression $\log(r_{h}/r_{c})=0.6c-0.4$, 
where $c$ is the concentration index. Our best-fit King
core radius is $r_{c}\sim0\farcm31$\, which is similar to previous determinations 
\citep{siegel01a, cote02a}. We conclude that our larger effective radius is due to
the fact that a King profile is not a good description of Palomar~13's unusual
surface brightness profile distribution, which
is better matched by a S\'ersic profile with a larger effective radius. 

\subsection{Diagnostic Tools for Program Objects}
\label{sec:diag}

The stellar catalogs generated from our wide-field $gr$ images form the basis of our maximum likelihood analysis and
can be used not only to estimate photometric and structural parameters, but also to help us understand the 
nature of each program object. Here, we briefly describe the diagnostic tools used to analyze each
satellite.

\subsubsection{Isodensity Contour Maps and One-Dimensional Density Profiles}

To examine the two-dimensional morphologies of our program objects in a systematic 
way, we created smoothed isodensity contour maps using the same
``cleaned" photometric catalogs that were used to derive structural parameters. These 
catalogs consist of star-like objects identified by applying DAOPHOT
$\chi$ and $sharp$ cuts and selected to fall in regions of the CMD delineated by
the adopted best-fit isochrone. For objects with more than one star formation
episode, like some of the classical dwarf galaxies, we also included stars that fall in other
regions of the CMD that were not covered by the best fitting isochrone.

\begin{figure}
\plotone{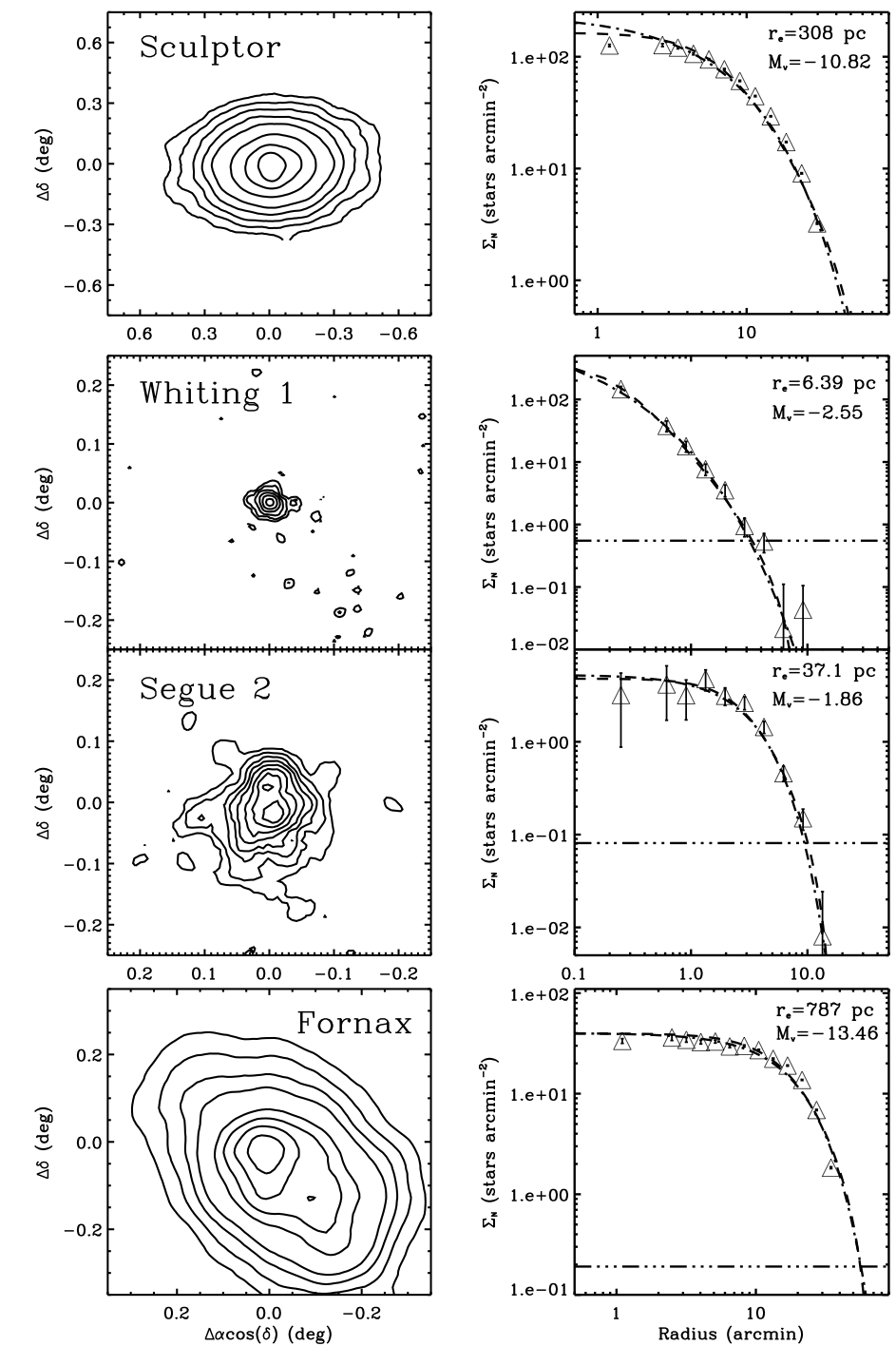}
\caption{Isodensity contour maps {\it (left panels)} and radial number density profiles {\it (right panels)} for four of 
our program objects: Sculptor, Whiting~1, Segue~2 and Fornax. For the number density profiles, 
the dashed and dotted-dashed curves show the best-fit King and S\'ersic models, while the horizontal line in each
panel shows the fitted background level. Note that these one-dimensional
profiles were produced using the best-fit parameters from the maximum likelihood analysis and do not
represent the best-fit parameters for the binned number density profiles.}
\label{dens1}
\end{figure}

\begin{figure}
\plotone{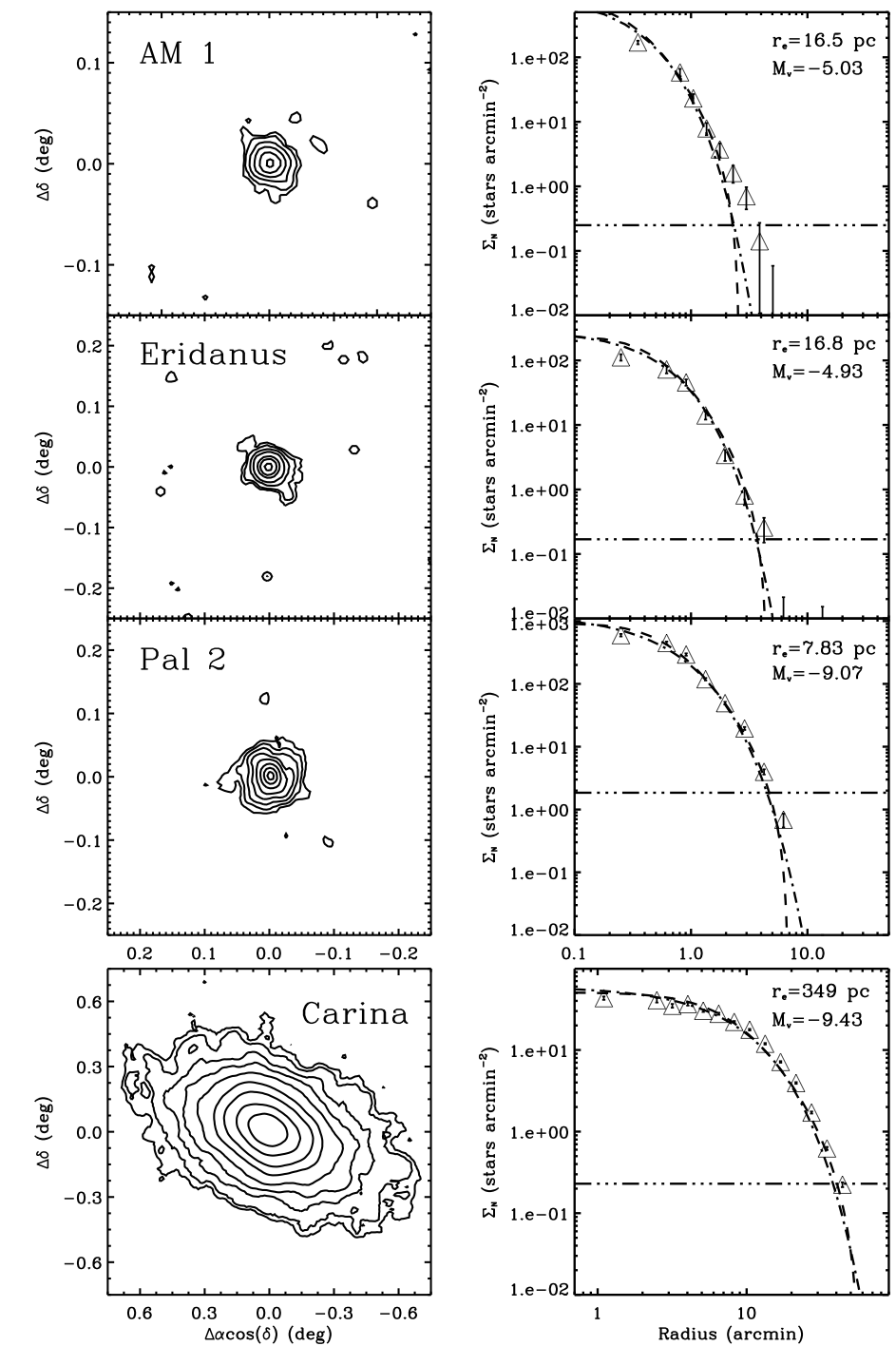}
\caption{Same as Figure~\ref{dens1}, except for AM~1, Eridanus, Palomar~2 and Carina.}
\label{dens2}
\end{figure}

\begin{figure}
\plotone{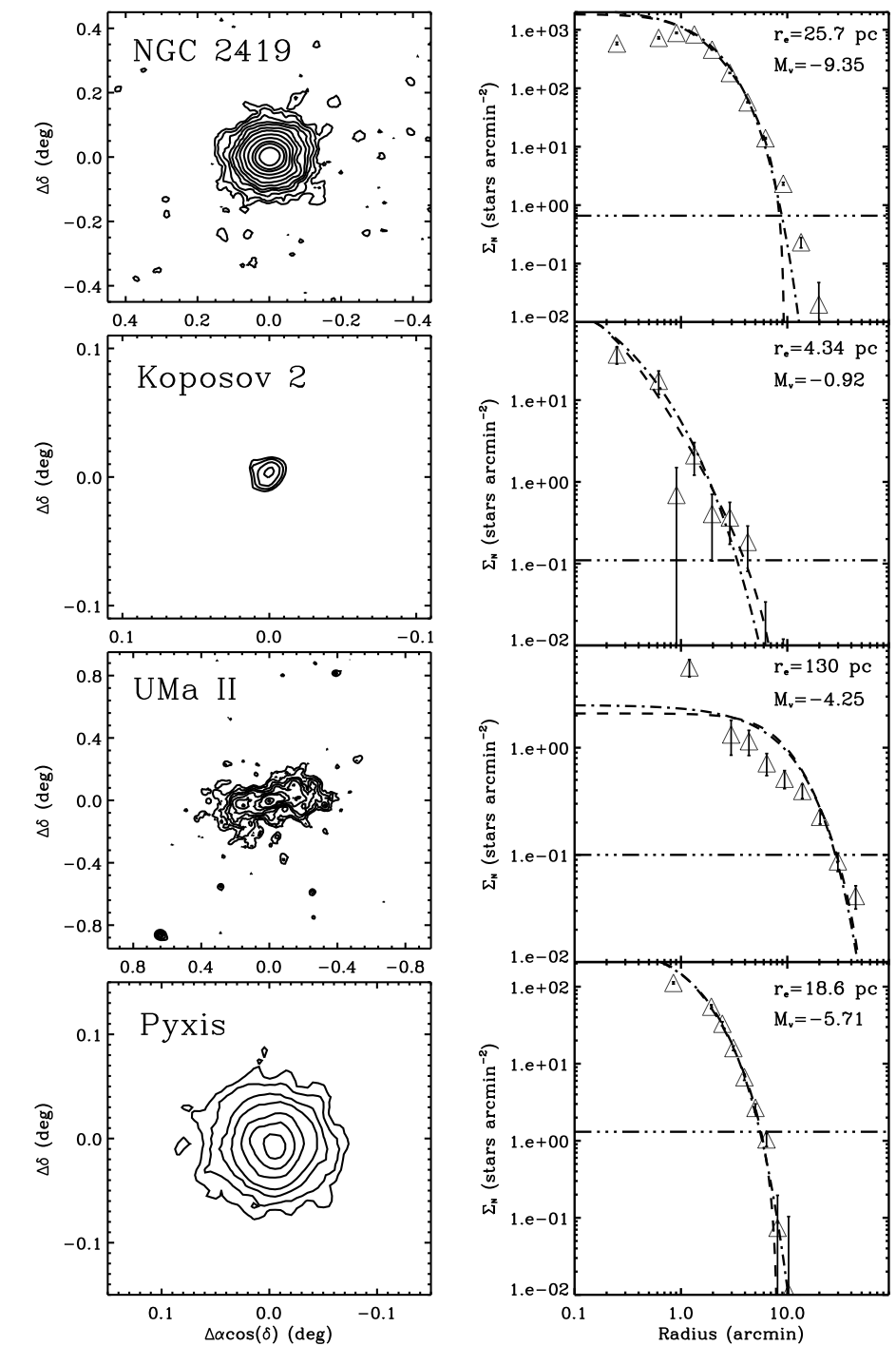}
\caption{Same as Figure~\ref{dens1}, except for NGC2419, Koposov~2, Ursa Major~II and Pyxis.}
\label{dens3}
\end{figure}

\begin{figure}
\plotone{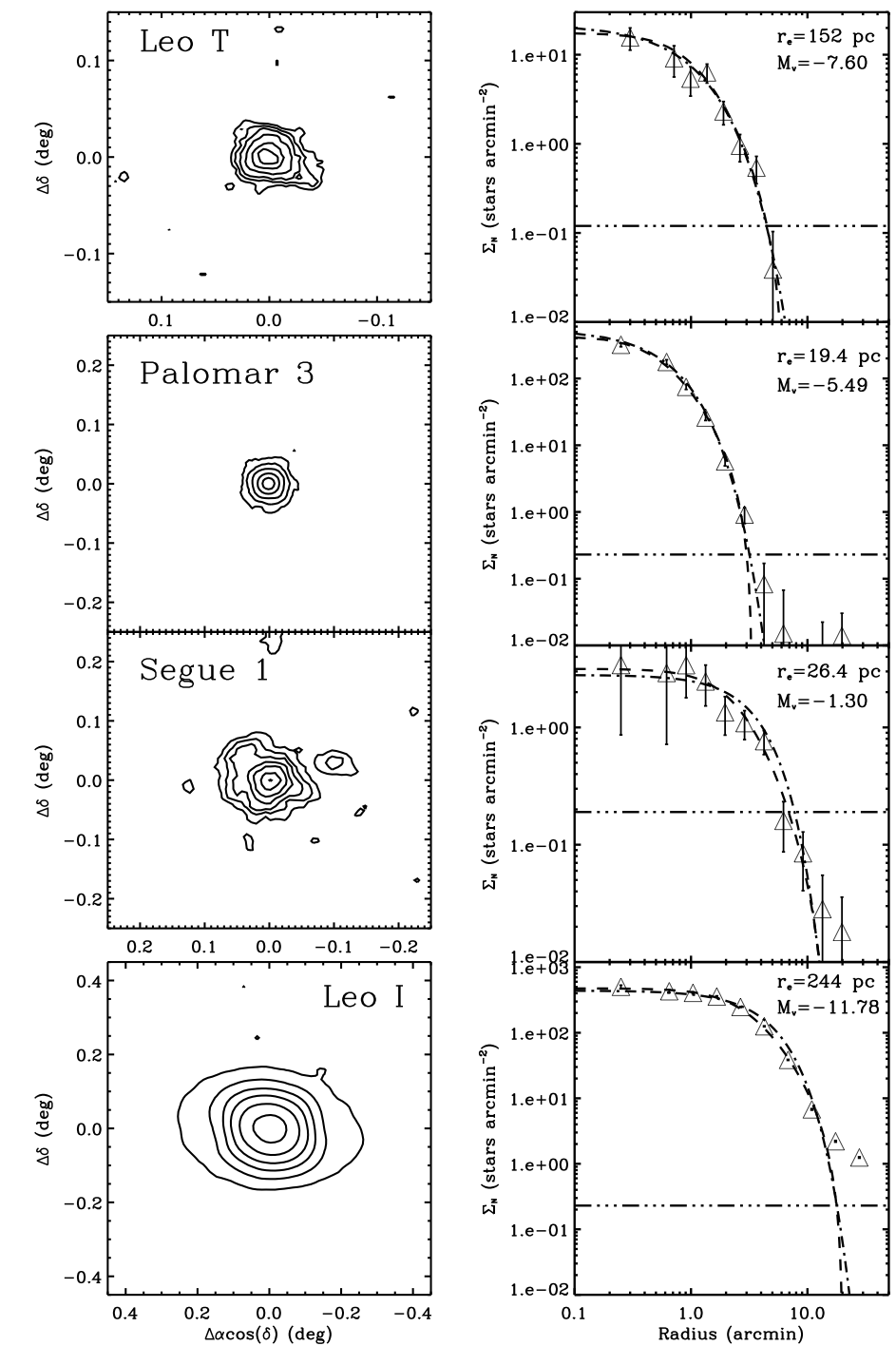}
\caption{Same as Figure~\ref{dens1}, except for Leo~T, Palomar~3, Segue~1 and Leo~I.}
\label{dens4}
\end{figure}

\begin{figure}
\plotone{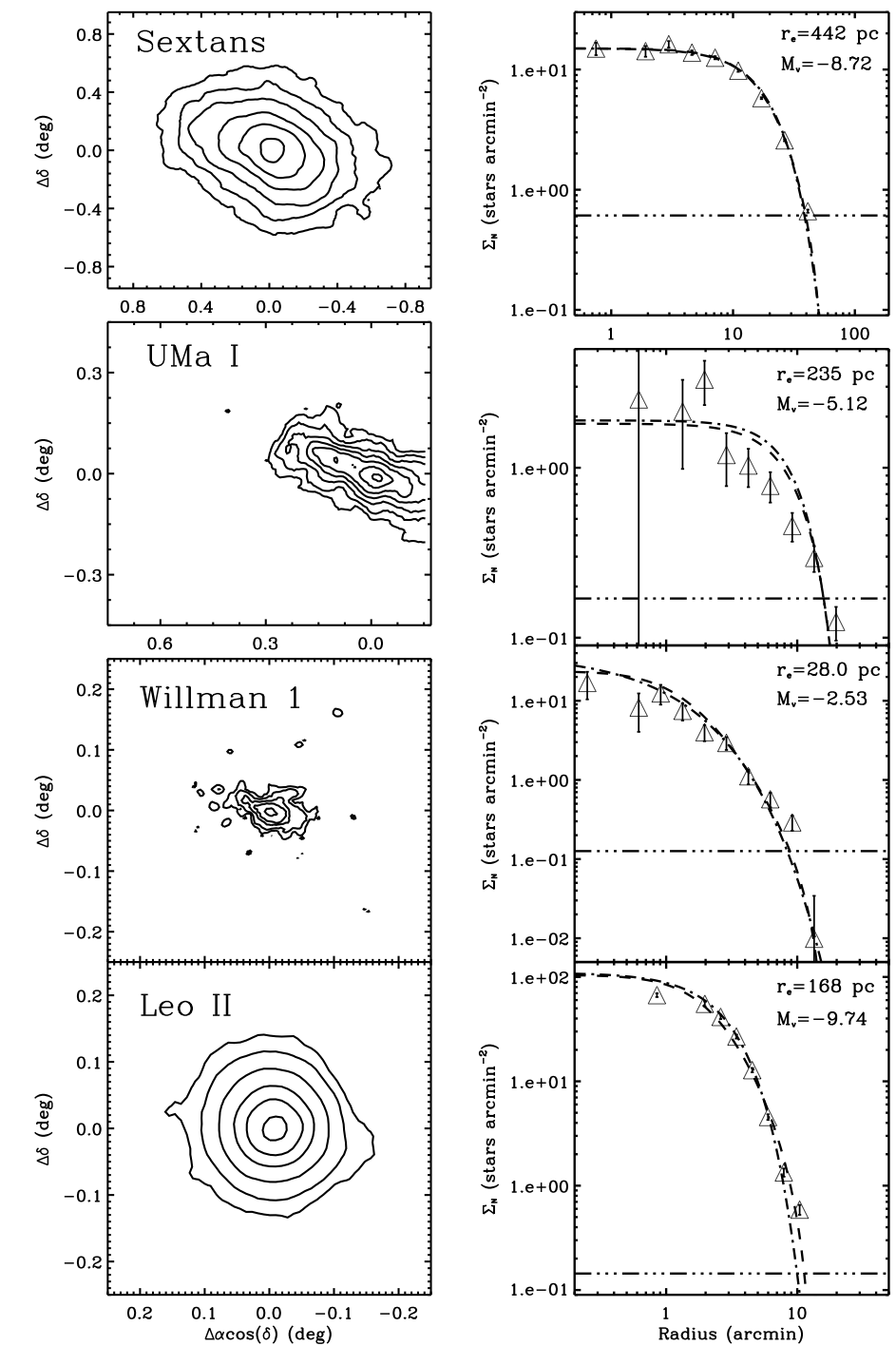}
\caption{Same as Figure~\ref{dens1}, except for Sextans, Ursa Major~I, Willman~1 and Leo~II.}
\label{dens5}
\end{figure}

\begin{figure}
\plotone{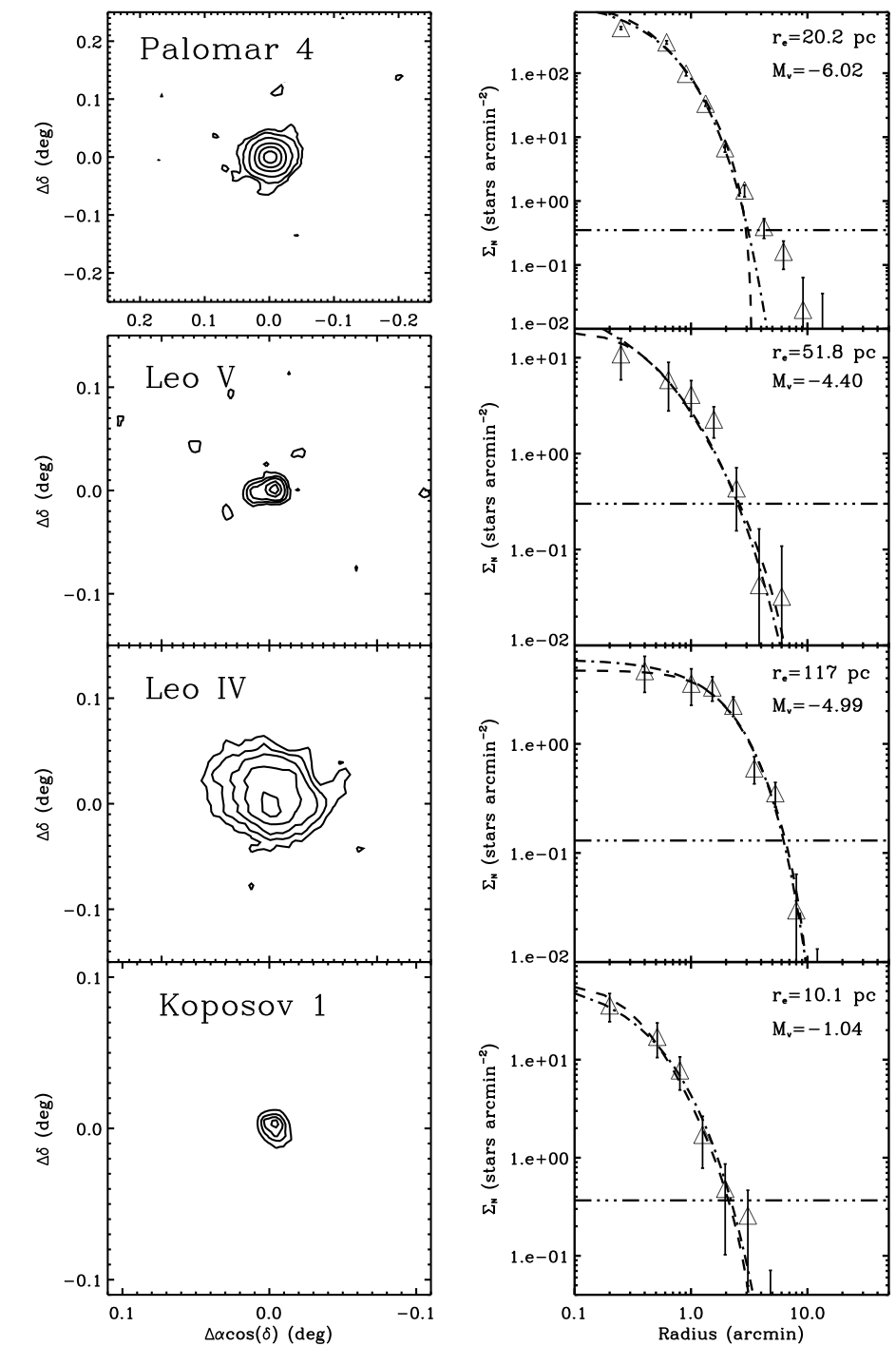}
\caption{Same as Figure~\ref{dens1}, except for Palomar~4, Leo~V, Leo~IV and Koposov~1.}
\label{dens6}
\end{figure}

\begin{figure}
\plotone{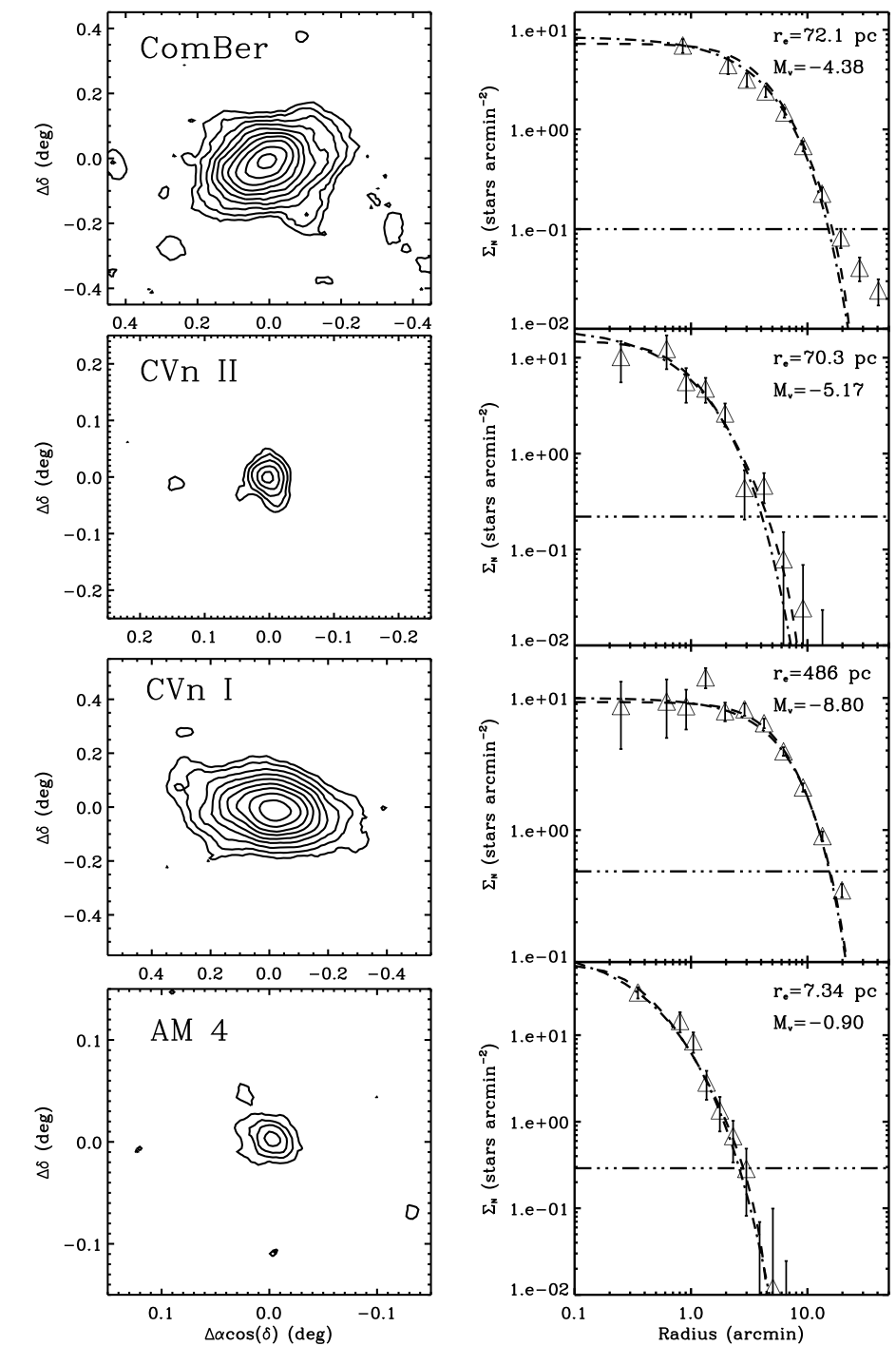}
\caption{Same as Figure~\ref{dens1}, except for Coma Berenices, Canes Venatici~II, Canes Venatici~I and AM~4.}
\label{dens7}
\end{figure}

\begin{figure}
\plotone{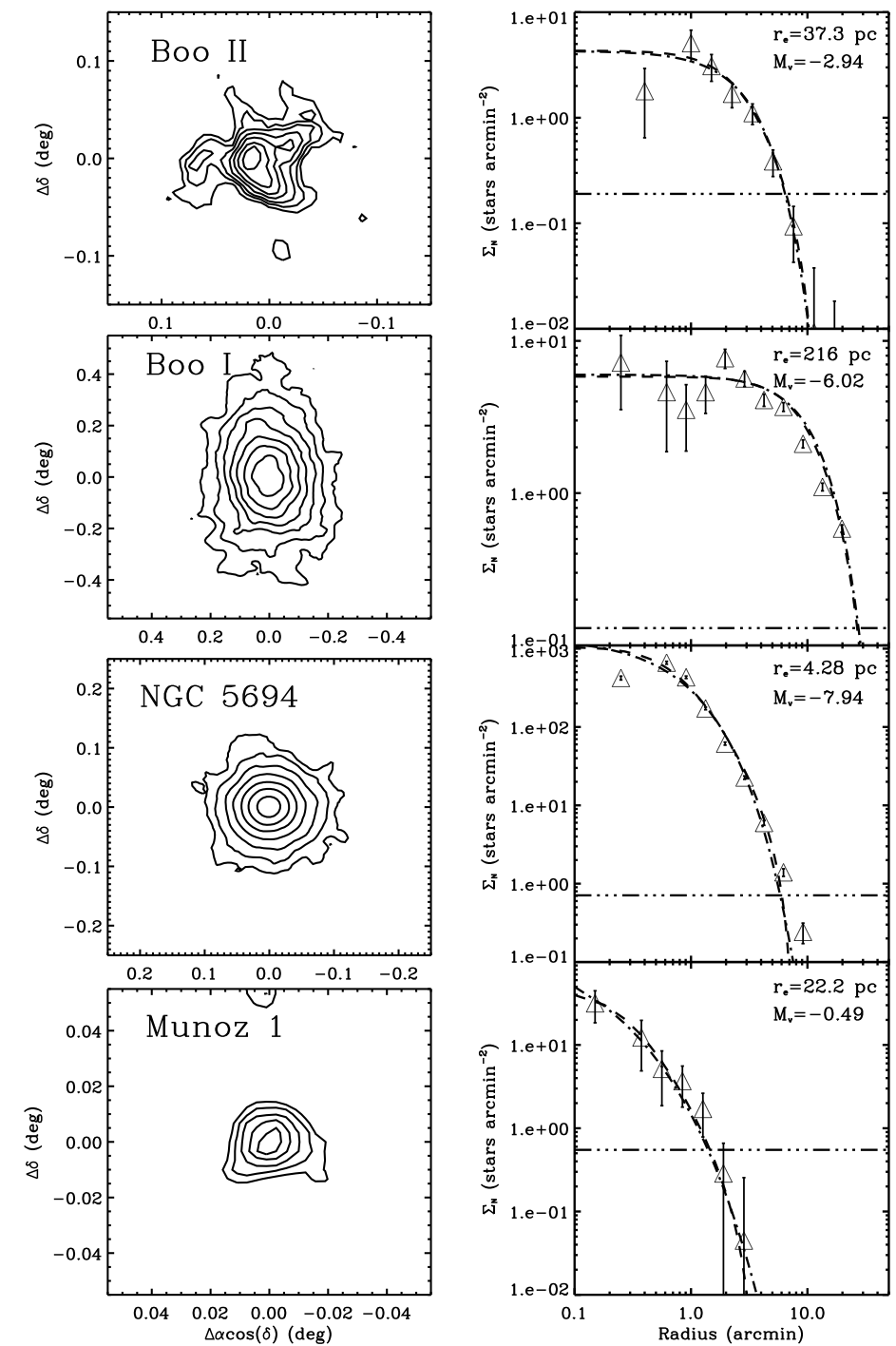}
\caption{Same as Figure~\ref{dens1}, except for Bootes~II, Bootes~I , NGC5694 and Mu\~noz~1.}
\label{dens8}
\end{figure}

\begin{figure}
\plotone{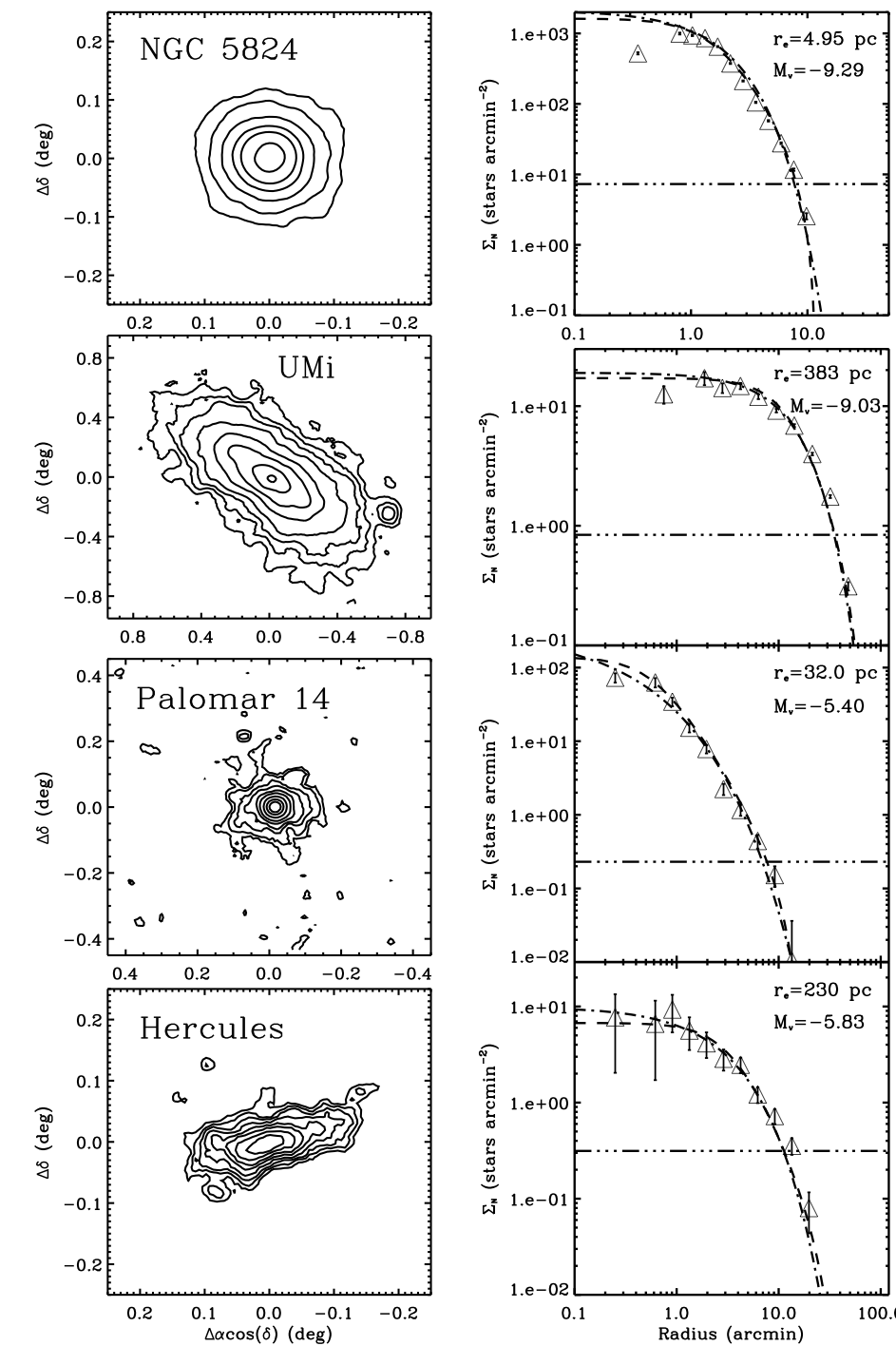}
\caption{Same as Figure~\ref{dens1}, except for NGC5824, Ursa Minor, Palomar~14 and Hercules.}
\label{dens9}
\end{figure}

\begin{figure}
\plotone{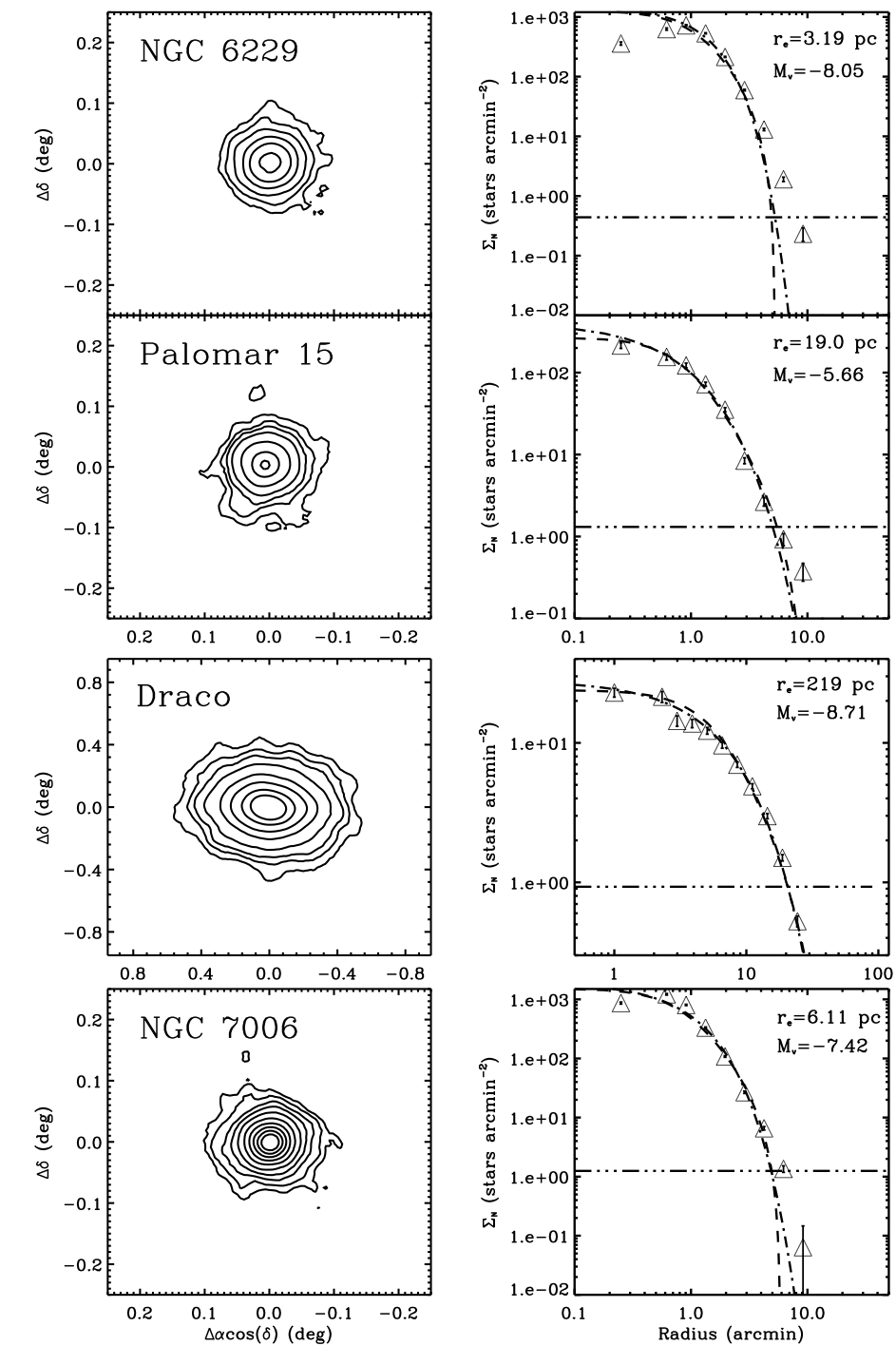}
\caption{Same as Figure~\ref{dens1}, except for NGC6229, Palomar~15, Draco and NGC7006.}
\label{dens10}
\end{figure}

\begin{figure}
\plotone{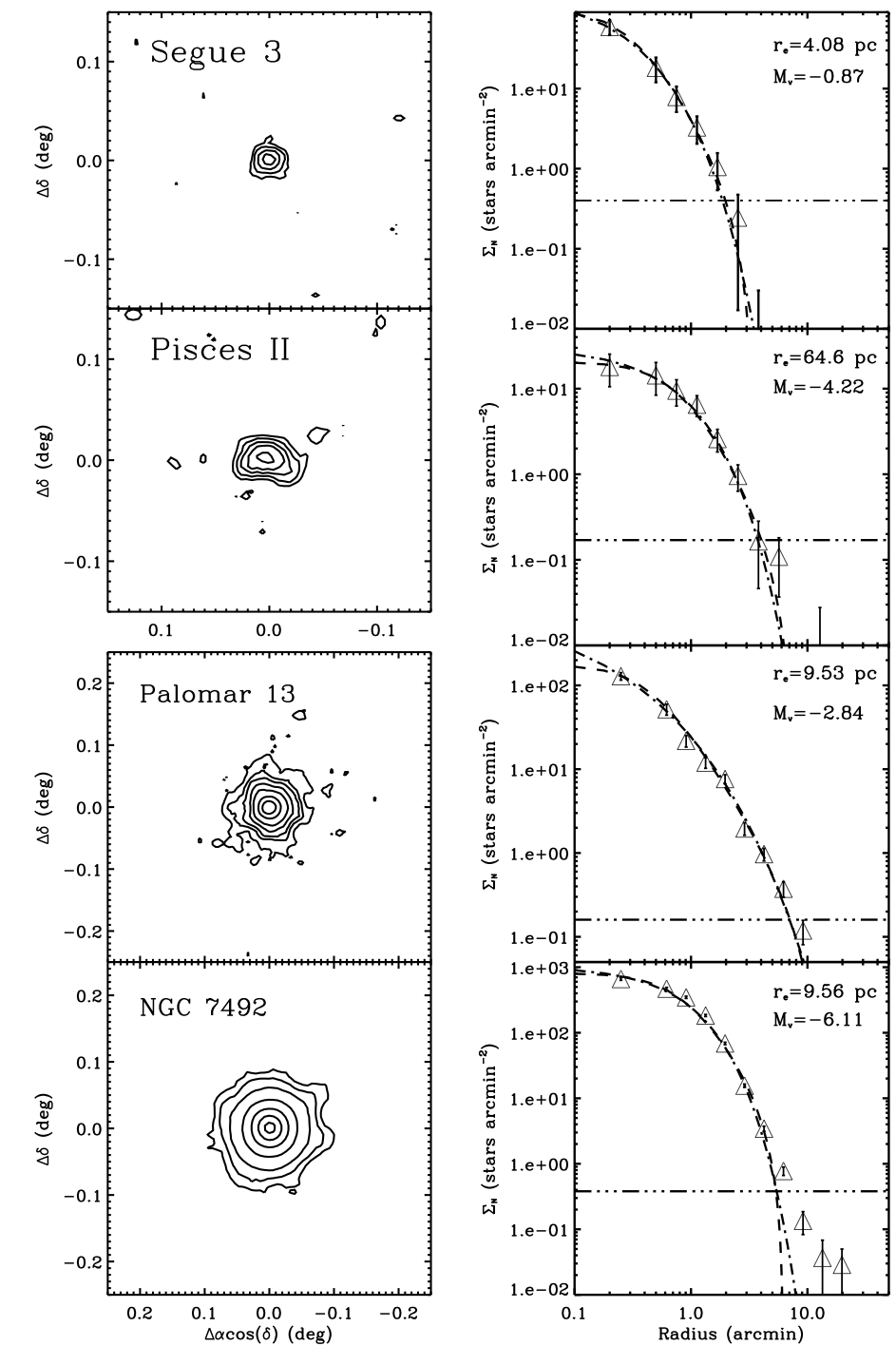}
\caption{Same as Figure~\ref{dens1}, except for Segue~3, Pisces~II, Palomar~13 and NGC7492.}
\label{dens11}
\end{figure}

To construct density maps, we count stars in bins that are subsequently spatially
smoothed with an exponential filter. The bin size and scale of the smoothing
exponential vary depending on the angular size of the object, and range from
$40\arcsec \times40\arcsec$ bins for the classical dwarfs to $10\arcsec \times10\arcsec$ bins
for the most compact satellites. The resulting contour maps are shown in the left panels of 
Figures~\ref{dens1} through \ref{dens11} and are discussed on a case-by-case basis in
\S\ref{sec:individ}. 

Using the photometric and structural parameters derived in \S\ref{sec:method}, we can also
generate radial stellar density profiles for all objects.  The stellar densities were measured by 
counting stars within concentric annuli of fixed ellipticity  centered around $\alpha_{\rm 0}$ 
and $\delta_{\rm 0}$ and oriented according to the derived position angle $\theta$. Uncertainties 
were calculated from Poisson statistics. The right panels of Figures~\ref{dens1} through \ref{dens11}
show background-subtracted, one-dimensional density profiles for all objects as well as the 
respective best-fit King and S\'ersic models. We remind the reader that these are not profiles 
fitted to the binned densities, but generated with the parameters derived using all of the detected stars.

\subsubsection{Star Count Maps and Color-Magnitude Diagrams}

For completeness, we show in the left panels of  Figure~\ref{cmds1}--\ref{cmds11} star count 
maps for the individual satellites. Although the isodensity contour maps shown in 
Figures~\ref{dens1} through \ref{dens11} are based on star count maps, the latter have
not been smoothed and so offer a different perspective on the two-dimensional structure
of each satellite. In the right panels of Figures~\ref{cmds1}--\ref{cmds11}, we show CMDs
for our program objects constructed from the catalog of star-like sources identified in
each object. 

\bigskip

\subsection{Analysis of Individual Systems}
\label{sec:individ}

In the following subsections, we briefly summarize the results shown in Figures~\ref{dens1} 
through \ref{cmds11} for each of the 44 objects in our primary sample, highlighting some 
relevant aspects of the individual targets. The systems are ordered in terms of increasing 
right ascension.

\subsubsection{Sculptor}

Sculptor was the first of the Milky Way's dSph galaxies to be discovered \citep{shapley38a}.
\citet{hodge61a} carried out 
the first detailed star-count analysis of this satellite, finding a smooth, slightly elongated distribution of 
stars 
that did not follow the models 
traditionally used to describe the luminosity profiles of elliptical galaxies.

\begin{figure}
\plotone{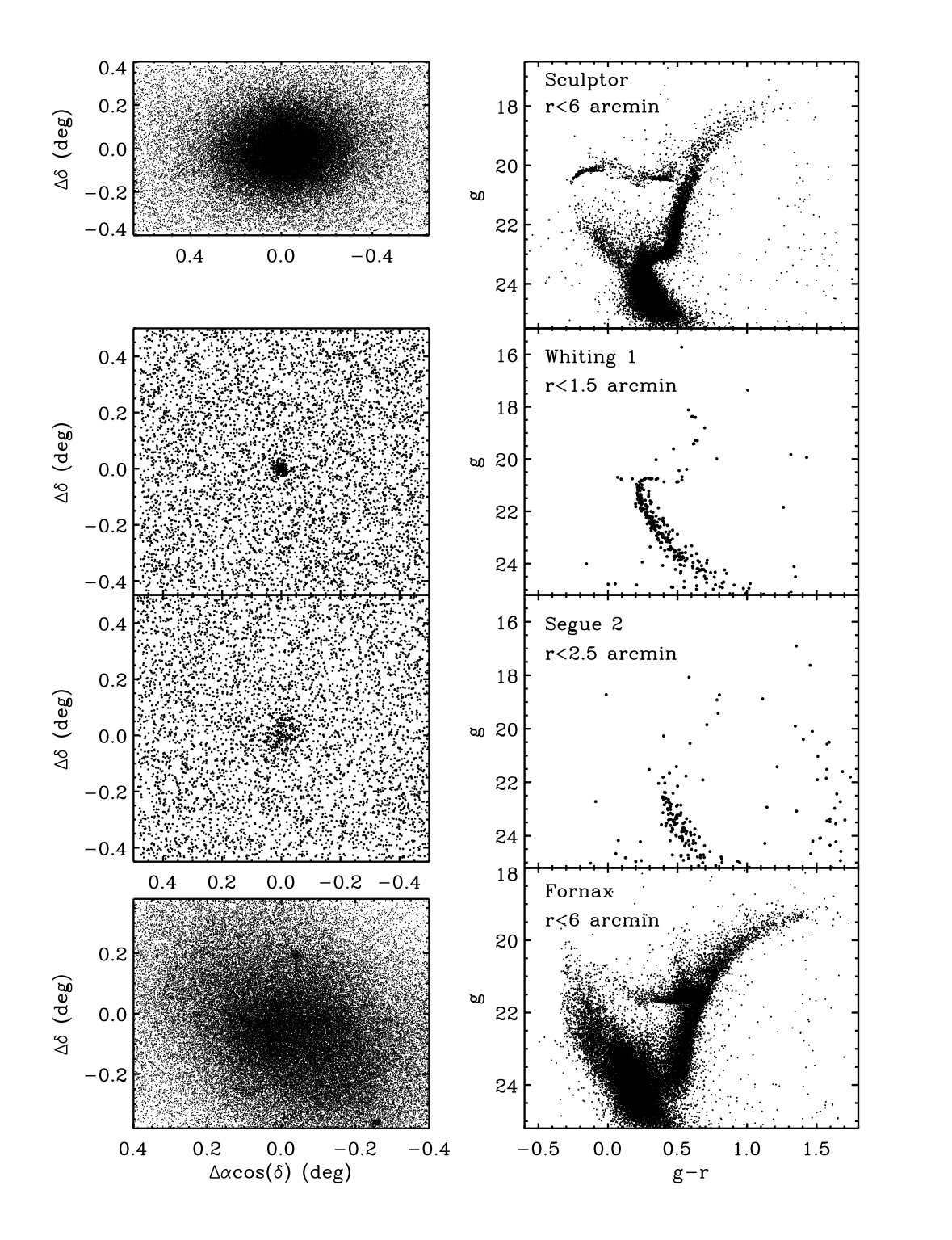}
\caption{Star count maps ({\it left panels}) and color-magnitude diagrams ({\it right panels}) for four of 
our program objects: Sculptor, Whiting~1, Segue~2 and Fornax.}
\label{cmds1}
\end{figure}

\begin{figure}
\plotone{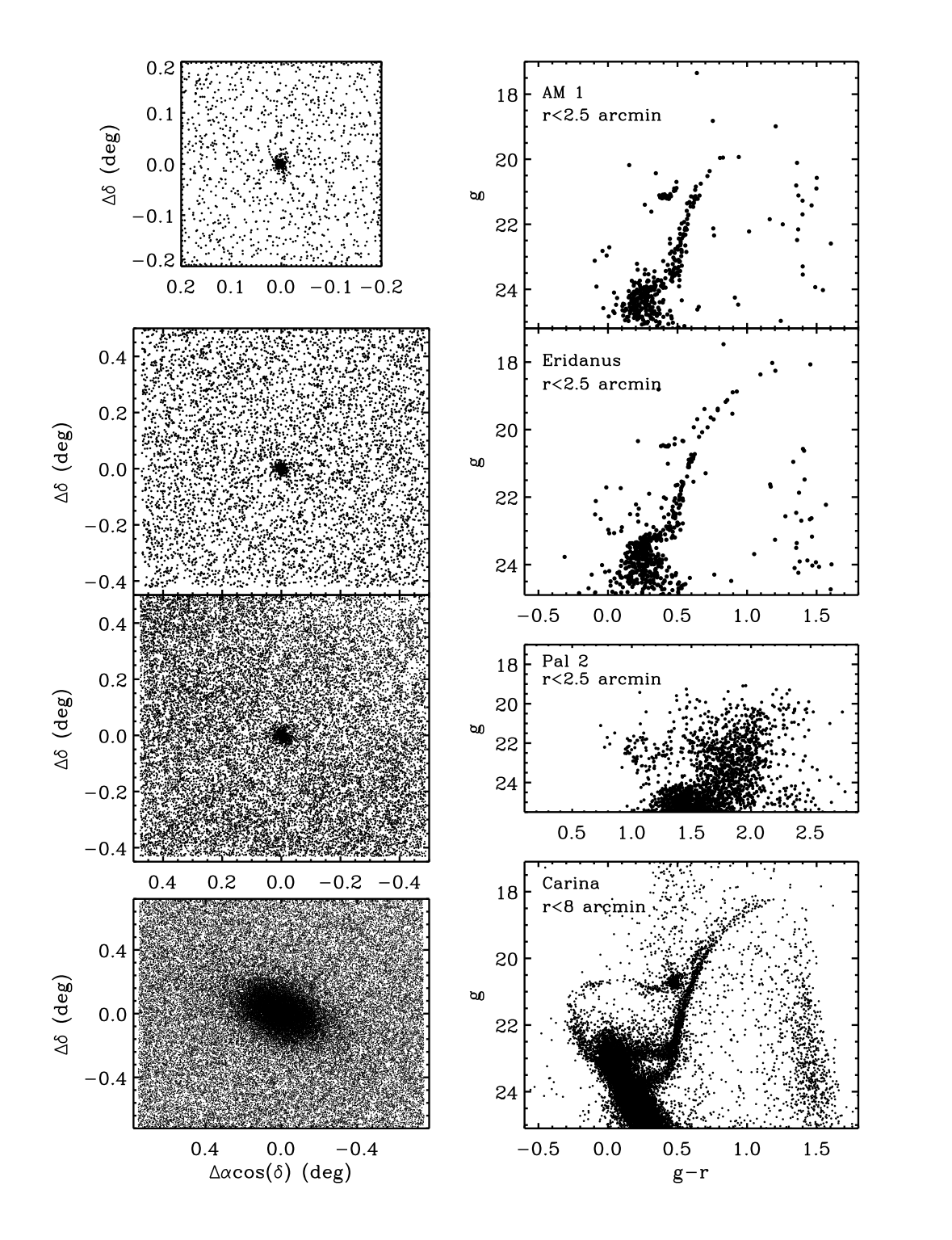}
\caption{Same as Figure~\ref{cmds1}, except for AM~1, Eridanus, Palomar~2 and Carina.}
\label{cmds2}
\end{figure}

\begin{figure}
\plotone{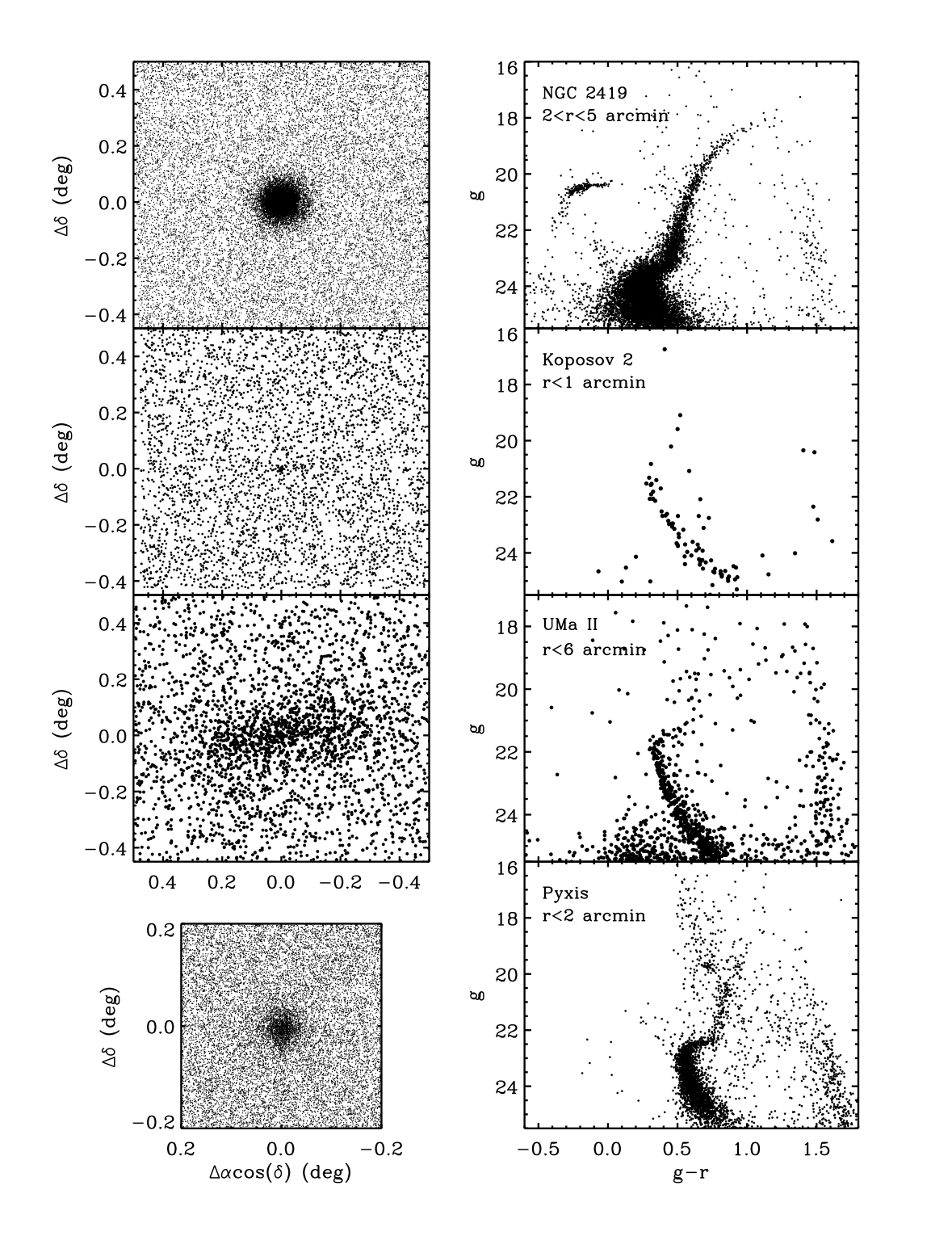}
\caption{Same as Figure~\ref{cmds1}, except for NGC2419, Koposov~2, Ursa Major~2 and Pyxis.}
\label{cmds3}
\end{figure}

\begin{figure}
\plotone{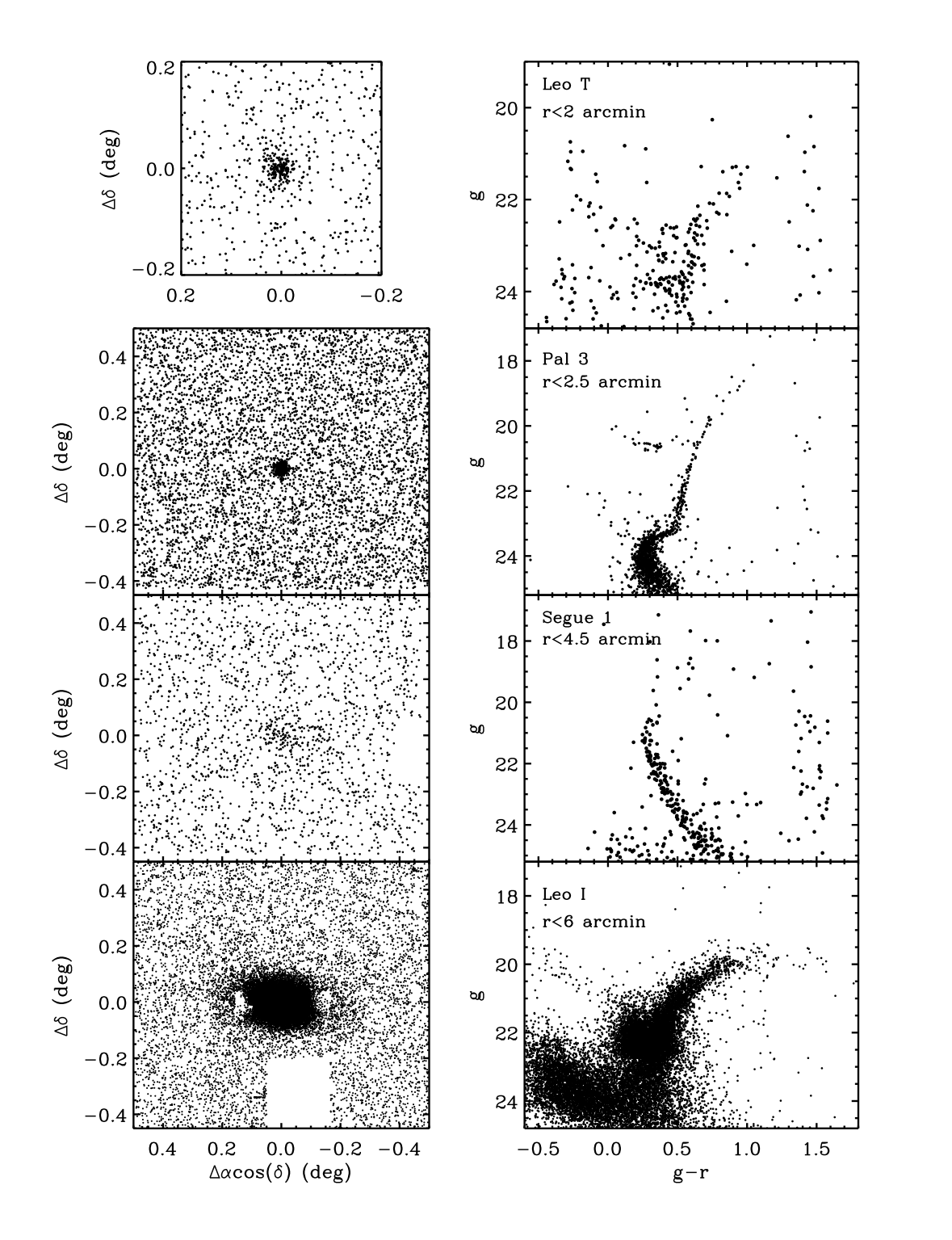}
\caption{Same as Figure~\ref{cmds1}, except for Leo~T, Palomar~3, Segue~1 and Leo~I.}
\label{cmds4}
\end{figure}

\begin{figure}
\plotone{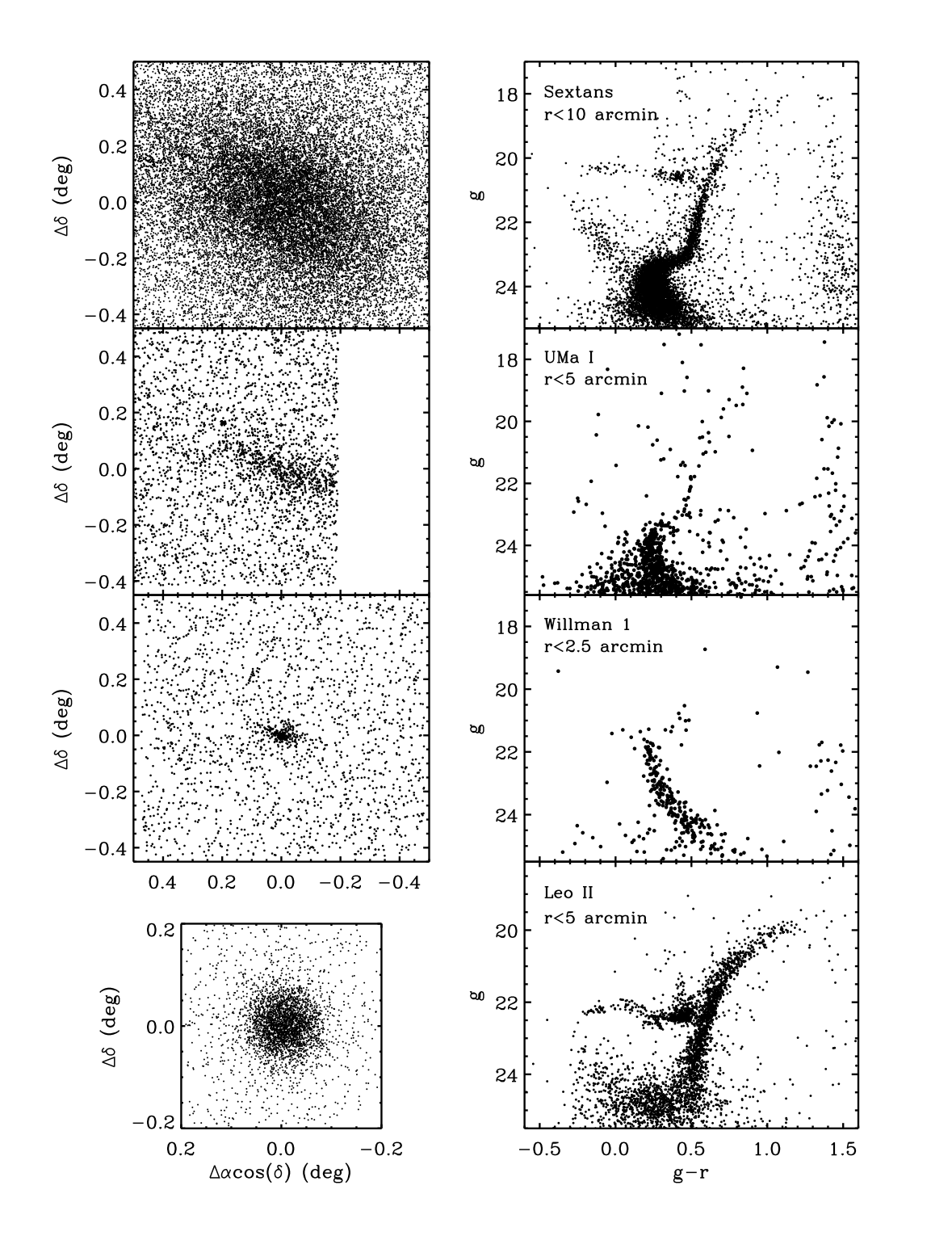}
\caption{Same as Figure~\ref{cmds1}, except for Sextans, Ursa Major~1, Willman~1 and Leo~II.}
\label{cmds5}
\end{figure}

\begin{figure}
\plotone{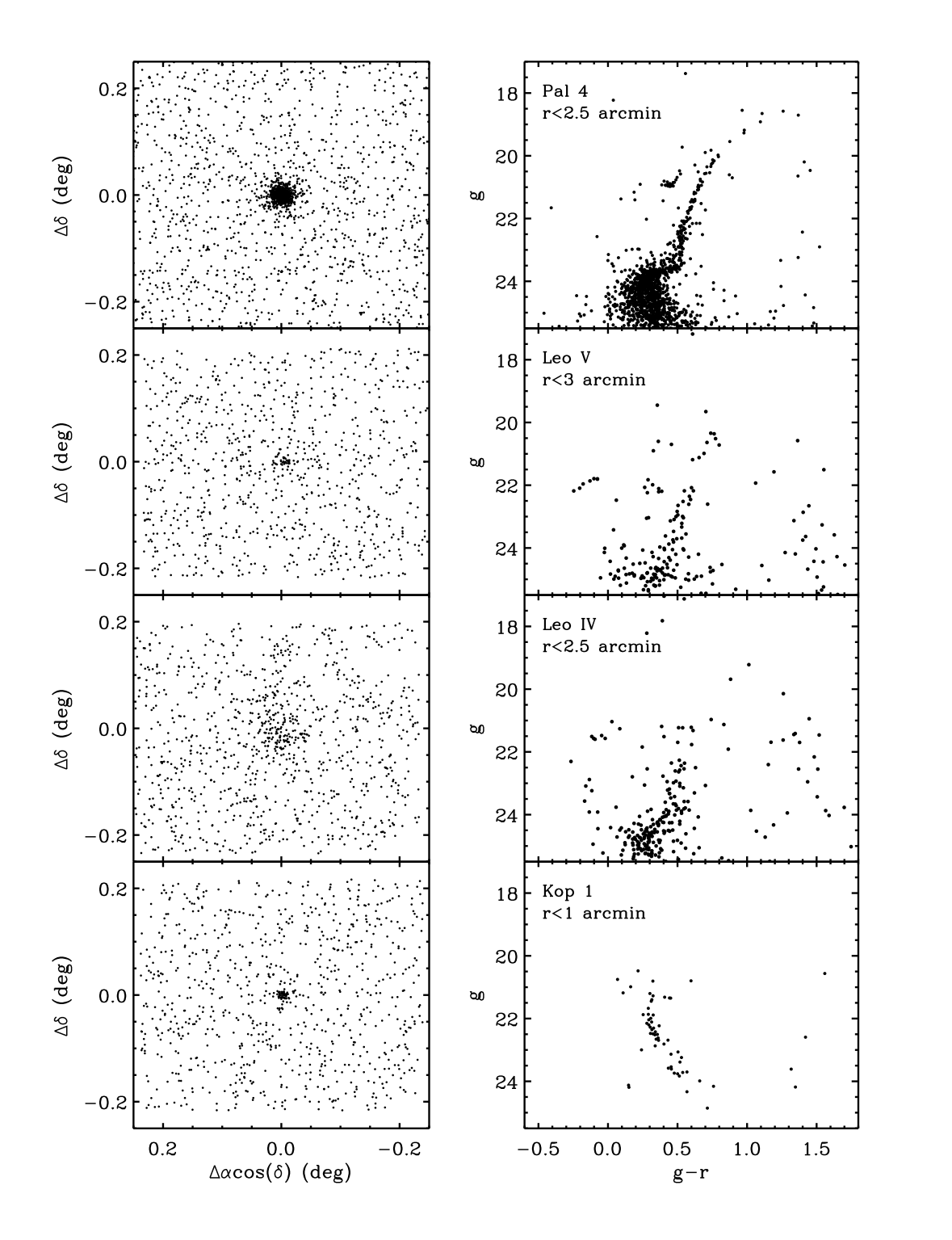}
\caption{Same as Figure~\ref{cmds1}, except for Palomar~4, Leo~V, Leo~IV and Koposov~1.}
\label{cmds6}
\end{figure}

\begin{figure}
\plotone{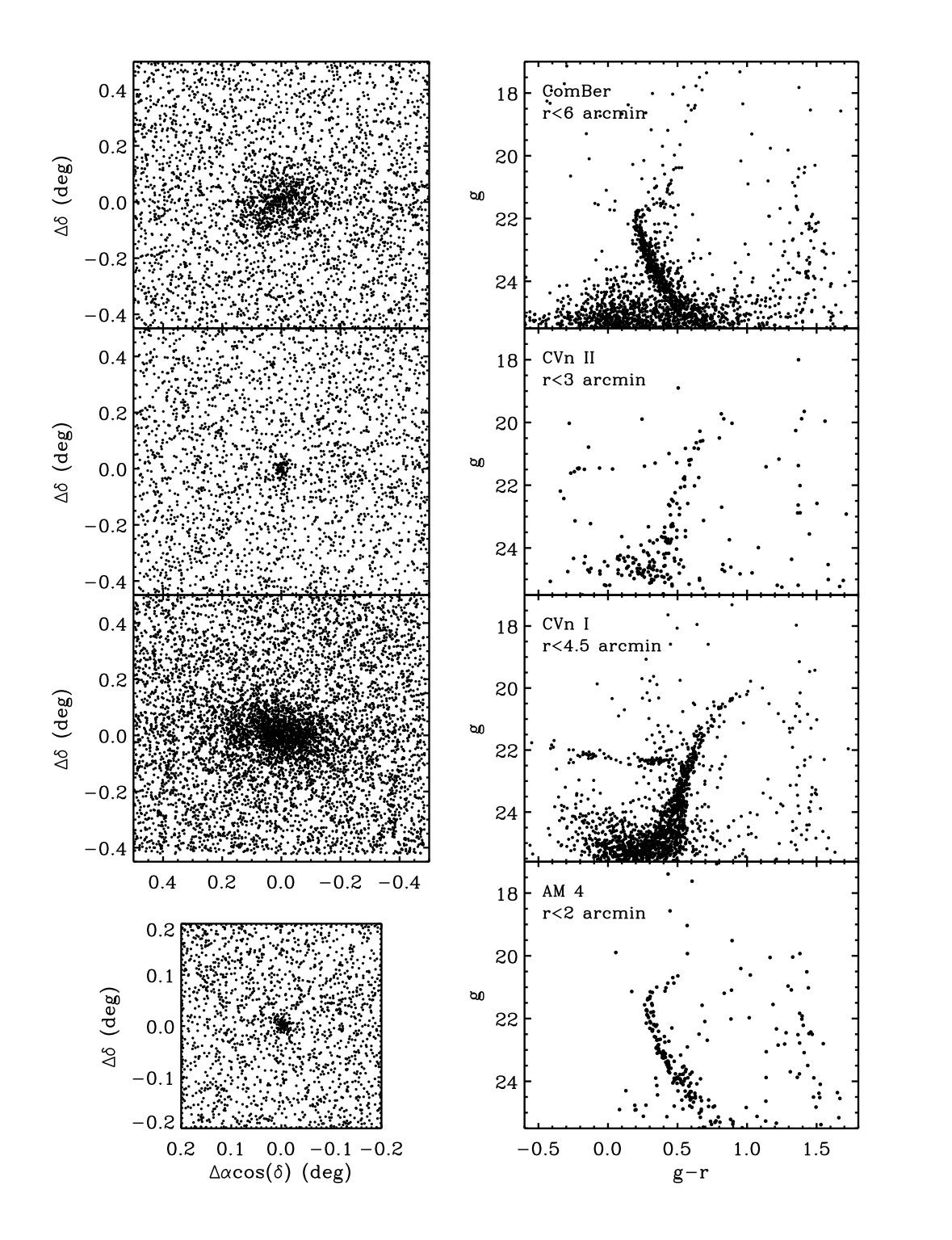}
\caption{Same as Figure~\ref{cmds1}, except for Coma Berenices, Canes Venatici~II, Canes Venatici~I and AM~4.}
\label{cmds7}
\end{figure}

\begin{figure}
\plotone{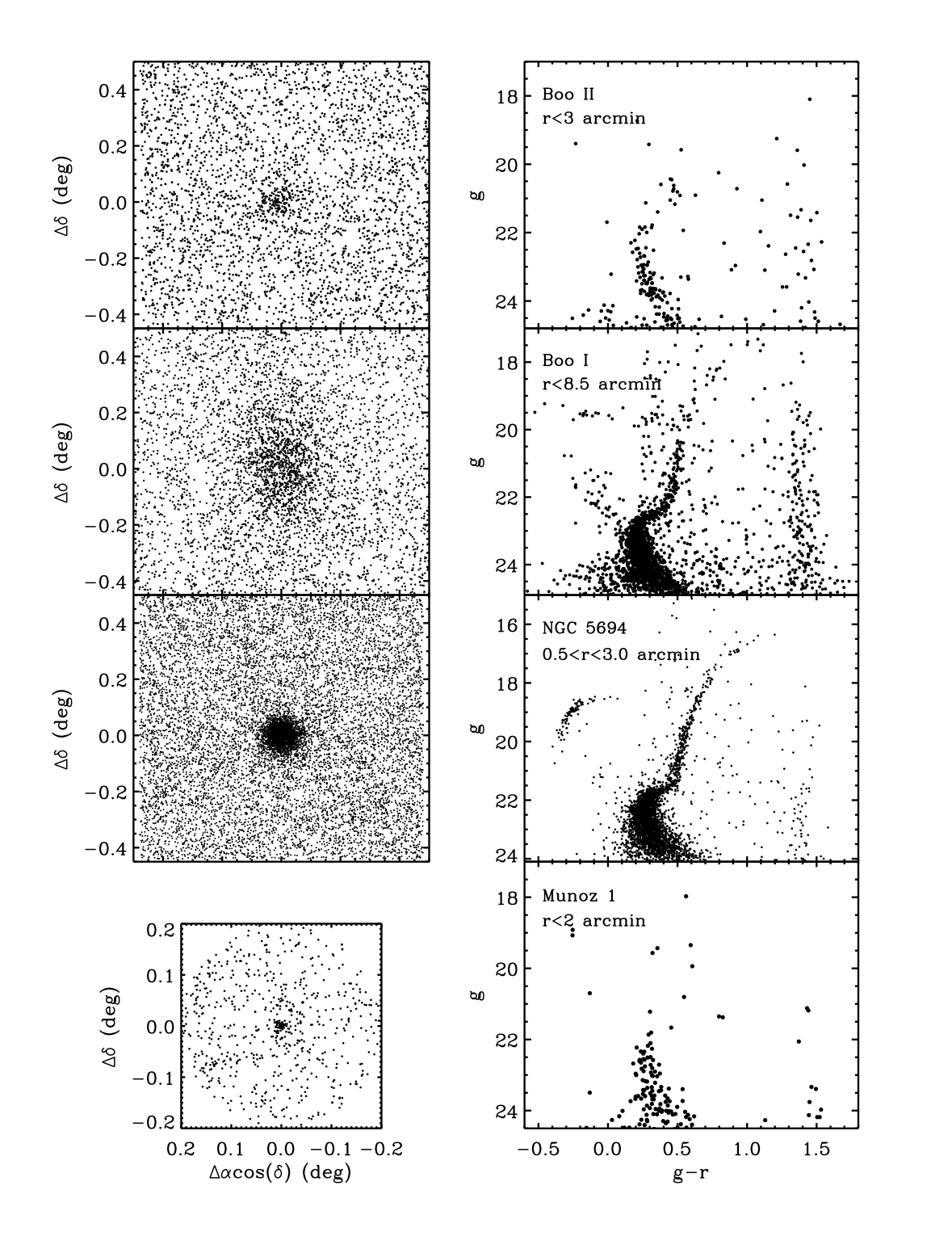}
\caption{Same as Figure~\ref{cmds1}, except for Bootes~II, Bootes~I , NGC5694 and Mu\~noz~1.}
\label{cmds8}
\end{figure}

\begin{figure}
\plotone{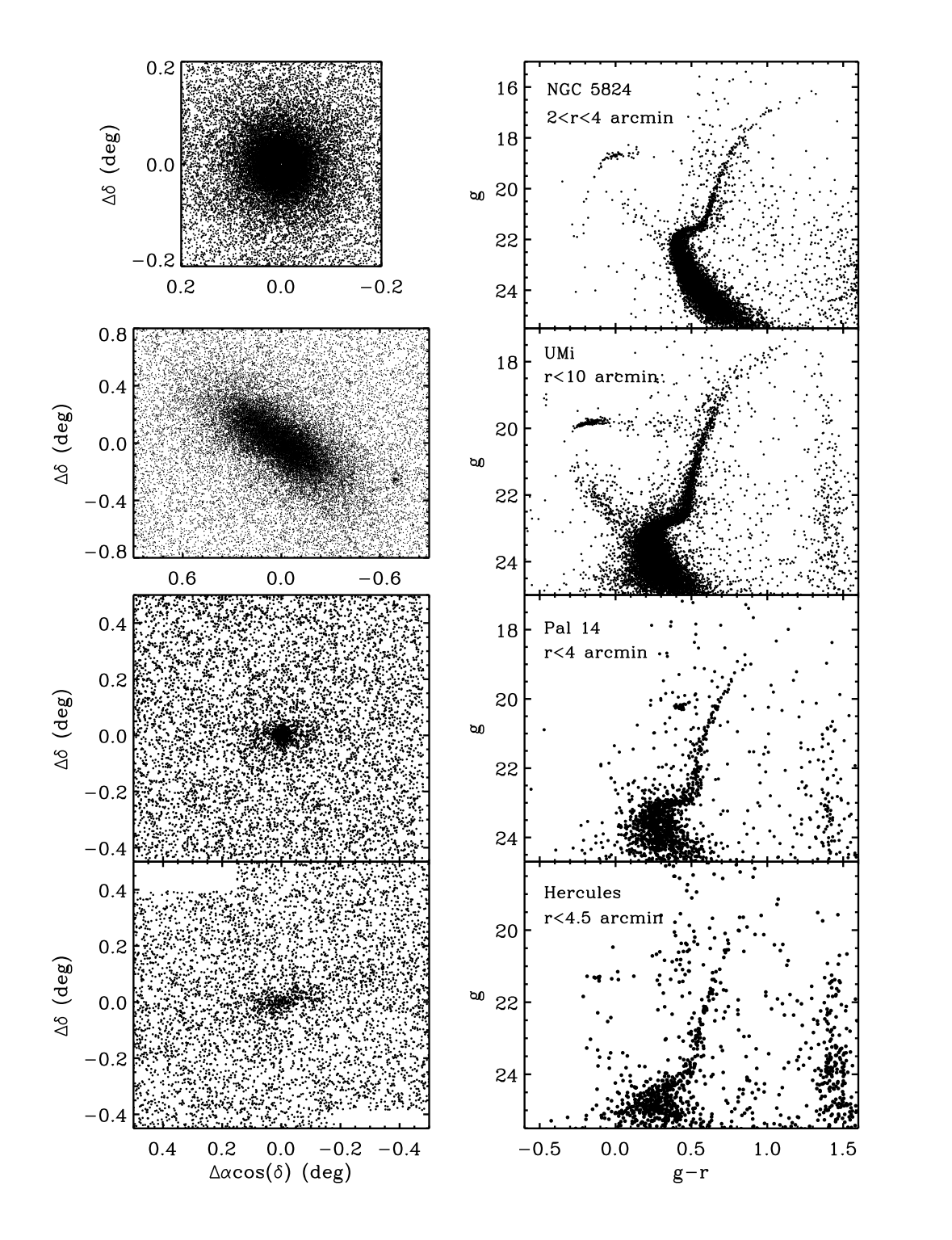}
\caption{Same as Figure~\ref{cmds1}, except for NGC5824, Ursa Minor, Palomar~14 and Hercules.}
\label{cmds9}
\end{figure}

\begin{figure}
\plotone{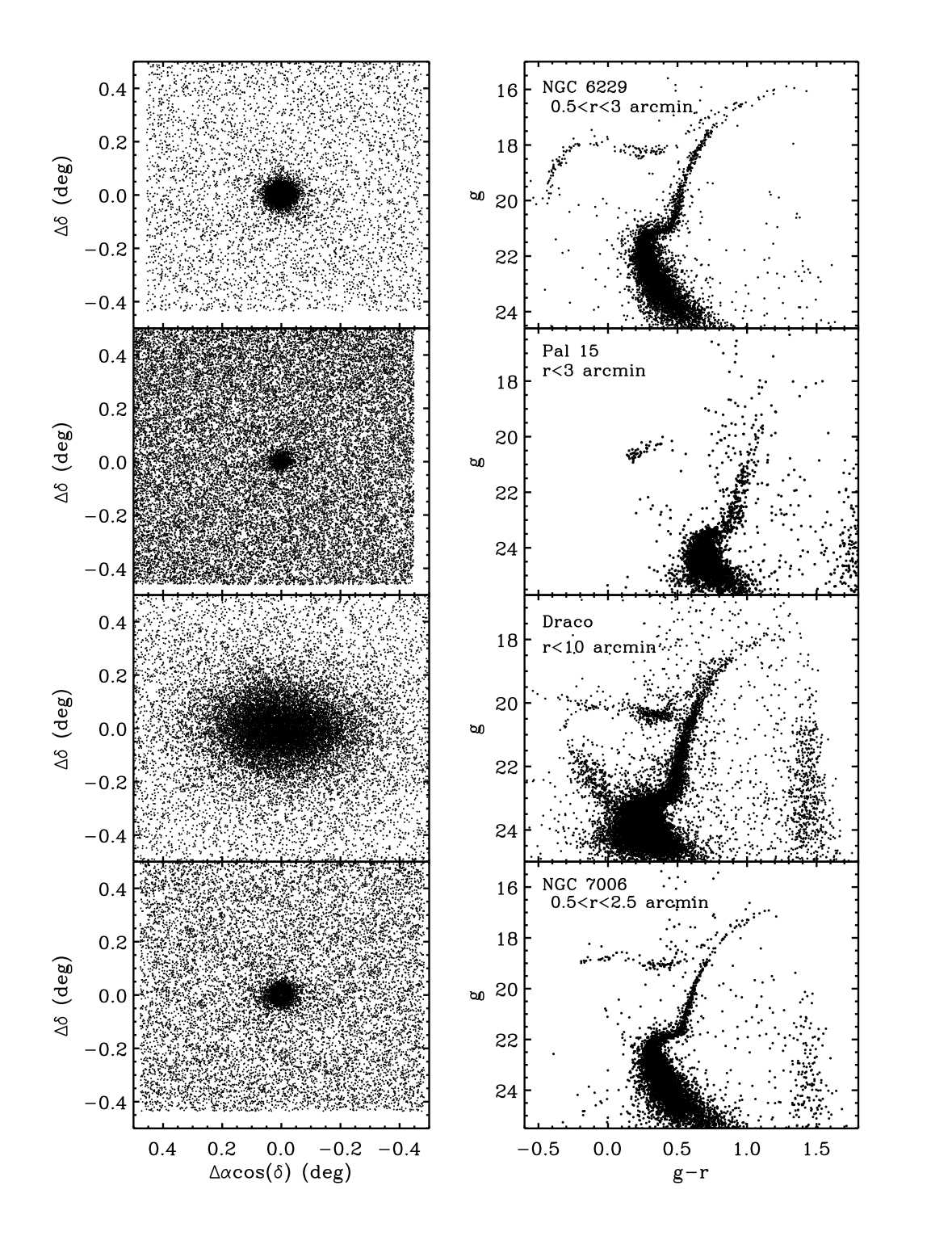}
\caption{Same as Figure~\ref{cmds1}, except for NGC6229, Palomar~15, Draco and NGC7006.}
\label{cmds10}
\end{figure}

\begin{figure}
\plotone{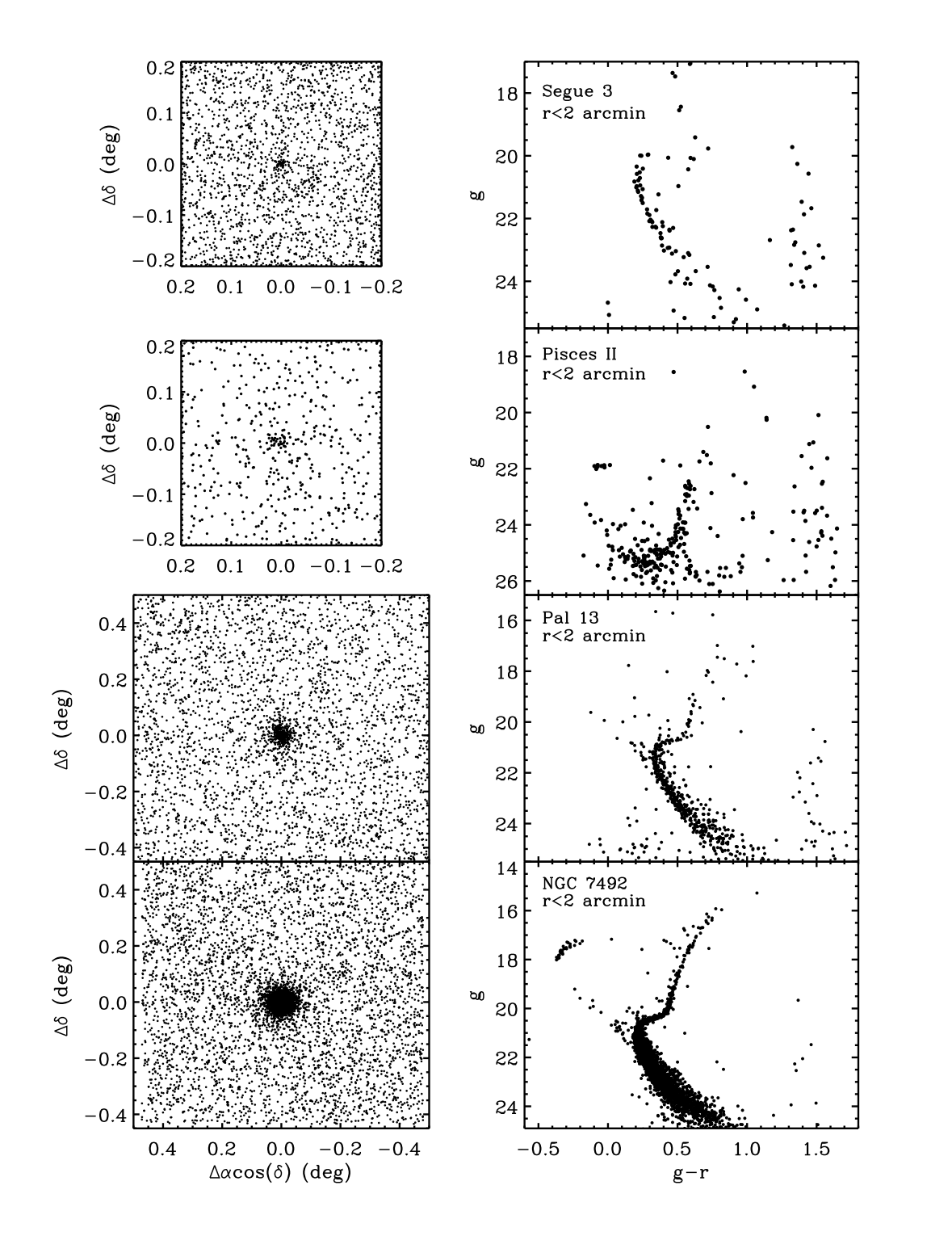}
\caption{Same as Figure~\ref{cmds1}, except for Segue~3, Pisces~II, Palomar~13 and NGC7492.}
\label{cmds11}
\end{figure}

More recently, \citet{coleman05a} carried out a $3\fdg1\times3\fdg1$ wide-field survey centered on Sculptor and 
studied its outer structure finding at most a mild level of tidal interaction with the Milky Way.
In our survey, Sculptor was imaged using Clay-MegaCam in a $2\times4$ mosaic configuration
yielding a total field of view of $0\fdg8\times1\fdg6$ (see Figure~\ref{cmds1}). 
This coverage barely reaches to the system's King tidal radius which, as shown by
\citet{munoz12a}, limits our ability to derive reliable structural parameters using a 
maximum likelihood method (since the background number density is one of the fitted parameters). 
Nevertheless, we measure structural parameters consistent with the values previously published for this 
satellite \citep[e.g.,][]{irwin95a}. Figure~\ref{cmds1} also shows the CMD of Sculptor's 
inner $10$\,arcmin, which reaches to $g \simeq 26$, or more than two magnitudes below 
the MSTO. Combined with our spatial coverage, this represents the deepest and widest photometry 
currently available for this galaxy.

The upper right panel of Figure~\ref{cmds1} shows the number density profile for Sculptor while
Figure~\ref{dens1} shows its isodensity contour map. Its one-dimensional profile is well described by a S\'ersic
profile of index $n_{s}=0.74$. Our measured ellipticity of $\epsilon=0.37$ (for a S\'ersic model) is somewhat higher 
than the value of $0.32$ from \citet{irwin95a}. It is worth noting that, in the structural study by
\citet{westfall06a}, the ellipticity was found to range between 0.23 and 0.49, depending
on the adopted stellar subsample, with deeper samples leading to lower ellipticities. 
In terms of its two-dimensional morphology, Sculptor shows an 
elongated but otherwise regular morphology, with no obvious signs of internal substructure 
or evidence of significant Galactic tidal disturbance. This finding is in line with previous studies.

\subsubsection{Whiting~1} 

Discovered by \citet{whiting02a}, Whiting~1 is the youngest outer halo globular cluster
thought to be associated with the Sgr dSph galaxy \citep{carraro05a,carraro07a, law10a}.
Sparsely populated and originally classified as an open cluster \citep{dias02a}, Whiting~1 is
located at $R_{\sun} \sim30$\,kpc with an estimated age and metallicity of
$\sim6$\,Gyr and [Fe/H]~$\sim-0.5$\,dex, respectively \citep{carraro05a,valcheva15a}.

In our photometry, the subgiant branch is clearly delineated, which allows us to confirm the system's 
young age through isochrone fitting (i.e., $\sim6.5$\,Gyr; see Figure~\ref{cmds1}). 
This satellite is one of the faintest globular clusters known, with a total luminosity of 
just $\sim10^{3}$\,$L_{\odot}$. Whiting~1 is surrounded by
debris from the Sgr dSph, which complicates any analysis of its potential disturbed
morphology (\citeauthor*{munoz18a}, \citealt{carballo17a}). Our measurement of its effective radius, $r_{e,s}\simeq6.4\pm0.6$\,pc, 
makes it one of the smallest outer halo satellites.

\subsubsection{Segue~2}  

With a total luminosity of just $\sim$~500\,$L_{\odot}$ and a central surface brightness of 
$\mu_{0,V} \sim~28.5$\, mag arcsec$^{-2}$, Segue~2 is one of the faintest and most diffuse of 
the ultra-faint satellites discovered in the SDSS \citep{belokurov09a}. It is usually considered to be 
a low-mass galaxy, primarily because of its large internal metallicity spread: i.e., spectroscopic
measurements for member stars range between [Fe/H] $\simeq$ $-2.85$ and $-1.33$\,dex 
\citep{kirby13a}.

Our data, reaching more than three magnitudes below the MSTO, reveal a sparsely populated
object with $r_{e,s}=37\pm3$\,pc and $M_{V}=-1.86\pm0.88$ (Figure~\ref{cmds1}). Because the
subgiant branch is so poorly defined, it is difficult to measure the object's age and distance accurately
from isochrone fitting.

Interestingly, Segue~2 does not appear to follow the familiar luminosity-metallicity relation obeyed 
by the Galaxy's other dwarf satellites, i.e., its metallicity is on the high side for its (low) luminosity. This may 
indicate that Segue~2 was once a more luminous object, perhaps similar to Ursa Minor, that underwent 
significant tidal stripping \citep{kirby13a}. From our photometry, the isodensity contour map (Figure~\ref{dens1}) 
shows a highly irregular morphology suggestive of tidal stripping, although its number 
density profile does not  show an excess of stars usually associated with the presence of tidal debris. 
We emphasize that the small number of member stars introduces significant shot-noise, so conclusions
on potential tidal interactions should be viewed with caution.

\subsubsection{Fornax}  

The Fornax dSph galaxy was discovered by \citet{shapley38b} using plates from the 
24-inch telescope at Boyden Observatory. 
\citet{baade39a} used plates from the 100-inch Mount Wilson telescope to 
establish many of its key photometric and morphological properties, reporting a
major-axis diameter of $\sim50\arcmin$, an ellipticity of $\epsilon \sim0.3$, a distance
of $R_{h}=188$\,kpc, and an absolute magnitude of $M=-11.9$.

Fornax has an apparent diameter that is among the largest of outer halo satellites
($r_t \sim$ 1\fdg2; \citealt{battaglia06a}). Our Clay imaging, which covers an area of 
$\sim0.6$ deg$^2$ in a $2\times2$ grid, does not reach the edge of the 
galaxy. For this reason, Fornax is the only object in our primary survey for which the maximum 
likelihood method failed to converge. To estimate its structural parameters, we therefore
resorted to the more traditional approach of fitting a density model to the binned number density profile. 
We find Fornax to be the 
largest and brightest object in our survey with $r_{e,s}=787\pm9$\,pc and 
$M_{V}=-13.46\pm0.14$. Its global ellipticity is measured to be $\epsilon=0.28\pm0.01$.

Fornax contains multiple stellar populations (i.e., a dominant
intermediate-age population, as well as both old and young stars; see, e.g., \citealt{deboer12a}) and these
different components are readily apparent in its complex CMD (Figure~\ref{cmds1}).
It is also known that the spatial distributions of these populations vary, with the younger
population being more centrally concentrated and showing a clumpy morphology, while
the older stars are more smoothly distributed. This complexity is evident in the isodensity map
shown in Figure~\ref{dens1}. Note that several of Fornax's globular 
clusters are visible in the lower left panel of Figure~\ref{cmds1}.

\subsubsection{AM~1}  

At a Galactocentric distance of $R_{\rm GC}\simeq125$\,kpc, AM~1 is one of the most remote 
Galactic globular clusters currently known. It was discovered by \citet{lauberts76a} while inspecting plates 
from the ESO Schmidt telescope in Chile.
AM~1 is an intermediate-luminosity 
($M_{V}=-5.2$) and metal-poor ([Fe/H] $=-1.7$\,dex) cluster whose angular proximity to the Large 
Magellanic Cloud --- it is located just $\sim15^{\circ}$ from the LMC center --- prompted early 
speculation that it may be physically associated with the LMC. However, the latest distance estimates 
suggest an association is quite unlikely.

Figure~\ref{cmds2} shows the CMD for the inner $2\farcm5$ of AM~1. 
Like most outer halo clusters, AM~1 shows a clear second parameter effect:
i.e., it has a red horizontal branch despite its relatively low metallicity. 
This effect has been recognized since the 1970s, most notably by 
\citet{harris76a},  \citet{searle78a} and \citet{zinn80a, zinn80b}, who pointed 
to an age spread among the outer halo clusters as a possible explanation for the
unusual horizontal branch morphologies. In the case of AM~1, \citet{dotter08b} 
used {\it HST} photometry to conclude that this cluster is indeed younger than the inner halo 
clusters, by $1.5-2$\,Gyr.  As is the case with other remote halo clusters,
AM~1's effective radius, $r_{e,s}=16.5\pm1.1$\,pc, is large compared to the 
inner halo clusters. The density contour map for AM~1 shows a fairly round and 
regular morphology while its number density profile (Figure~\ref{dens2}) 
extends beyond both King and S\'ersic profiles.

\subsubsection{Eridanus} 

Also discovered on plates from the ESO Schmidt telescope \citep{cesarsky77a}, Eridanus is among 
the most distant clusters in the outer halo ($R_{\rm GC}\simeq95$\,kpc)
as well as a sparse system whose photometric properties --- an effective radius of  
$r_{e,s}=16.8\pm1.1$\,pc and an absolute magnitude of $M_{V}=-4.9$ --- are nearly identical to 
those of AM~1. Like most outer halo clusters, Eridanus' CMD (see Figure~\ref{cmds2}) 
exhibits a clear second parameter effect with a red horizontal branch. Using WFPC2 data, 
\citet{stetson99a} studied in detail the morphology of the cluster subgiant branch and determined
that Eridanus --- like AM~1, Palomar~3 and Palomar~4 --- is $1.5-2$\,Gyr younger than the inner halo 
clusters M3 and M15 (assuming that the element abundance ratios have been 
estimated correctly).

Recently, \citet{myeong17a} reported the discovery of tidal tails
around this cluster. In
Figure~\ref{dens2} we show our density contour map for Eridanus. Morphologically, the map shows 
no remarkable or unusual features, and its number density profile is well fitted by either a  
King or S\'ersic model and thus we cannot confirm the presence of a tidal
structure around Eridanus, although we note that \citet{myeong17a}'s data are at 
least one magnitude deeper than ours.

\subsubsection{Palomar~2}  

One of 13 globular clusters discovered by \citet{abell55a}, Palomar~2 is located at 
low latitude in a heavily obscured field in the direction of the
Galactic anticenter ($l = 171^{\circ},~b = -9^{\circ}$). It has been seldom studied 
since its discovery. \citet{peterson76a} used a single KPNO 2.1-m
telescope $B$-band plate to carry out a star count analysis of the cluster and estimated its 
King core and tidal radii to be $r_{c}<0\farcm08$ and $r_{t}=4\farcm7$, respectively.
\citet{harris80a} revised these values to $r_{c}=0\farcm14\pm0\farcm03$ and 
$r_{t}=3\farcm16\pm0\farcm35$, using a deep $V$ photographic plate, and estimated a tentative 
heliocentric distance of $17\pm4$\,kpc. 

The first CMD for Palomar~2 was published by \citet{harris97a}
using data from the CFHT UH8K camera. This CMD revealed the cluster's
main evolutionary sequences, albeit with significant absorption and differential reddening.
Their improved distance put Palomar~2 at a Galactocentric distance of 
$R_{\rm GC}\sim34$\,kpc, with an absolute magnitude of $M_{V}=-7.9$.
\citet{harris97a} also examined the spatial distribution of stars and found 
the bright giants to have a different distribution than the
fainter, main-sequence stars. We see this effect as well, probably
a consequence of a dust lane crossing the cluster in the {\tt N-S} direction and
causing differential reddening that affects the fainter stars more noticeably
than the brighter giants.

Figure~\ref{cmds2} shows the CMD based on our new CFHT imaging.
Despite the heavy obscuration and differential reddening, the brighter evolutionary sequences are readily 
discernible, meaning that it is possible to identify candidate member stars and measure reliable structural parameters. 
Our effective radius, $r_{e,s}=7.8\pm0.2$\,pc, is larger than published values while our 
absolute magnitude, $M_{V}=-9.07\pm0.07$\footnote{The low uncertainty in our
measurement does not include the effect of differential reddening which are difficult to assess.}, is significantly brighter than
the value of \citet{harris97a}, indicating that Palomar~2 
is one of the brightest outer halo clusters. 

\subsubsection{Carina}  

Carina was discovered by \citet{cannon77a} while inspecting plates from the 
ESO/SRC Southern Sky Survey. 
Using CCD images taken with the CTIO 1m telescope, 
\citet{smecker-hane94a} published wide-field photometry for Carina
revealing a well-defined horizontal branch
with two distinct components --- direct evidence for two populations of differing
age.
From the color of the RGB, \citet{smecker-hane94a} obtained a mean metallicity 
of [Fe/H]\,$\sim-2.1$\,dex. Deeper photometric studies, based on imaging from {\it HST}/WFPC2 and 
the CTIO 4m telescope, confirmed these findings \citep[e.g.,][]{mighell97a,hurley-keller98a}. 

In our survey, Carina was observed in a $4\times4$ mosaic with the Clay telescope,
covering a total area of $\simeq 2.2$~deg$^2$. This coverage extends well beyond the 
system's nominal tidal radius (e.g., $R_{t}=28\farcm8$\,; \citealt{mateo98a}) and
our CMD, which reaches to $g\sim 26$, reveals in detail the full 
complexity of  Carina stellar populations (Figure~\ref{cmds2}). The combination of 
depth and field coverage makes our photometric catalogue the most extensive currently 
available for this system. Indeed, \citet{santana16a} recently used these data
to derive a detailed star formation history for the galaxy. In our maximum likelihood analysis, we
measure an effective radius of $r_{e,s}=313\pm3$\,pc and an absolute 
magnitude of $-9.43\pm0.05$. Thus, Carina appears to be significantly 
larger than reported in some previous  studies (although still consistent with early reports of 
likely member stars beyond the nominal tidal radius). Our
number density profile does not show the previously reported ``break" at $\sim20\arcmin$ 
\citep{majewski00a, majewski05a, munoz06a},
and is well fitted over its full extent by a S\'ersic profile with $n_{s}=0.84\pm0.02$.
The galaxy is moderately flattened, with $\epsilon = 0.37\pm0.01$, but 
its isodensity contour map show no clear signs of tidal interaction (Figure~\ref{dens2}).

\subsubsection{NGC2419}  

By far the brightest of the outer halo clusters, NGC2419 is located in the direction of the Galactic 
anticenter. It was discovered by W. Herschel in 1788 and recognized as a globular cluster almost a 
century and a half later by C.O. Lampland on plates from the Lowell Observatory \citep{baade35a}. 
\citet{racine75a} produced the first modern CMD for the cluster using $BI$ plates taken with the
Palomar $200-$inch telescope. Their diagram reached $V\sim22$, slightly more
than four magnitudes below the tip of the RGB, and revealed an extended and predominantly
blue horizontal branch unlike most other remote halo clusters. From a 
comparison to M92, these authors concluded that NGC2419 is equally metal poor.
 
NGC2419 is interesting because of its rather unusual nature: i.e., it is much 
brighter than the other remote halo cluster and it does not show a second
parameter anomaly in its horizontal branch. It also appears to be unique in its chemical 
properties \citep{cohen10a, cohen12a}.

We imaged NGC2419 with CFHT in a single pointing roughly
centered on the cluster.  Figure~\ref{cmds3} shows our CMD for the cluster, which reaches
one magnitude below the MSTO. A blue, extended horizontal branch is clearly visible.
We measure an absolute magnitude of $M_{V}=-9.35\pm0.03$.
With an effective radius of $r_{e,s}\sim25.7\pm0.2$\,pc and King limiting radius of 
$r_{k,t}=227$\,pc it is also the largest of the star clusters in our sample.
Its isodensity map reveals a round and regular morphology, 
showing no obvious signs of tidal interaction with the Milky Way (Figure~\ref{dens3}).
We measure an ellipticity of $\epsilon_s=0.05\pm0.01$. 

\subsubsection{Koposov~1 and 2}  
 
Koposov~1 and 2 were initially identified by \citet{koposov08a} using SDSS data and subsequently 
confirmed as Galactic satellites by the same group using imaging from Calar Alto.
Both systems were originally classified as globular clusters but later reported to be
old open clusters by \citet{paust14a}. These authors 
found both satellites to be significantly younger than other outer halo clusters. They also proposed
that both Koposov~1 and 2 were originally born as part of the Sagittarius dSph and 
later removed by the Galactic tidal field.
 
From our Clay imaging, we determined a very faint absolute magnitude of 
$M_{V}=-1.0\pm0.7$ for Koposov~1 (slightly brighter than the value of \citealt{paust14a}) and 
an effective radius of $r_{e}=10.1\pm2.5$\,pc, similar to earlier measurements. 
As is the case with Koposov~2 and Mu\~noz~1, the measured ellipticity is high ($\epsilon=0.55\pm0.15$) 
but poorly constrained due to the small number of member stars. Indeed, the
CMD of Koposov~1 is sparsely populated, with no stars visible on the RGB
and only a handful of potential sub-giant stars (Figure~\ref{cmds6}). If these are bonafide
Koposov~1 members, then isochrone fitting points to an age between $8$ and $10$\,Gyr, 
consistent with the conclusions of \citet{paust14a}.

Koposov~2 has a nearly identical low luminosity as Koposov~1. We measure
an absolute magnitude of $M_{V}=-0.9\pm0.8$ and an effective radius of 
$r_{e,s}=4.3\pm0.9$\,pc. Its CMD has no stars on the upper RGB
and the upper main sequence is scarcely populated as well (Figure~\ref{cmds3}). Despite the depth 
of our photometry, 
which reaches to $g=25$, the number of detected stars remains low, translating
into rather large uncertainties in the measured structural parameters, particularly in the case of the King core and tidal
radii where the uncertainties reach 65\%. Its measured
ellipticity is among the highest obtained for a globular cluster, at $\epsilon=0.48\pm0.15$, 
but given the low number of stars, this may be an artifact of shot noise.

\subsubsection{Ursa Major~II}  

Ursa Major~II was one of the first ultra-faint dwarfs  discovered in the SDSS \citep{zucker06b}. 
Follow-up images from the Subaru telescope acquired by the same authors 
revealed a stellar system with a highly elongated morphology,
a size of $\sim$ 250\,pc $\times$ 125\,pc, and an 
absolute magnitude of $M_{V}\sim-3.8$.  Its unusual morphology prompted
\citet{fellhauer07a} to suggest that Ursa Major~II may be the disrupted progenitor
of the Orphan Stream \citep{belokurov06b}.

Using the same CFHT images described here, \citet{munoz10a} carried out a morphological 
analysis and argued that UMa~II is probably a system that is undergoing severe tidal disruption. 
This view is supported by the observed number density profile and isodensity contour map (Figure~\ref{dens3}). 
The former is poorly fitted by each of the density laws considered in this study and the latter shows an 
unusually elongated structure. If this interpretation is correct, then previous reports of high mass-to-light 
ratios (which were obtained under the assumption of dynamical equilibrium) should be viewed with 
caution. Kinematic measurements covering the full extent of UMa~II will be needed to
conclusively determine its dynamical state.

Our CMD for Ursa Major~II reaches three magnitudes below the well defined MSTO (Figure~\ref{cmds3}).
The sub-giant branch is clearly visible, although the RGB is only sparsely populated. 
Using our maximum-likelihood method, we revisit this system's structural parameters and
measure an absolute magnitude of $M_{V}=-4.2\pm0.3$ and an effective
radius of $r_{e,s}=130\pm4$\,pc. We find its mean ellipticity to be $\epsilon=0.56\pm0.03$ although,
as noted by \citet{munoz10a}, this value increases inwards.

\subsubsection{Pyxis} 

Located at a Galactocentric distance of $R_{\rm GC}\simeq41$\,kpc, Pyxis was detected by 
\citet{weinberger95a} as a stellar overdensity, and quickly confirmed as a globular cluster by 
\citet{dacosta95b} and \citet{irwin95b}. 

Pyxis is the cluster in our sample with the lowest Galactic latitude, ($l=26^{\circ},~b=7^{\circ}$),
and its CMD (Figure~\ref{cmds3}) is significantly contaminated by foreground disk stars. 
Nevertheless, the cluster sequences are readily apparent and, unlike Palomar~2, 
are only somewhat broadened.
In line with most outer halo clusters, 
Pyxis has a predominantly red horizontal branch and thus exhibits the well known
second parameter effect. In this context, \citet{dotter11a}, used {\it HST}/ACS data
to measure an age of $11.5\pm1.0$\,Gyr, consistent with Pyxis being somewhat younger than
the inner halo systems. 
 
Our analysis suggests an effective radius of $r_{e,s}=18.6\pm0.5$\,pc, which is typical for outer halo clusters, 
and an absolute magnitude of $M_{V}=-5.7\pm0.2$, also consistent with published values. Its number 
density profile is well fitted by either a King or S\'ersic model and its $2-$D morphology is round and regular, 
with a low ellipticity of $\epsilon=0.04\pm0.02$.

\subsubsection{Leo~T}  
 
At a Galactocentric distance of $R_{\rm GC}=420$\,kpc, or roughly half the distance to M31, 
Leo~T is easily the most distant satellite in our survey. It was discovered  by \citet{irwin07a} as a stellar 
overdensity in SDSS DR5 data. Because of its large distance, the original SDSS CMD reached 
only a few magnitudes below the tip of the RGB, but was sufficient to reveal two key features: a well 
defined RGB and a bluer sequence likely due to the presence of young stars ($\lesssim1$\,Gyr).
\citet{irwin07a} also detected HI associated with Leo~T, classifying this object as a
transition-type dwarf galaxy. 
Based on deep Large Binocular Telescope data, \citet{dejong08a} 
found that Leo~T has been forming stars at least as recently as
a few hundred
Myr ago, making Leo~T the faintest star-forming galaxy known at this time.

In our Clay imaging, we detect stars in the Leo~T's CMD only down to the red clump level 
(Figure~\ref{cmds4}). We find an absolute magnitude of $M_{V}=-7.6\pm1.0$ and 
an effective radius of $r_{e,s}=151\pm17$\,pc, both similar to published values. In terms of 
morphology, Leo~T is one of the roundest galaxies in our sample, with $\epsilon=0.23\pm0.09$. 
Its number density profile is well fitted by either a King or S\'ersic model and 
its morphology appears undisturbed (Figure~\ref{dens4}).

\subsubsection{Palomar~3}  

Palomar~3 was discovered  by \citet{wilson55a} and \citet{abell55a} using plates from the  National 
Geographic Society-Palomar Observatory Sky Survey. At  $R_{\rm GC}\simeq92.5$\,kpc, it is, with AM~1, 
Eridanus, NGC2419, Palomar~4 and Palomar~14, among the most remote Galactic globular clusters. 
The CMD based on our CFHT imaging (see Figure~\ref{cmds4})  reveals a definite second parameter 
effect, with a horizontal branch that is composed almost exclusively of red stars despite its low metallicity 
of [Fe/H]$\simeq-1.6$\,dex \citep{koch09b}.  

Our analysis reveals Palomar~3 to be a relatively faint ($M_{V}=-5.5\pm0.2$) and
extended ($r_{e,s}=19.4\pm0.5$\,pc) cluster, similar in size to the smallest of the ultra-faint dwarf 
galaxies, Willman~1 and Segue~1. However, unlike these objects, the cluster's two-dimensional morphology 
(see Figure~\ref{dens4}) is unremarkable, with an ellipticity of $\epsilon=0.07\pm0.03$,
and shows the regular density contours typical of globular clusters. 

\subsubsection{Segue~1}  

Among the Galactic satellites discovered during the last decade, Segue~1 \citep{belokurov07a} 
is arguably one of the most fascinating objects. With an absolute magnitude of $M_{V}=-1.3\pm0.7$,
it has been described as the faintest Milky Way satellite galaxy discovered to date \citep{geha09a, simon11a},
although it was originally classified as a diffuse globular cluster by \citet{belokurov07a}. 
The CMD based on our CFHT imaging (Figure~\ref{cmds4}) traces the main sequence roughly 
four magnitudes below the MSTO, and yet only a handful of subgiant stars
are visible, highlighting  the very low luminosity of the object ($\simeq 280\,L_{\odot}$).

Its low luminosity, modest ellipticity, $\epsilon=0.31\pm013$, and compact 
size ($r_{e,s}=26\pm4$\,pc)
has prompted some researchers to suggest that Segue~1 may be a dark matter-free, 
disrupted cluster \citep{dominguez16a}. Based on our imaging, we see no signs 
that Segue~1 is being affected by tides. Its density contour map  (see Figure~\ref{dens4}) shows 
a round central structure 
and the system's kinematic and chemical properties clearly favor a dwarf galaxy classification.

\subsubsection{Leo~I}  

At a Galatocentric distance of $R_{\rm GC}=258$\,kpc, Leo~I is one of the most 
distant stellar system that is likely bound to the Milky Way. It was discovered, 
together with Leo~II,  by \citet{harrington50a} while inspecting plates from the
Palomar Sky Survey. 
\citet{hodge63a} carried out a star count analysis using a variety of plate material, and found Leo~I
to be elliptical ($\epsilon \sim0.3$) and almost perfectly symmetrical in structure.
Assuming a distance similar to that of Leo~II, \citet{hodge63a} estimated
a linear cutoff radius along the major axis of $950\pm70$\,pc. He also reported
an absolute magnitude of $M_{V}\simeq-11.4$ for his adopted distance of $230$\,kpc.
  
\citet{gallart99a, gallart99b} presented what is, to date, the deepest photometric 
analysis of Leo~I. Their {\it HST}/WFPC2 $VI$ imaging for a central field reached 
more than two magnitudes deeper than the study of \citet{lee93a}, barely detecting the 
old MSTO. \citet{gallart99b} used these data to measure a detailed
star formation history based on comparison with synthetic CMDs. They concluded
that Leo~I is dominated by an intermediate-age population with the majority of
the stars having formed between $7$ and $1$\,Gyr ago, at which point star formation 
abruptly ceased.

Our CFHT photometry is visibly affected by the presence of Regulus, as can be seen in 
Figure~\ref{cmds4}. Our CMD reaches $g\sim25$, albeit with strongly varying completeness 
across the field. Nevertheless, we are able to estimate useful structural parameters, finding 
$M_{V}=-11.8\pm0.3$, $r_{e,s}=244\pm2$\,pc and $\epsilon=0.30\pm0.01$. The morphology
of Leo~I appears to be fairly regular, with no sign of obvious
tidal features (Figure~\ref{dens4}).

\subsubsection{Sextans}  

Sextans is one of the faintest and most diffuse of the Galaxy's classical dSph galaxies.
Due to its low surface brightness, and the significant foreground contamination
arising from its low Galactic latitude ($b\simeq 8^\circ$), it had escaped detection
until \citet{irwin90a} discovered it in an analysis of plates taken 
with the 1.2m UK Schmidt telescope. 
From the mean magnitude of the horizontal branch, the authors estimated a distance of 
$85\pm5$\,kpc. Fitting a  \citet{king62a} model to the surface density profile yielded a 
King limiting radius of $\sim2$\,kpc and an absolute magnitude of $M_{B}\sim-8$.

Several deep CCD photometric surveys covering a large fraction of the galaxy 
have now been published. \citet{lee03a} used the CFH12K camera at 
CFHT to produce a CMD that reaches $V\sim24$\, about 
one magnitude below the MSTO. The CMD was found to be consistent with a mean metallicity 
of [Fe/H]$=-2.1\pm0.1$\,dex. \citet{okamoto08a} used Subaru imaging to produce a
CMD that reached to $V\sim25$, the deepest to date. Their analysis
showed that the red horizontal branch stars seem to be more concentrated than
the blue HB stars, consistent with the metallicity gradient reported by \citet{battaglia11a}.

In our survey, Sextans was imaged in four CFHT pointings, arranged in a
$2\times2$ grid that covers an area of nearly 4 deg$^2$. Our photometry reaches to a depth 
of $g\sim25.5$, a little more than two magnitudes below the MSTO. This makes our survey the 
most extensive to date in terms of depth and spatial coverage. Our CMD shows a narrow RGB 
as well as an extended, but predominantly red, horizontal branch (Figure~\ref{cmds5}). A blue 
straggler sequence is also readily apparent. We measure an effective radius of 
$r_{e,s}\sim442\pm4$\,pc, making Sexans one of the largest satellites included in this catalog.
We also find an absolute magnitude of $M_{V}=-8.7\pm0.06$, in good agreement with 
previous measurements (but see \S\ref{subsec:prev}). The isodensity contour map in
Figure~\ref{dens5} shows a fairly  regular morphology with no obvious signs of tidal features.

\subsubsection{Ursa Major~I}  

Ursa Major~I was one of the first ultra-faint dwarfs to be discovered in automated searches for substructures in 
the SDSS \citep{willman05b}. Located at a Galactocentric distance of $R_{\rm GC} \simeq102$\,kpc, early
estimates of its photometric properties \citep[$M_{V}\sim-6.75$, $r_{h}=250$\,pc;][]{willman05b} were 
consistent with a dwarf galaxy classification, and its CMD appeared to be similar to that of Sextans, 
albeit with more sparsely populated evolutionary sequences. 

\citet{okamoto08a} obtained deep Suprime-Cam imaging with the  Subaru telescope, revising Ursa 
Major~I's distance and showing it to be dominated by an old, metal-poor population more closely 
resembling that of a globular cluster than a typical dSph. \citet{brown12a, brown14a} included 
Ursa Major~I in their {\it HST}/ACS survey of ultra-faint dwarfs. From their $VI$ photometry, which reached
four magnitudes below the MSTO, they concluded that Ursa Major~I appears to be a ``fossil" galaxy, 
in the sense that it hosts an ancient and metal-poor population with all star formation having ceased 
$\sim\,11.6$\,Gyr ago. It terms of its morphology, \citet{okamoto08a} found Ursa Major~I to be both
elongated and disturbed, and suggested that it has suffered (or continues to suffer) significant tidal stripping.

Although we covered this satellite with two CFHT pointings, the {\tt SW} field was unfortunately taken in 
conditions of poor seeing, resulting in a significantly shallower photometry. We therefore restricted our analysis
to just the deeper {\tt NE} field (see Figure~\ref{cmds5}). Like its neighbor Ursa Major~II, we find Ursa Major~I to be highly 
elongated, with an overall ellipticity of $\epsilon=0.57\pm0.03$, and irregular isodensity contours resembling 
those of Ursa Major~II (Figure~\ref{dens5}).  We measure an absolute magnitude of $M_{V}=-5.1\pm0.4$
and an effective radius of $r_{e,s}=235\pm10$\,pc.

\subsubsection{Willman~1}  

The first of the ultra-faint stellar systems discovered in the SDSS \citep{willman05a}, 
Willman~1 is also, given its complex photometric and kinematical properties, one of the 
more intriguing Galactic satellites \citep[e.g.,][]{willman11a}. It occupies 
a region in the luminosity-size diagram ($M_{V}=-2.5\pm0.7$, $r_{e,s}=28\pm2$\,pc) 
that includes Segue~1, Segue~2 and and Bo\"otes~II, all of which have been
classified as ultra-faint dwarf galaxies. Its size, however, is more comparable to that of 
a remote halo cluster. It should be noted that all clusters having a similar size are 
brighter than Willman~1 by an order of magnitude or more.  

Our CFHT photometry, which reaches to $g\simeq25$, reveals an irregular morphology 
that hints at tidal interactions. In particular, the isodensity map seems to have an unusual three-tailed 
structure, as noted by \citet{willman06a} (see Figure~\ref{dens5}). Unfortunately, the 
sparsely populated CMD renders this finding tentative.

\subsubsection{Leo~II}   

The Leo~II satellite was discovered, along with Leo~I, by \citet{harrington50a} 
while examining plates taken with the 48-inch Palomar Schmidt telescope for 
the National Geographic Society-Palomar Observatory Sky Survey. 
\citet{hodge62a,hodge71a} carried out a star count analysis of the galaxy using plates
from multiple telescopes, finding a (low) ellipticity of $\epsilon=0.01\pm0.10$ and a 
limiting radius of $r_r= 11\farcm9$\,, corresponding to $\simeq800$\,pc at a 
distance of $230$\,kpc. 

\citet{mighell96a} presented deep {\it HST}/WFPC2 photometry for Leo~II reaching
$V\sim27.4$ and $I\sim26.6$. Their CMD, reaching about three magnitudes
below the MSTO, suggested a mean metallicity of 
[Fe/H]\,$=-1.60\pm0.25$\,dex and an age of $9\pm1$\,Gyr. 
More recently, \citet{coleman07a} used SDSS to explore the outer 
structure of the galaxy, concluding that the influence of the Galactic gravitational field 
on the structure of the galaxy has been relatively mild.

In our survey, we imaged Leo~II in a $2\times2$ grid pattern with the Clay telescope, 
covering  an area of $\sim0.6$~deg$^2$. Our CMD reaches $g\sim25.5$, covering 
about three magnitudes below the horizontal branch (Figure~\ref{cmds5}).
We measure an absolute magnitude of $M_{V}=-9.7\pm0.04$ and
an effective radius of $r_{e,s}=168\pm2$\,pc, consistent with 
previous estimates. Leo~II is the roundest of the dwarf satellites in our survey with an overall
ellipticity of $\epsilon=0.07\pm0.02$, similar to many of the globular clusters in our sample.
Its morphology is found to be quite regular, with no signs of tidal features (Figure~\ref{dens5}).

\subsubsection{Palomar~4}   

The second most distant of the Galactic globular clusters ($R_{\rm GC}=110$\,kpc), 
Palomar~4 was discovered by \citet{abell55a} on plates from the
National Geographic Society-Palomar Observatory Sky Survey.
\citet{stetson99a} used {\it HST}/WFPC2 data to show that
Palomar~4 is  a second parameter cluster, with an age 
$1.5-2$\,Gyr younger than M3 and M5.

Our CFHT imaging reaches $\sim$\,1.5 magnitudes below the MSTO as shown 
in Figure~\ref{cmds6}. A red horizontal branch is clearly visible despite the cluster's 
relatively low metallicity (e.g., [FeH] $\simeq -1.4$\,dex; \citealt{koch10a}).
From our data, we find an absolute magnitude of $M_{V}=-6.0\pm0.2$ and an effective
radius of $r_{e,s}=20\pm0.6$\,pc, values that are in line with those of its outer halo 
counterparts,  Palomar~3, AM~1 and Eridanus. Isochrone fitting suggests an age a 
few Gyr younger than inner halo clusters \citep{stetson99a}. Its surface density profile shows a possible 
excess of stars at large radii with respect to the fitted King, or even S\'ersic, model, although 
its two-dimensional morphology appears round and regular (Figure~\ref{dens6}). We 
see no evidence for a tidal tail in the Galactic anticenter direction, as
reported by \citet{sohn03a}.

\subsubsection{Leo~IV and V} 

This pair of low-luminosity satellites was discovered in SDSS DR5 and DR6, respectively, 
by \citet{belokurov07a,belokurov08a}  The two systems are located at large, and rather similar, 
heliocentric distances: $R_{\odot}\simeq160$ (Leo~IV) and $180$\,kpc (Leo~V). 
In the discovery papers, Leo~IV was found to be the larger and brighter of the two systems, 
with $r_{h}\sim160$\,pc and $M_{V}=-5.1\pm0.6$,
compared to $r_{h}\sim40$\,pc and $M_{V}=-4.3\pm0.5$ for Leo~V.
The two systems are separated by just $\sim$\,3$^\circ$
on the sky and are receding with similar heliocentric radial velocities: 
$v_{r}=132$ and $173$\,km s$^{-1}$, respectively \citep{simon07a, belokurov08a}. 
Taken together, these properties might suggest a physical connection.

\citet{dejong10a} used imaging from the 3.5m Calar Alto telescope
to analyzed the distribution of RGB and horizontal branch stars
and found both galaxies to be larger than initially reported, 
with $r_{h}=206\pm36$ and $133\pm31$\,pc for Leo~IV and Leo~V, respectively.
Both galaxies were found to be highly elongated ($\epsilon>0.5$).
Later imaging studies of Leo~IV \citep{sand10a, okamoto12a} reexamined its 
structural properties, reporting values more similar to the initial estimates than to those 
of \citet{dejong10a}.  
\citet{brown14a} presented {\it HST}/ACS photometry
for Leo~IV and argued that the system is composed exclusively
of old, metal-poor stars: i.e., the system formed more than $80$\% of its stars
by $z=6$. Leo~IV is thus another example of a fossil galaxy in the Galactic halo.

In Figure~\ref{cmds6}, we show CMDs for these systems based on our Clay imaging. 
Our photometry reaches just slightly below the MSTO in each galaxy. Although this is
similar in depth to the photometric study of \citet[][]{dejong10a}, 
our measured structural parameters are significantly smaller for both objects. We find 
$r_{e,s}=117\pm14$ and $52\pm11$\,pc, and 
$M_{V}=-5.0\pm0.3$ and $-4.4\pm0.4$, respectively, for Leo~IV and V. We are
unsure of the reason for this discrepancy but note that our results are 
similar to most previously published values using both shallower and deeper photometry 
\citep{belokurov08a,martin08a,sand09a,okamoto12a}.

\subsubsection{Coma Berenices} 

The discovery of Coma Berenices and four other faint Galactic satellites was 
reported by \citet{belokurov07a}. The discovery itself was based on SDSS DR5 imaging, 
but included deeper follow-up observations from Suprime-Cam on the Subaru telescope.
These deeper data were 
used to examine Coma Berenices' stellar content and 
derive structural parameters. On the 
basis of their estimated half-light radius, $r_{h}\sim70$\,pc, these authors classified the 
object as an ultra-faint dwarf galaxy.

Coma Berenices has previously been studied, along with Ursa Major~II, by 
\citet{munoz10a} using the same CFHT imaging included in this study. The
CMD shown in Figure~\ref{cmds7} reaches almost four magnitudes below the MSTO
and shows a well defined subgiant branch as well as a hint of an RGB. \citet{munoz10a}
confirmed that Coma Berenices is a faint, compact dwarf galaxy of modest luminosity. We
find an absolute magnitude of $M_{V}=-4.4\pm0.3$ and an effective radius of
$r_{e,s}=72\pm4$\,pc. Its number density profile is well described by any of the 
density models explored in our analysis, including King and S\'ersic laws (Figure~\ref{dens7}).
Unlike what was seen in the Subaru data, our isodensity map reveals a regular morphology 
with an overall elongation of $\epsilon=0.37\pm0.05$, similar to most
dSph galaxies, with no signs of tidal stripping.

\subsubsection{Canes Venatici~II}   

One of five ultra-faint satellites discovered by \citet{belokurov07a} using SDSS DR5 data, 
Canes Venatici~II is a faint, compact system located $R_{\rm GC} \sim\,161$\,kpc from the 
Galactic center. 

To date, the deepest photometry published for Canes Venatici~II is the Subaru $VI$ data of
\citet{okamoto12a}. In their CMD, a sparsely populated RGB was visible, as well as a few blue HB 
star candidates. Our $gr$ CMD looks nearly identical to the earlier CMD from Okamoto
(Figure~\ref{cmds7}).  Although our photometry is slightly shallower, we cover an area that is
four times larger. Based on our CFHT photometry, we measure an effective radius of $r_{e,s}=70\pm11$\,pc, 
slightly smaller than the \citet{okamoto12a} value, and an absolute magnitude of $M_{V}=-4.9\pm0.4$.
We find an overall ellipticity $\epsilon=0.46\pm0.11$, significantly larger than the value of
$\epsilon=0.23$ reported by \citet{okamoto12a}. The two-dimensional morphology of Canes Venatici~II 
shows no obvious irregularities or perturbations (Figure~\ref{dens7}), although the number of stars at
large radii is low.

\subsubsection{Canes Venatici~I}  

Canes Venatici~I was discovered by \citet{zucker06a} using SDSS DR5 data. At a Galactocentric 
distance of $R_{\rm GC}\sim220$\,kpc, it is one of  the most remote Galactic dwarf satellites.
The original estimate of its absolute magnitude, $M_{V}\sim~-7.9$, placed Canes Venatici~I 
at the edge of the region of the size-luminosity diagram occupied by classical dSph galaxies.
\citet{martin08b} acquired deep $BV$ imaging reaching the level of the MSTO, with the Large Binocular 
Telescope. They found a complex star formation history with at least two populations: 
a spatially extended, old ($>10$\,Gyr) and metal-poor population that dominates ($95$\,\%) the stellar 
mass, and a younger ($\sim1.4$-$2.0$\,Gyr), more metal-rich, and more spatially concentrated
population. 

More recently, \citet{okamoto12a} published the deepest CMD to date, reaching $V\sim26$,
using imaging collected with the Suprime-Cam on the Subaru telescope. These authors
measured a distance of $216\pm8$\,kpc from isochrone fitting, and noted that the system's
horizontal branch morphology and RGB look remarkably similar to those of Draco.

Our CFHT images are nearly as deep as those of Okamoto (see Figure~\ref{cmds7}) but cover a 
roughly four-fold larger area. We measure an effective radius of $r_{e,s}=486\pm14$\,pc and 
an absolute magnitude of $M_{V}=-8.5\pm0.1$. Our isodensity map reveals a morphology that
is quite elongated in the {\tt EW} direction, with an overall ellipticity of $\epsilon=0.46\pm0.2$.
The contours look fairly regular with a possible twist in the outer regions, 
but at a low statistical significance (Figure~\ref{dens7}).

\subsubsection{AM~4}  

AM~4 was discovered by \citet{madore82a} as a sparse stellar overdensity in the ESO/SRC 
Southern Sky Survey. The first CMD for the object was produced by \citet{inman87a} using 
data from the 1.5m telescope at CTIO. Their CMD showed  AM~4 to be an extremely low 
luminosity system with a striking lack of evolved stars. Prior to the discovery of the ultra-faint 
satellites in the SDSS, AM~4 was by far the faintest globular cluster known in the Galaxy.
\citet{carraro09a} obtained new imaging for AM~4 using the 1m telescope at Las Campanas 
Observatory  and estimated an age of $\sim9$\,Gyr. This is similar to the globular cluster Terzan~7,
which is known to be associated with the Sagittarius dwarf galaxy. Due to its extremely low luminosity and 
relatively young age, \citet{carraro09a} postulated that AM~4 may also be associated with 
Sagittarius, although its current location in the Galaxy appears 
inconsistent with that of other Sagittarius debris.

From our Clay imaging, we calculate an absolute magnitude of 
$M_{V}=-0.9\pm0.8$ and an effective radius of $r_{e,s}=7.3\pm1.4$\,pc. 
Its extreme luminosity and fairly compact size are comparable to those of Koposov~1 and 2, 
Mu\~noz~1 and Segue~3, all of which have effective radii smaller than $10$\,pc. 
As with these ultra-faint counterparts, AM~4 shows a  decided lack of stars in the RGB region, which
prevent us from estimating its metallicity photometrically (see Figure~\ref{cmds7}). 
However, the SGB region is well defined and from isochrone fitting we conclude 
that AM~4 is indeed an ancient ($>12$\,Gyr) stellar system.

\subsubsection{Bo\"otes~II}  

Discovered in SDSS DR5 imaging by \citet{walsh07a}, Bo\"otes~II lies just $1\fdg7$ from 
the Bo\"otes~I satellite (discovered one year earlier). Unlike the case of Leo~IV and~V, 
the two systems do not appear to form a physical pair. Bo\"otes~II is located closer to the Galactic 
center than Bo\"otes~I with a distance difference comparable to that of the Leo~IV and V pair:  
($R_{\rm GC} \simeq$ 42 and $60$\,kpc, respectively; \citealt{walsh08a}). 
However, Bo\"otes~II is moving with a mean heliocentric radial velocity that differs by nearly $200$\,km s$^{-1}$ 
from that of Bo\"otes~I, and in the opposite direction \citep{koch09a}. A physical association between the 
two satellites thus seems unlikely.

From our CFHT data, we measure an absolute magnitude of $M_{V}=-2.9\pm0.7$ and
an effective radius of $r_{e,s}=37\pm6$\,pc. These values are slightly smaller than previous estimates 
\citep{walsh08a, koch09a}, and make Bo\"otes~II one of the smallest and faintest of the ultra-faint satellites. Our 
imaging, which reaches $\sim$\,3 magnitudes below the MSTO, shows the
CMD (Figure~\ref{cmds8}) to be sparsely populated, with only a handful of 
RGB star candidates. Despite its relatively low overall ellipticity, $\epsilon=0.24\pm0.12$,  
Bo\"otes~II is one of the most distorted of the ultra-faint systems. Its two-dimensional morphology 
shows an irregular structure (Figure~\ref{dens8}) with multiple 
tidal features that resemble those of Willman~1 --- a system that is similar in size and luminosity, and
located at a comparable Galactocentric distance of $R_{\rm GC}\sim40$\,kpc.

It has been suggested in the past that Bo\"otes~II may be associated with the Sagittarius dwarf galaxy
\citep{koch09a}. Our CFHT photometry shows that a second main sequence is clearly present 
in this field --- well beyond the extent of the dwarf and likely associated with Sagittarius \citep{law10a}. From 
isochrone fitting, we find the population responsible for this second sequence to be located
at heliocentric distance of $\sim55$\,kpc. Bo\"otes~II itself is located in the foreground,
at a distance of $\sim42$\,kpc.

\subsubsection{Bo\"otes~I}  

The Bo\"otes~I dwarf galaxy was one of the first ultra-faint satellites to be discovered in 
SDSS \citep{belokurov06a}. The discovery article placed it at a Galactocentric distance of 
$R_{\rm GC}\sim 60$\,kpc and estimated the half-light radius to be $r_h\sim220$\,pc,
similar in size to classical dSph galaxies. \citet{munoz06b} used the WIYN telescope to 
carry out the first spectroscopic 
study of this system, and found its velocity dispersion of $6.6\pm2.3$\,km s$^{-1}$ to translate 
into a mass-to-light ratio between $130$ and $610$ in solar units. They also reported a
mean metallicity of [Fe/H]\,$\sim-2.5$\,dex, making Bo\"otes~I the darkest and 
least chemically evolved dwarf galaxy known at that time.

Our CFHT imaging (covering a $2^\circ\times^\circ1$ region) reaches $\sim$\,2.5 magnitudes 
below the MSTO. The subgiant branch and lower RGB are clearly defined and a blue
horizontal branch is discernible, although it is sparsely populated. A clear
blue straggler sequence is also visible (Figure~\ref{cmds8}).
We measure an absolute magnitude of $M_{V}=-6.0\pm0.2$, 
an effective radius of $r_{e,s}=216\pm5$\,pc, and an overall ellipticity of 
$\epsilon=0.25\pm0.02$, consistent with previous estimates. Quite recently,
\citet{roderick16a} have presented a deep imaging survey carried out
with the DECam imager on the CTIO 4m telescope. They found
a large, extended stellar substructure surrounding the galaxy and 
argued that this system may have undergone significant tidal disruption. Our
isodensity contour map (Figure~\ref{dens8}) shows an outer
structure that is consistent with the findings of \citet{roderick16a}.
  
\subsubsection{NGC5694}   
 
Located at a Galactocentric radius of $R_{\rm GC}\simeq29$\,kpc, NGC5694 was discovered in 1784 by
Herschel. It was first resolved into stars and confirmed to be a globular cluster by \citet{lampland32a} using 
plates taken with the 13-inch Lawrence-Lowell telescope at Lowell Observatory.  
 An unusual feature of NGC5694 is its large spatial extent. \citet{correnti11a} reported the discovery of 
a low surface brightness halo surrounding the cluster. They were able to trace this feature out to a distance of 
at least $\sim9\arcmin$\ ($20r_{e}$). This is well beyond the King limiting radius of 
$4\farcm28$ estimated by \citet{trager95a}.
In the same vein, \citet{bellazzini15a} measured the velocity
dispersion profile of NGC5694 and found it to decrease, and then flatten, out to a distance of
$14r_{h}$. They argued that NGC5694 is a cluster that has yet to fill
its Roche Lobe, remaining tidally undisturbed after evolving in isolation.  They
also noted that this seems to be the case for a number of outer halo
clusters, including Eridanus, Palomar~2, NGC5824, Palomar~4, NGC6229, NGC7006 
and NGC7492.

From our CFHT data, we measure an effective radius of $r_{e,s}=4.3\pm0.1$\,pc and an absolute 
magnitude of $M_{V}=-7.9\pm0.1$. In agreement with \citet{correnti11a}, we detect a main 
sequence population out to at least $9\arcmin (\simeq$ 90\,pc). The cluster is fairly round and regular in 
structure, with $\epsilon=0.06\pm0.02$, and its CMD shows an extended blue horizontal branch (Figure~\ref{cmds8}). 
At a Galactocentric distance of $R_{GC}=29$\,kpc, it is located close to NGC5824, an outer halo cluster 
that is similar in size, luminosity and horizontal branch morphology.  Overall, NGC5694 differs markedly 
from most of the  halo clusters beyond $R_{\rm GC}\sim80$\,kpc in being brighter, more concentrated,
and with a blue horizontal branch.

\subsubsection{Mu\~noz~1} 

Mu\~noz~1 is the lone object in our survey to be discovered from our own imaging \citep{munoz12b}. 
Identified on the CFHT images acquired for the Ursa Minor dSph galaxy, which were arranged in a 
$4\times4$ grid, Mu\~noz~1 was detected as a stellar overdensity located $\sim$\,45\arcmin~ to the 
{\tt SW} of the Ursa Minor photocenter (and inside the latter's tidal radius). Indeed, Mu\~noz~1 is visible in 
the isodensity contour map for Ursa Minor as the round, compact feature in Figure~\ref{dens9}. 
Spectra taken with Keck/DEIMOS yielded a systemic radial velocity of $v_{r}=-137\pm4$\,km s$^{-1}$, 
which is lower than that of Ursa Minor by more than $100$\,km s$^{-1}$. This fact, along with a 
line-of-sight  distance difference of $\sim30$\,kpc (in the sense that Mu\~noz~1 is the closer of the 
two systems), rules out any physical association.

Our CMD shows a clear MSTO, although the system is so faint that almost no stars are visible in either the 
subgiant or RGB regions (Figure~\ref{cmds8}). We measure an effective radius of $r_{e,s}=22\pm5$\,pc and an 
absolute magnitude of $M_{V}=-0.5\pm0.9$, making Mu\~noz~1 one of the faintest of known
Galactic satellites.

\subsubsection{NGC5824}  

At a Galactocentric distance of $R_{\rm GC}=26$\,kpc, and with an absolute magnitude (from this
work) of $M_{V}=-9.3\pm0.04$, NGC5824 is the second brightest of the outer halo clusters 
(after NGC2419). It was discovered in the 19th century but not studied in detail for roughly a century. 

From our data, we measure an effective radius of  $r_{e,s}=4.9\pm0.1$\,pc and a 
fairly round shape, $\epsilon=0.03\pm0.01$, with no obvious signs of morphological perturbation. 
Our CMD reaches more than three magnitudes below the MSTO (Figure~\ref{cmds9}) 
and clearly 
shows an extended blue horizontal branch. This is similar to NGC5694 but 
unlike other extended clusters at $R_{\rm GC}\sim100$\,kpc.  Our 
photometry is consistent with an old age ($>12$\,Gyr), as was previously noted by \citet{sanna14a} 
from {\it HST}/WFPC2 data. As is the case with NGC5694, 
the clusters is not well described by a King profile \citep{carballo12a}, but instead shows 
a number density profile that is well fitted by a power law \citep{sanna14a} to a
distance of more than $30r_{e,s}$ (Figure~\ref{dens9}). 
This is a remarkable result, as it means that we are able to detect NGC5824 member stars 
out to a distance of at least $\sim180$\,pc from its center.

\subsubsection{Ursa Minor}  

The Ursa Minor galaxy was discovered by \citet{wilson55a} using photographic plates
from the 48-inch Palomar Schmidt telescope taken for the National Geographic Society-Palomar 
Observatory Sky Survey. 
\citet{hodge64a} used plates taken with the 120-inch Lick and the 48-inch Palomar Schmidt 
telescopes to carry out a star count analysis of Ursa~Minor. He found an ellipticity
of $\epsilon=0.55\pm0.10$ and fitted a \citet{king62a} model to measure 
a physical major-axis radius of $1.5\pm0.5$\,kpc ($75\arcmin\pm25\arcmin$).

In a pioneering study, \citet{aaronson87a} measured a velocity dispersion of $\sigma_{v_{r}}=11\pm3$\,km s$^{-1}$, 
for Ursa Minor, prompting many subsequent investigations into the dark matter content of this
and other Galactic dSph galaxies \citep[e.g.,][]{pryor90a, lake90a}. \citet{hargreaves94a},
\citet{olszewski95a, olszewski96a}, \citet{wilkinson04a} and \citet{munoz05a}
led subsequent spectroscopic campaigns, collecting hundreds of
radial velocity measurements over the following decade. 

More recent photometric studies have focused on the galaxy's morphological structure, including 
a survey of main sequence and blue horizontal branch stars by \citet{martinez01a} 
and K giants by \citet{palma03a}. Both studies reported the detection of probable member
stars well beyond the King tidal radius. Based on this finding, and the elongated 
shape of the galaxy, these authors have argued that Ursa Minor has 
experienced significant tidal heating.

In our program, we used CFHT to cover the galaxy in a $2\times2$ mosaic, an area of nearly 
4 deg$^2$. Our CMD reaches roughly three magnitudes below the MSTO, revealing in great
detail all of the galaxy's evolutionary sequences including a large population of
blue straggler stars (Figure~\ref{cmds9}).
We measure the effective radius to be $r_{e,s}=383\pm2$\,pc. Although this is significantly larger than the 
value of $r_{h}=180$\,pc from \citet{irwin95a}, it is consistent with the measurement of $r_{h}\sim390$\,pc 
from \citet{bellazzini02a}. We find an absolute magnitude of  $M_{V}=-9.0\pm0.05$. In 
agreement with previous studies, the galaxy is found to be fairly elongated, with a global ellipticity 
of $\epsilon=0.55\pm0.10$.  On the other hand, its isodensity contour map (Figure~\ref{dens9}) 
shows no obvious secondary peak in the 
density distribution, contrary to several previous reports  
\citep[e.g.,][]{kleyna98a, bellazzini02a,palma03a,pace14a}.  

\subsubsection{Palomar~14} 

Palomar~14 was discovered by
\citet{arp60a} from an inspection of Palomar Sky Survey plates. It is a sparsely populated cluster 
located at a Galactocentric distance of $R_{\rm GC}\simeq71$\,kpc. 
Palomar~14 exhibits the 
second parameter effect like most of its outer halo counterparts \citep{dacosta82a, harris84a}.
\citet{sarajedini97a} was the first to recognize that Palomar~14 was indeed younger than the inner halo
clusters by $\sim$\,3--4\,Gyr, a typical value for remote halo clusters. The age difference was later revised to 1.5--2\,Gyr 
by \citet{dotter08a} using high-quality {\it HST}/WFPC2 photometry.

We observed Palomar~14 in a single CFHT pointing, reaching $\sim$\,1.5 magnitudes 
below the level of the MSTO (Figure~\ref{cmds9}).
From our data, we measure an absolute magnitude of $M_{V}=-5.4\pm0.2$ and
an effective radius of $r_{e,s}=32\pm1$\,pc, leading to a
surface brightness that is among the faintest for clusters in our sample. 
We also find a relatively high ellipticity (compared to other clusters) of $\epsilon=0.11\pm0.01$.
It has been reported that the cluster shows some evidence for tidal interaction 
\citep{jordi10a,sollima11a}. In these previous studies, the morphology appears distorted, 
especially in its outer regions, suggesting the presence of tidal debris.
Our isodensity map (Figure~\ref{dens9}) shows a possible elongation in the {\tt E-W}
direction, similar to published results and consistent with the presence of nascent tidal tails.

\subsubsection{Hercules}  

\citet{belokurov07a} reported the discovery of this remote, low-luminosity satellite in SDSS DR5 data. 
Soon after its discovery, \citet{coleman07a} presented deeper follow-up imaging taken with the 
Large Binocular Telescope. Their CMD reached $\sim1.5$\,magnitudes below the MSTO, 
revealing a highly elongated structure. Indeed, these authors noted that Hercules is the most 
elongated of the ultra-faint satellites, with an axial ratio of $\sim$ 3:1, and also one of the
largest, with a half-light radius of $r_{h}\sim170$\,pc.  

Despite its large distance, Hercules is likely to be among the most tidally stripped of satellites,
given its elongated shape and unusual morphology. Evidence for tidal disruption --- both 
photometric and spectroscopic --- has been reported in the past 
\citep[e.g.,][]{coleman07a,aden09a,sand09a,martin10a,deason12a,blana15a,garling18a}.
\citet{martin10a} found the spatial orientation of 
Hercules' elongation to be consistent with an orbit that would bring the satellite as close
as $6^{+9}_{-2}$\,kpc from the Galactic center, making the tidal disruption scenario
a plausible hypothesis.

Our CFHT imaging for Hercules (covering a $2^\circ\times1^\circ$ field) reaches almost a full 
magnitude below the MSTO (Figure~\ref{cmds9}). Although its CMD is scarcely populated,
the evolved sequences are clearly visible. A potential blue straggler population is also apparent, 
as was seen in the deep {\it HST}/ACS photometry of \citet{brown12a}.
We measure the overall ellipticity of Hercules to be $\epsilon=0.69\pm0.04$,
although the ellipticity in the inner parts can be even larger. The estimated
absolute magnitude is $M_{V}=-5.2\pm0.4$ and its effective
radius $r_{e,s}=230\pm23$\,pc, making Hercules somewhat fainter and larger
than the original estimates from \citet{belokurov07a} and \citet{coleman07a}.
As expected, our data shows an unusually elongated morphology
(Figure~\ref{dens9}), consistent with earlier studies.

\subsubsection{NGC6229}  

NGC6229, discovered by Herschel in 1787, is located at a Galactocentric distance of 
$R_{\rm GC}\simeq30$\,kpc. 
NGC6229 is
another halo cluster that exhibits the second parameter effect. There is, however, no direct
evidence to date that NGC6229 is younger than its inner halo counterparts.

Our CFHT photometry reaches $\sim$\,3.5 magnitudes below the MSTO and 
reveals a well populated horizontal branch that covers nearly a full magnitude in $(g-r)$ 
color and exhibits an apparent bifurcation at its red edge (see Figure~\ref{cmds10}). From 
isochrone fitting, we estimate that
NGC6229's age is consistent with being old (i.e., $>13$\,Gyr).
We measure an absolute magnitude of $M_{V}=-8.0\pm0.2$ and 
an effective radius of $r_{e,s}=3.19\pm0.09$\,pc. We are able to
detect probable member stars out to $25r_{e}$, similar to other bright
halo clusters in the range $30<R_{GC}<40$\,kpc.
Our isodensity map reveals the cluster to be fairly round in shape,
($\epsilon=0.02\pm0.01$, with a regular morphology and no signs of tidal 
distortion (Figure~\ref{dens10}).  As expected due to crowding, its number density 
profile produced only from star counts is poorly fit by the parameters derived from the
surface brightness analysis described in \S 3.3.

\subsubsection{Palomar~15}  

This diffuse and somewhat poorly studied cluster was discovered by \citet{zwicky59a}.
It is located in the Galactic anticenter direction, at a relatively low 
Galactic latitude, ($l =19^{\circ},~b = 24^{\circ}$), which results in a significantly reddened CMD.
\citet{dotter11a} used imaging from {\it HST}/ACS to derive an age 
of $13\pm1.5$\,Gyr.

Our CFHT images reach nearly two magnitudes fainter than the MSTO, as shown
in Figure~\ref{cmds10}.  The cluster sequences, including its blue horizontal branch, appear 
somewhat broader than usual for sparse halo clusters, probably owing to differential reddening 
across the field. From our data, we derived an absolute magnitude of $M_{V}=-5.6\pm0.2$ 
and an effective radius of $r_{e,s}=19.0\pm0.4$\,pc. Palomar~15's overall ellipticity is 
$\epsilon=0.05\pm0.02$ and its two-dimensional morphology shows no  irregularities or signs 
of tidal interaction (Figure~\ref{dens10}). We note that \citet{myeong17a}, using photometry
taken with the DECam imager on the 4m Blanco telescope at CTIO, reports the presence of tidal-like 
substructure on the outskirts of Palomar~15.

\subsubsection{Draco}  

The Draco dSph galaxy was discovered by \citet{wilson55a} from an inspection of plates taken 
with the 48-inch Palomar Schmidt telescope for the National Geographic Society-Palomar Observatory
Sky Survey. 
\citet{hodge64b} used 200-inch Palomar plates,
among others, to perform star counts in the galaxy. His analysis showed that the galaxy
resembled other ``Sculptor-type" systems: i.e., it had an elliptical
appearance, with $\epsilon=0.29\pm0.04$, and was well represented by a \citet{king62a} model 
having a limiting radius of $510\pm40$\,pc (for his adopted distance 
of $68$\,kpc). 

During the last decade,  attention has been devoted to the possibility that
Draco may be tidally influenced by the Milky Way. Using the radial velocity
data of \citet{kleyna02a}, \citet{munoz05a} reanalyzed Draco's velocity
dispersion profile. They found it to remain flat beyond the nominal tidal
radius, suggesting tidal stripping as a possible explanation. On the other hand,
several deep photometric studies 
\citep[e.g.,][]{odenkirchen01a, klessen03a, segall07a} have been unable to detect the
morphological features expected for tidally disrupting systems, such as 
isophotal twisting or tidal tails. 

In our survey, we used CFHT to image Draco in a  $2\times2$\, grid pattern, covering an area of nearly 
4 deg$^2$ and reaching more than two magnitudes below the MSTO.
The combination of depth and areal coverage makes our imaging survey the most
extensive to date for this galaxy. Our CMD (Figure~\ref{cmds10}) reveals Draco's extended
horizontal branch as well as its very sizable population of blue stragglers.
We measure an effective radius of $r_{e,s}=219\pm2$\,pc, an absolute magnitude of 
$M_{V}=-8.7\pm0.1$, and a mean ellipticity of $\epsilon=0.30\pm0.01$,
consistent with previous determinations. 
With our photometry, we are able to study Draco's two-dimensional morphology to a fainter
surface brightness limit than any previous study. However, the galaxy's isodensity contour 
map shows no sign of irregularities (Figure~\ref{dens10}) that would suggest it has been strongly
perturbed by the Galactic tidal field.

\subsubsection{NGC7006}  

NGC7006 is a low-latitude globular cluster, $(l=64^{\circ},~b=-19^{\circ})$, located at a 
Galactocentric distance of $R_{\rm GC}\simeq38$\,kpc. Several early studies placed it on the far
side of the Galaxy, at distances greater than $R_{\rm GC}\sim50$\,kpc
\citep[e.g.,][]{shapley21a,shapley20a,shapley30a,baade35a}. As a result, it attracted
considerable attention in studies of the spatial extent of the Galaxy: i.e.,
prior to 1950, it was considered the second most distant cluster in the 
Milky Way, after NGC2419.

Our CFHT photometry reaches $\sim$\,4 magnitudes below the level of the MSTO (Figure~\ref{cmds10}). 
We measure an absolute magnitude of $M_{V}=-7.4\pm0.1$ and an effective radius of 
$r_{e,s}=6.11\pm0.12$\,pc. Although its surface brightness profile is not as extended
as those of NGC5694 or NGC5824, we are able to detect likely member stars out to $\sim\,16r_{e}$. 
Like most clusters in our sample, its morphology is round and regular (Figure~\ref{dens10}), with an 
overall ellipticity of $\epsilon=0.07\pm0.01$ and no signs of tidal distortion. We see no evidence for
the extratidal halo reported by \citet{jordi10a} from an analysis of SDSS images.

\subsubsection{Segue~3}  

This extremely faint satellite was discovered by \citet{belokurov10a} using SDSS DR7 
data. From follow-up imaging acquired at the 4m KPNO telescope, these authors estimated 
a half-light radius of $r_{h}\sim3$\,pc and an absolute magnitude of $M_{V}=-1.2$, making
Segue~3 one of the faintest Galactic satellites currently known. Based on its compact size, 
they classified the system as a faint globular cluster.

\citet{fadely11a} used deep imaging taken with the 6.5m Baade telescope to conclude that 
Segue~3 is an old halo cluster located at Galactocentric distance of $R_{\rm GC}\sim17$\,kpc.
Soon afterwards, \citet{ortolani13a} used $BVI$ imaging from the Telescopio Nazionale Galileo
to argue that Segue~3 is, in fact, a much younger system, with an age of just $\sim3.2$\,Gyr, that 
is located at a distance of $R_{\rm GC}\sim29$\,kpc. Recently, \citet{boettcher13a} 
have carried out a search for RR Lyrae stars in the Segue~3 field, but were unable to
identify any promising candidates.

From our Clay imaging, we measure an absolute magnitude of  $M_{V}=-0.9\pm0.7$ and 
an effective radius of $r_{e,s}=4.1\pm0.7$\,pc, making Segue~3
somewhat fainter and larger than the estimates of \citet{belokurov10a}.
Age, distance and metallicity estimates from isochrone fitting are critically 
dependent on the presence of subgiant stars. In our CMD 
(Figure~\ref{cmds11}), a handful of stars are found in the subgiant branch region; if these are bona fide 
cluster members, then our isochrone fitting results are consistent with those findings of \citet{ortolani13a}, 
supporting the observation that Segue~3 is (by far) the youngest of outer halo clusters. 
However, given the extremely sparse nature of the cluster's upper sequences, 
definitive measurements of age, metallicity and distance are probably not possible at this time.
Spectroscopic identification of member stars will be needed to settle this discrepancy.

\subsubsection{Pisces~II}  

The discovery of Pisces~II was reported simultaneously with that of Segue~3 by
\citet{belokurov10a}, who detected it as an overdensity in SDSS DR7 data. 
Based on follow-up imaging acquired with the KPNO 4m telescope, the authors 
calculated an absolute magnitude of $M_{V}=-5.0$, a half-light radius of 
$r_{h}=58$\,pc and a distance of $\sim182$\,kpc. \citet{sand12a} obtained
deeper imaging with the Clay telescope, revising the absolute magnitude to 
$M_{V}=-4.1\pm0.4$ and confirming the size and distance measurements of 
\citet{belokurov10a}.

From our Clay imaging, we measure an effective radius of $r_{e,s}=64\pm10$\,pc, an
absolute magnitude of $M_{V}=-4.2\pm0.4$ and an overall ellipticity of 
$\epsilon=0.40\pm0.10$ . These values are all consistent with the results of \citet{sand12a}.
The CMD of Pisces~II (Figure~\ref{cmds11}) shows a sparsely populated but well defined 
blue horizontal branch. Its two-dimensional morphology shows no sign of tidal stripping.

\subsubsection{Palomar~13}  

The sparsely populated globular cluster Palomar~13 was discovered by \citet{wilson55a}
using plates from the National Geographic Society-Palomar Observatory Sky Survey. 
Based on {\it HST}/WFC3 imaging that reached $m_{\rm F606W}\sim27.2$,
\citet{hamren13a} reported an age of $13.4\pm0.5$\,Gyr and a metallicity of [Fe/H]\,$\sim-1.6$\,dex.

We observed Palomar~13 in a single pointing with CFHT. Our photometry
reaches to $g\sim25$, roughly four magnitudes below the  MSTO 
(Figure~\ref{cmds11}). The CMD reveals a prominent blue straggler population, consistent
with previous findings \citep{cote02a,clark04a}.
From our data, we measure an effective radius of $r_{e}=9.5\pm0.7$\,pc, 
and an absolute magnitude of $M_{V}=-2.8\pm0.6$. The overall ellipticity is measured to
be $\epsilon=0.10\pm0.06$, on the high side for globular clusters.
  
\subsubsection{NGC7492}  

NGC7492 is a sparse and relatively nearby ($R_{\rm GC}\simeq25$\,kpc) outer halo cluster.
The first CCD study of the cluster was carried out by \citet{buonanno87a}, whose CMD
reached below the level of the MSTO. From isochrone fitting, these authors estimated a metallicity 
of [Fe/H]\,$=-1.51\pm0.20$\,dex. 
\citet{cote91a} presented a deep CMD, 
confirming the
metallicity estimate of \citet{buonanno87a} and refining their distance estimates.
 
Figure~\ref{cmds11} shows the CMD based on our CFHT imaging, which reaches about 
four magnitudes below the level of the MSTO. We measure an effective radius of 
$r_{e,s}=9.6\pm0.1$\,pc and an absolute magnitude of  $M_{V}=-6.10\pm0.05$. 
We see no indication of surrounding tidal debris, contrary to the claims of \citet{lee04a}.
Its isodensity contour map (Figure~\ref{dens11}) shows a regular and almost perfectly
round morphology with $\epsilon=0.02\pm0.02$. However, we do note that an additional population of 
main sequence stars
seem to be present in the NGC7492 field, located at a similar line-of-sight distance. In a companion paper, we
show that this new population probably corresponds to debris from the Sagittarius dwarf galaxy 
and it is likely not physically associated with the cluster (\citeauthor*{munoz18a}) .

\subsection{Comments on Secondary Targets}
\label{subsec:prevcom}

Basic data for the 14 satellites that make up our secondary sample are presented in
Table~\ref{t:cat1}, and our best-fit exponential, Plummer, King and S\'ersic parameters for these systems are 
summarized in Tables~\ref{t:structural1}, \ref{t:structural2} and \ref{t:structural3}. Like our primary objects, these
secondary targets include a mixture of ultra-faint dwarf galaxies, diffuse star clusters, and low-luminosity objects 
of an as-yet-undetermined nature.  The full sample consists of: Balbinot~1 \citep{balbinot13a}; Laevens~1 and 
2 \citep{laevens14a,laevens15a};  Kim~1 \citep{kim15a}; Hor~II \citep{kim15c}; Hyd~II \citep{martin15a}; Gru~I \citep{koposov15a};
Ind~I \citep{kim15b,bechtol15a,koposov15a}; and Eri~III, Hor~I, Ret~II, Eri~II, Pic~I and Pho~II \citep{bechtol15a,koposov15a}

Our analysis of these objects is based on the same data used in the discovery papers: i.e., either our own analysis 
of the original images retrieved from the archive, or photometric catalogs kindly provided by the authors (see \S\ref{sec:secondary}).
Thus, we expect no dramatic differences between our measurements and those available in the literature. Still, our analysis 
of these systems allows us to: (1) carry out an independent check on our methodology (which can be important for faint, diffuse
objects observed against a background of contaminating sources); and (2) report photometric and
structural parameters for these objects that were measured in an identical manner as those for our primary sample.

\begin{figure}
\plotone{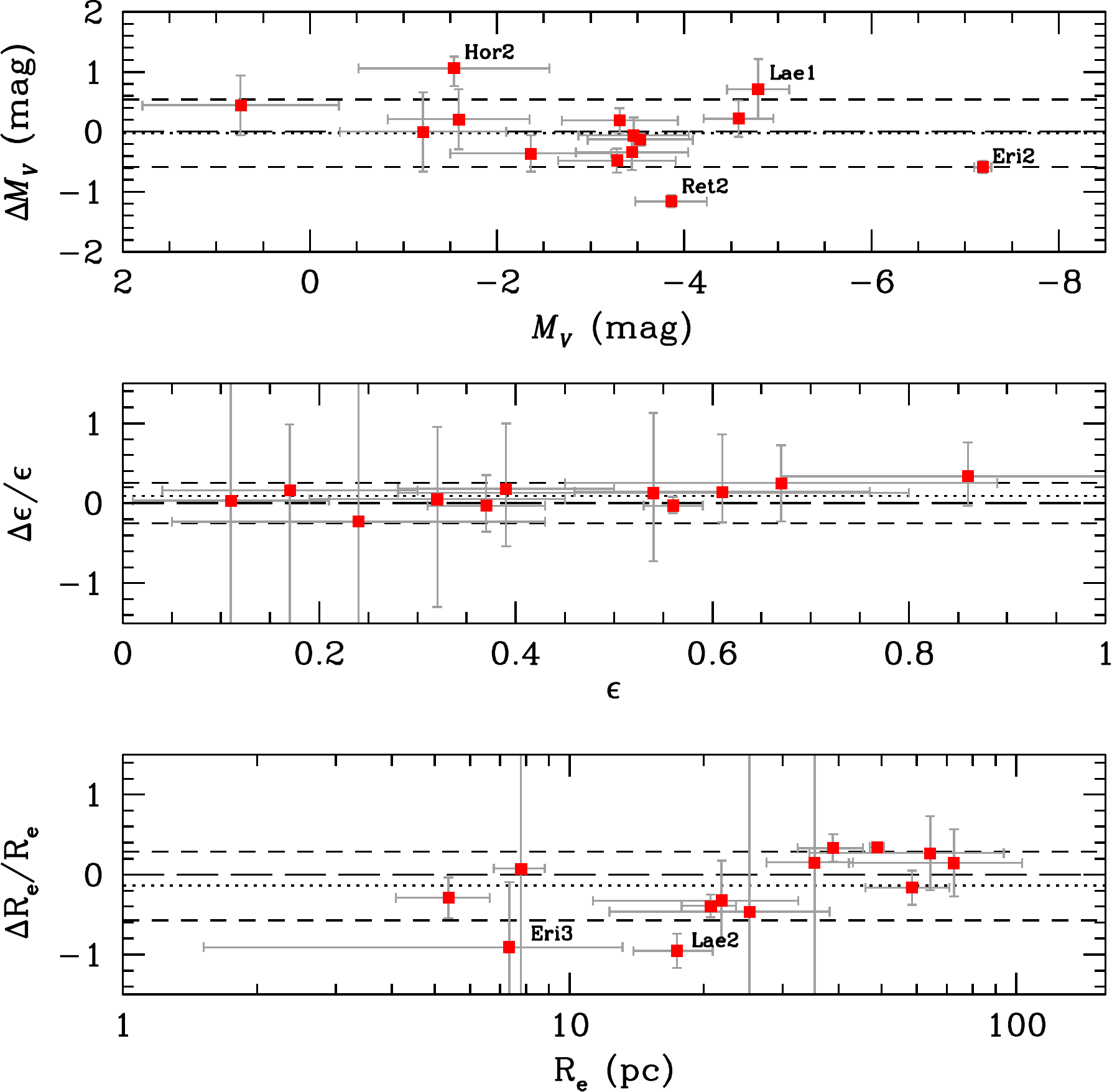}
\caption{
Comparison of photometric and structural parameters, relative to the literature values, for the 14 satellites in our
secondary sample. From top to bottom, the panels show comparisons of absolute magnitude, ellipticity and 
effective radius. In all cases, the plotted residuals are in the sense our values minus those in the literature. 
} \label{comp3}
\end{figure}

Figure~\ref{comp3} presents a comparison between our measured absolute magnitudes, ellipticities and effective radii
with those reported in the above papers; in all cases, residuals are in the sense of our values minus those in the literature. The 
horizontal dotted and dashed lines show the mean differences and $\pm$1$\sigma$ scatter in each case: i.e.,
\begin{equation}
\begin{array}{lcl}
\Delta{M_V} & = & -0.02\pm0.56 \\
\Delta\epsilon/\epsilon & = & +0.09\pm0.16 \\
{\Delta}R_e/R_e & = & -0.14\pm0.43 \\
\end{array}
\end{equation}

For most systems, there is very good agreement between our measurements and those in
the literature (e.g., Laevens~2, Eri~III, Hor~I, Hyd~II, Ind~I, Balbinot~1, Pic~I and Gru~I). In a few other cases,
there is marginal disagreement in one or more derived parameters:

\begin{enumerate}
  \item Hor~II: We measure $M_V = -1.54\pm1.02$ and $R_e = 64\pm30$~pc, slightly fainter and larger
  than reported by \citet{kim15c}, who give  $M_V = -2.60\pm0.20$ and $R_e = 47\pm10$~pc. We also find this 
  system to be more flattened ($\epsilon = 0.86_{-0.19}^{+0.14}$) than previously reported ($\epsilon=0.52^{+0.13}_{-0.17}$). 
  However, most parameters are consistent within their respective errors. 
  \item Ret~II: We find this system to be somewhat brighter ($M_V = -3.65\pm0.24$) and larger
   ($R_e = 49\pm2$~pc) than reported by \citet{koposov15a}, who found $M_V = -2.70\pm0.10$
   and $R_e = 32\pm2$~pc. However, our values are in good agreement with those ($M_V = -3.60\pm0.10$ 
   and $R_e = 55\pm5$~pc) quoted in \citet{bechtol15a}, and all three studies measure an
   ellipticity in the range 0.56 to 0.60.
  \item Eri~II: We measure $M_V = -7.19\pm0.09$ and $R_e = 200\pm19$~pc, slightly brighter and larger
  than reported by \citet{koposov15a}, who found  $M_V = -6.60\pm0.10$ and $R_e = 169\pm16$~pc. By
  contrast, \citet{bechtol15a} report $M_V = -7.40\pm0.10$ and $R_e = 172\pm57$~pc. \citet{crnojevic16a}
  published deeper photometry and reported $M_V = -7.1\pm0.3$ and $R_e = 277\pm14$~pc.
  \item Laevens~1: We measure $M_V = -4.62\pm0.22$ and $R_e = 20.7\pm2.9$~pc, making this 
  system somewhat fainter and smaller than reported by \citet{laevens14a}: $M_V = -5.50\pm0.50$ and 
  $R_e = 28.8\pm2.4$~pc. The respective ellipticities ($0.11\pm0.10$ and $0.08\pm0.08$) are in good
  agreement.
  \item Kim~1: Parameter estimation for this system is especially challenging given its very low
  luminosity (the lowest of any known Galactic satellite). We find $M_V = +0.74\pm1.05$, 
  $R_e = 5.4\pm1.3$~pc and $\epsilon = 0.67\pm0.22$. For comparison, \citet{kim15a} report
  $M_V = +0.30\pm0.50$, $R_e = 6.9\pm0.6$~pc and $\epsilon = 0.42\pm0.10$. However, all measurements
  are consistent within the uncertainties.
  \item Pho~II: Our measurements ($M_V = -3.28\pm0.63$ and $R_e = 38.9\pm6.5$\,pc) point to a somewhat
  brighter and more extended system than found by \citet{koposov15a}: $M_V = -2.80\pm0.20$ and 
  $R_e = 26.0\pm6.2$\,pc. In both cases, though, the measurements agree to within their respective errors. We note
  that the estimates of \citet{bechtol15a}, $M_V = -3.70\pm0.40$ and $R_e = 33^{+20}_{-11}$~pc, are 
  more in line with our measurements.
\end{enumerate}

\section{Discussion: The Magnitude-Size relationship}
\label{sec:discussion} 

As discussed in \S\ref{sec:introduction}, our photometric catalogs and structural parameters  have already formed 
the basis of several published studies \citep{bradford11a,munoz12a,santana13a,carballo15a,santana16a,carballo17a}, and will 
be used in a future paper to explore the scaling relations of outer halo satellites. For the time being, we use our 
measurements to examine the distribution of these objects in the size-luminosity plane --- a customary tool for studying the 
structural properties of stellar systems in the Milky Way, M31 and nearby galaxy groups and clusters \citep[e.g.,][]{kormendy90a,cote02a,willman05a,belokurov07a,dabringhausen08a,tollerud11a,huxor11a,misgeld11a,norris14a}.

\begin{figure}
\plotone{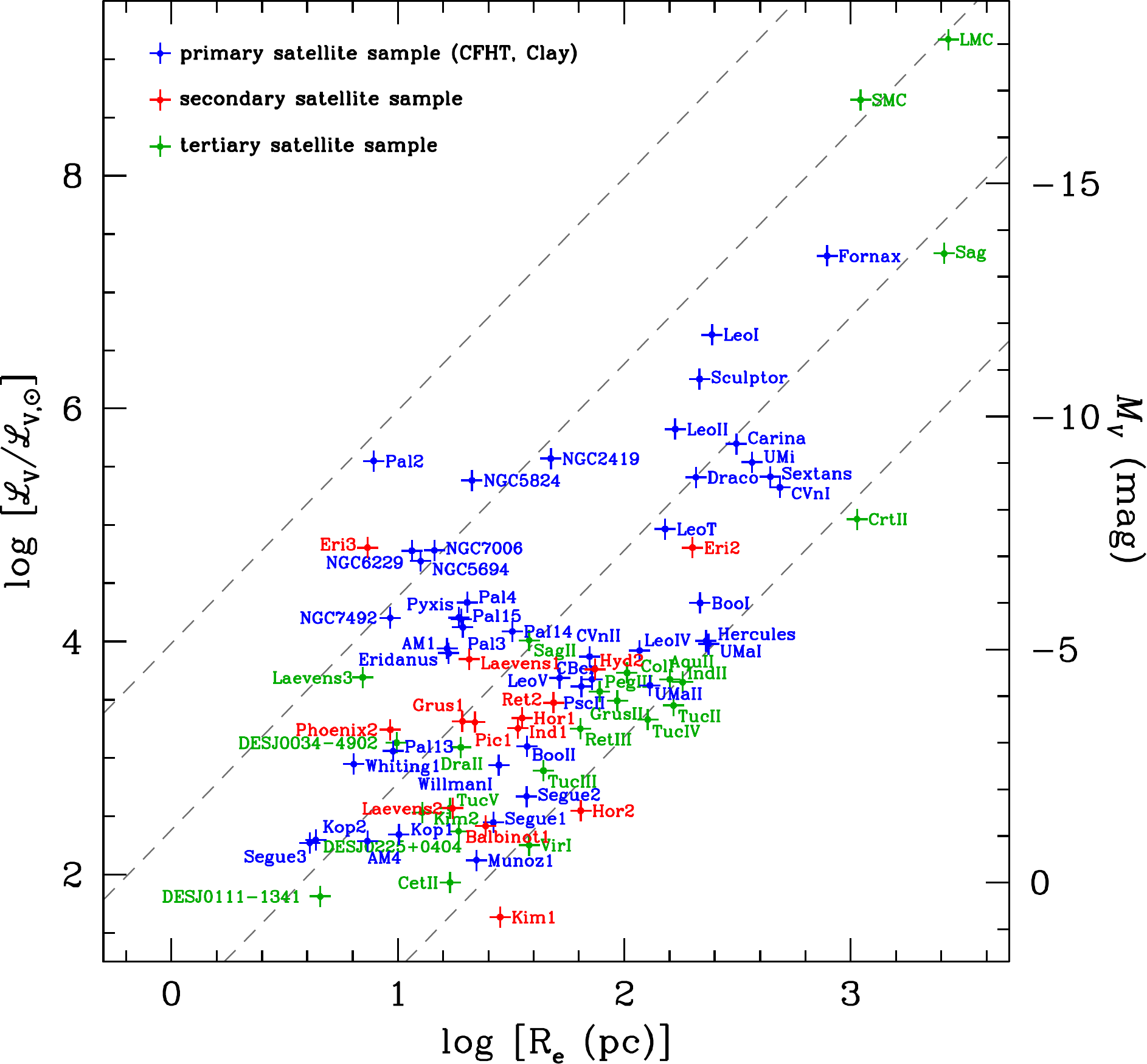}
\caption{Distribution of outer halo Milky Way satellites in the size-luminosity plane. The 58
objects belonging to our primary (44) and secondary (14) samples are shown in blue and red, respectively.
Green symbols show 19 satellites that were not included in our survey, most of which were
recently discovered. For these 19 objects, we show $M_V$ and $r_h$ measurements from the literature.
Note that four satellites shown here have Galactocentric distances less than 25~kpc, and thus do not strictly meet
our definition of ``outer halo" --- Sag ($R_{\rm GC}$ = 18.0 kpc), Kim~1 (19.2), Draco~II (22.0) and Tuc~III (23.0). 
The dashed curves shows lines of constant surface brightness: $\mu_V =$ 18, 22, 26 and 30 mag~arcsec$^{-2}$.
} \label{sizemag1}
\end{figure}

Figure~\ref{sizemag1} shows the distribution of 81 Galactic substructures in the $\log{R_e}$-$\log{L_{V}}$ 
diagram.\footnote{For completeness, we include four known or suspected dwarf satellites of the Milky Way (Sag, Kim~1, Draco~II 
and Tucana~III) that have $R_{\rm GC}$ < 25~kpc and are thus, strictly speaking, not members of the outer halo according to
the definition adopted here.} Objects have been labelled individually and color-coded according to the samples 
from which they are drawn, with primary, secondary and tertiary objects shown in blue, red and green, respectively. 
For comparison, we also show lines of constant surface brightness: $\mu_V$ = 18, 22, 26 and 30 mag~arcsec$^{-2}$.
These limits roughly bound the surface brightnesses of the outer halo substructures known at this time.

\begin{figure}
\plotone{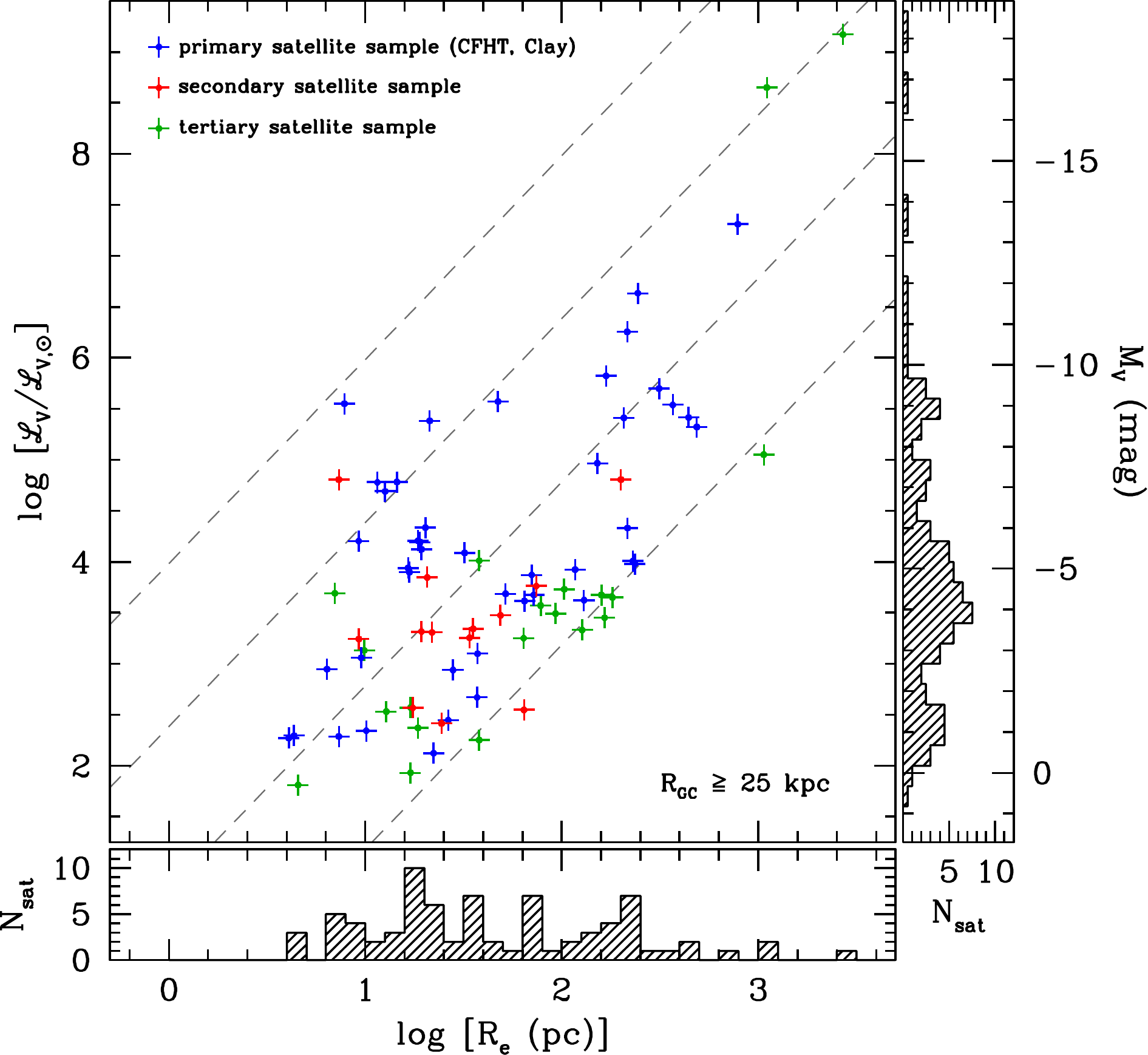}
\caption{Same as Figure~\ref{sizemag1}, but with labels removed and the four satellites with $R_{\rm GC} \le 25$ 
kpc removed. The histograms in the lower and right panels show the distribution of these 77 outer halo satellites 
in terms of effective  radius and absolute magnitude.
} \label{sizemag2}
\end{figure}

The same data are shown again in Figure~\ref{sizemag2}, but now excluding the four objects belonging to the inner halo. For
clarity, labels have been removed although the color coding remains the same. At the time of writing, this figure shows
the complete sample of 77 known outer halo substructures, irrespective of their classification as globular cluster, 
classical dwarf or ultra-faint dwarf galaxy.  The histograms in the lower and right panels show the projected distribution of 
these satellites in terms of effective radius and absolute magnitude.

Two points are worth noting in this figure. First,  the dramatic increase in the number of cataloged satellites
during the past two decades has been accompanied by a commensurate increase in their structural diversity. For
instance, the 77 objects shown in Figure~\ref{sizemag2} span factors of $\sim$10$^3$ in effective radius, $\sim$~10$^6$
in $V$-band luminosity and $\sim$10$^{4.8}$ in surface brightness. Clearly, these substructures represent a
remarkably diverse population. Second, with the benefit of an expanded sample
size, a homogeneous analysis based on high-quality imaging, and a uniform parameterization of the density profiles, the
separation of halo substructures into two distinct populations --- i.e., globular clusters and dwarf 
galaxies --- with fundamentally different formation paths has become difficult to support, at least on the basis of
photometric and structural parameters alone. Incompleteness in this diagram is notoriously
difficult to gauge, but the once-clear dichotomy in the sizes of globular clusters and dwarf galaxies has blurred
considerably during the past decade (although a paucity of satellites having effective radii of $\sim$ 30-100 pc and luminosities 
of $\sim$10$^5$\,L$_{\sun}$ may persist). In any event, Figure~\ref{sizemag2} demonstrates that spectroscopy for individual member stars
will be indispensable for establishing the true nature of new substructures, through dynamical mass estimates 
and measurements of star-to-star variations in chemical abundances.

\begin{figure}
\plotone{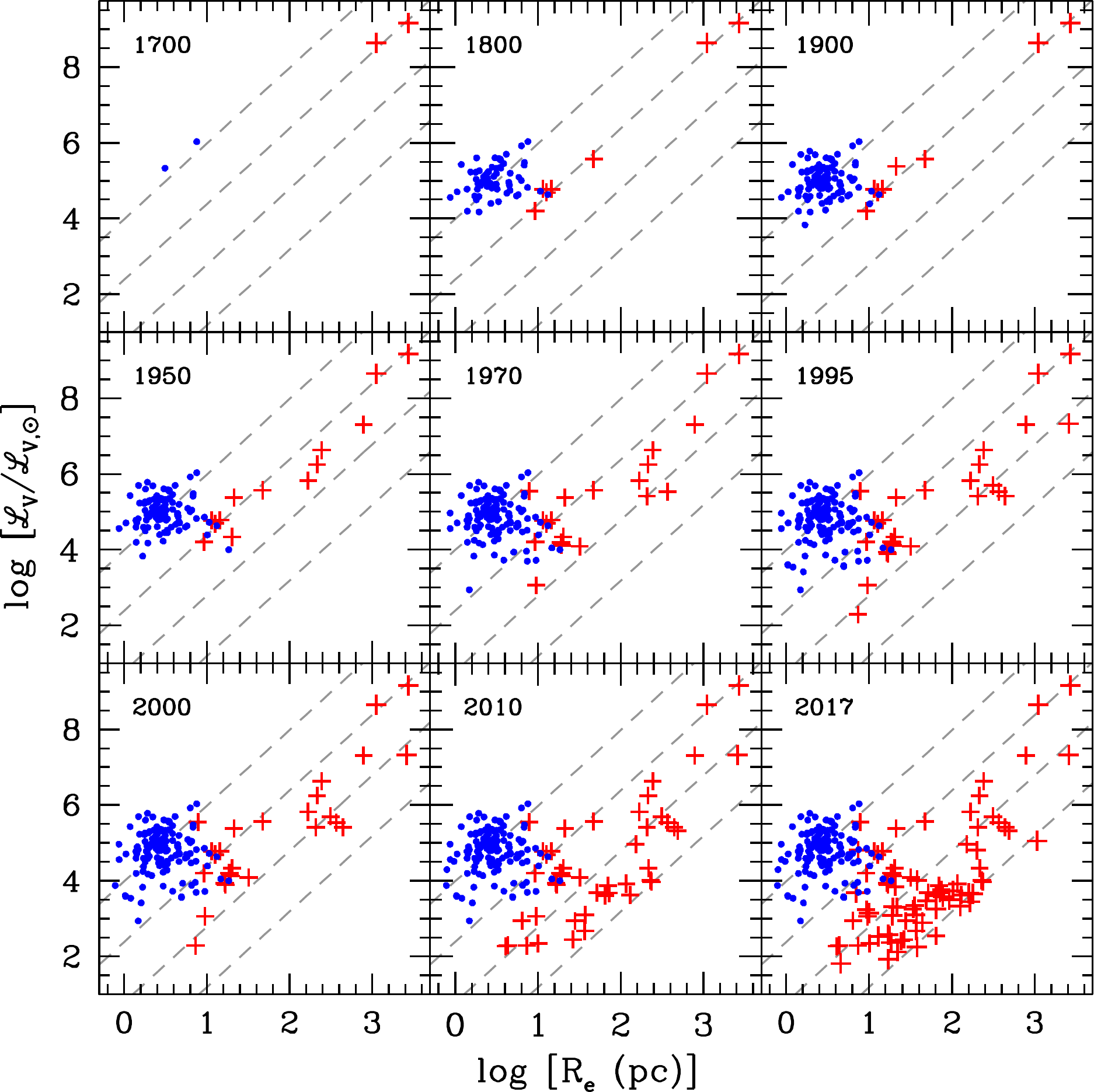}
\caption{The discovery of Galactic satellites as a function of time, illustrating the importance of selection effects when
identifying objects in the size-luminosity plane. The nine panels
show the population of Galactic satellites known at the years labelled in each panel. A total of 139 globular
clusters with $R_{\rm GC} < 25$~kpc are shown as blue points. Red crosses show the 81 remaining
satellites (i.e., globular clusters and galaxies) that are known at the present time, 77 of which reside in the outer 
halo,  $R_{\rm GC} \ge 25$~kpc.
} \label{historical}
\end{figure}

Finally, we reflect upon the discovery histories of Galactic globular clusters and dwarf galaxies, with the benefit of a 
baseline that now spans three and a half centuries. In Figure~\ref{historical}, we show how the
census of halo substructures has evolved with time. The nine panels in this figure shows the size-magnitude diagram of satellites
at the time indicated in each panel. We plot globular clusters from \citet{harris96a} with $R_{\rm GC} < 25$~kpc as blue dots and 
satellites in the outer halo as red crosses. It is striking to see the role that surface brightness has
played in the defining the census of known substructures, with the ensemble of known satellites at any time generally lying above a
well-defined threshold in surface brightness \citep[e.g.,][]{disney76a}. As a result, discoveries have historically been driven by improvements
in telescope technology, with the most dramatic gains in history (see Figure~1 of \citeauthor*{munoz18a}) coming on the heels
of surveys made with wide-field instruments (e.g., the surveys of W. Herschel and J. Dunlop, the Palomar Sky Survey
conducted with the 48-inch Oschin Schmidt telescope, and, most recently, with the SDSS, Pan-STARRS and the DES surveys).
The highly anticipated next step will come from LSST, which is expected to add hundreds of
new objects to our census of halo substructures \citep[e.g.,][]{tollerud08a}.

\section{Summary}
\label{sec:summary} 

In this paper, we have presented homogeneous photometric and structural parameters for a large sample of substructures in
the outer halo of the Milky Way. Our measurements are based on wide-field $gr$ images for 44 satellites obtained with the 
MegaCam instruments on the CFHT and Clay telescopes, supplemented by a reanalysis of $gr$ data for an additional 14 
satellites. Because we imposed no selection on the basis of morphology, our targets include a mixture of remote globular 
clusters, classic dSph galaxies, and ultra-faint dwarfs. 

Photometric and structural parameters were derived by fitting, using a two-dimensional maximum likelihood technique, 
four  different density laws: i.e., exponential, \citet{plummer11a}, \citet{king62a}  
and \citet{sersic68a} profiles. For seven high surface brightness targets, these four models were fitted to 
composite one-dimensional profiles obtained from a combination of star counts and surface photometry. 
We tabulate our best-fit photometric and structural parameters, including ellipticities, position 
angles, effective radii, S\'ersic indices, absolute magnitudes, and surface brightness measurements.
We compare our results to measurements  in the literature, and find generally good 
agreement for most systems. A critical evaluation of the fitted density 
laws suggests that the S\'ersic model is preferred parameterization for these substructures as it has the flexibility to fit the profiles
for satellites spanning a range in luminosity, surface brightness and morphology. 

We examine the isodensity contour maps and color magnitude diagrams for our targets, and present a careful 
comparison with  previous results for each object in our survey. As a rule, we find most of the globular 
clusters in our survey to have regular morphologies, with few signs of strong tidal interactions with the Milky Way. 
A notable exception is Palomar~14 which shows a faint, elongated distribution in its outer regions, consistent
with the presence of nascent tidal tails, as it has been also reported by \citet{sollima11a}. 
The classical dSphs in our survey are often elongated but otherwise 
show regular density contours, with no clear signs of tidal disruption; some systems, like UMi and Carina, for which claims
of tidal stripping exist in the literature, will need a closer analysis to make conclusive 
statements in this context. The situation for several of the ultra-faint galaxies is different, with a
number of systems, such as Hercules, UMa~I and II,  and Willman~1,
showing unusual morphologies and potential tidal features consistent with at least some degree of 
stripping.

Finally, we examine the distribution of outer halo satellites in the size-magnitude diagram using our catalog of
photometric and structural parameters. A wide diversity in structural parameters is
observed, with satellites spanning factors of $\sim$10$^3$ in effective radius, $\sim$~10$^6$
in $V$-band luminosity and $\sim$10$^{4.8}$ in surface brightness. Indeed, based on the available
sample and measured parameters, the separation of halo substructures into two distinct, non-overlapping 
populations --- i.e., globular clusters and dwarf  galaxies --- having fundamental different origins seems 
increasingly difficult to support. In the coming LSST era, when vast numbers of
satellites are expected to be discovered, spectroscopy for individual member stars
will prove essential for establishing the true nature of these systems, through dynamical mass measurements and 
elemental abundance ratios measured for individual stars.

\acknowledgements
This paper is based on observations obtained with MegaPrime/MegaCam, a joint project of CFHT and 
CEA/DAPNIA, at the Canada-France-Hawaii Telescope (CFHT) which is operated by the National Research 
Council (NRC) of Canada, the Institut National des Science de l'Univers of the Centre National de la 
Recherche Scientifique (CNRS) of France, and the University of Hawaii. 
This work was supported in part by the facilities and staff of the Yale University
Faculty of Arts and Sciences High Performance Computing Center.
The authors would like to thank an anonymous referee for helping us improve this article
and  Alan McConnachie for helpful discussions.
R.R.M. and F.A.S.~acknowledge partial support from project 
BASAL PFB-$06$. R.R.M.~also  acknowledges support from FONDECYT project N$^{\circ}1170364$. 
M.G.~acknowledges support from the National Science Foundation under
award number AST-0908752 and the Alfred P.~Sloan Foundation.
S.G.D. was supported in part by the NSF grants AST-1313422, AST-1413600, AST-1518308, and by the Ajax Foundation.

Facilities: CFHT, Magellan

\clearpage

\begin{deluxetable}{rlrrrrrrllrrcl}
\tablewidth{0pt}
\tablecaption{Adopted and Derived Parameters for Outer Halo Satellites}
\tablehead{
\colhead{No.} &
\colhead{Object} &
\colhead{$\alpha_{0}$\,(J2000)} &
\colhead{$\sigma_{\alpha}$} &
\colhead{$\delta_{0}$\,(J2000)} &
\colhead{$\sigma_{\delta}$} &
\colhead{$R_{\sun}$} &
\colhead{$R_{\rm GC}$} &
\colhead{[Fe/H]} &
\colhead{$\sigma_{\rm [Fe/H]}$} &
\colhead{$\overline{v}_{r}$} &
\colhead{$\sigma_{v_r}$} &
\colhead{$M/L_{V}$} &
\colhead{Sources} \\
\colhead{} &
\colhead{} &
\colhead{(deg)} &
\colhead{(sec)} &
\colhead{(deg)} &
\colhead{(\arcsec)} &
\colhead{(kpc)} &
\colhead{(kpc)} &
\colhead{(dex)} &
\colhead{(dex)} &
\colhead{(km~s$^{-1}$)} &
\colhead{(km~s$^{-1}$)} &
\colhead{($M_{\odot}/L_{V,{\odot}}$)} &
\colhead{} 
}
\startdata
&&&&&&&&&&&&\\
\multicolumn{14}{c}{{Primary Sample}} \\
&&&&&&&&&&&&\\
1 & Sculptor	  	&   15.0183 &   0.30 &  $-$33.7186 &   2.6 & 86.0 &	86.1	& $-1.68$	 	&	$0.46$ 	& $111.4$  &  $9.20\pm1.40$ 	& $12$ & 1,2 \\
2 & Whiting~1  		&   30.7372 &   0.18 &   $-$3.2519 &   3.0 & 30.1 &	34.9	& $-0.70$ 		&	 \nodata	& $-130.6$ & \nodata 		& \nodata	 & 3 \\ 
3 & Segue~2     	&   34.8226 &   0.95 &   $+$20.1624 &  16.5 & 35.0 & 41.2 & $-2.22$ 	&	 $0.43$	& $-40.2$	  &  $<2.6$		& $<500$ & 4 \\
4 & Fornax	  	&   39.9583 &   0.26 &  $-$34.4997 &   3.5 & 147.0 & 149.1  & $-1.04$ 	&	$0.33$ 	& $55.2$ 	  & $11.7\pm0.9$	& $5.7$  & 1, 2\\
5 & AM~1      	 	&   58.7608 &   0.14 &  $-$49.6152 &   1.5 &  123.3 	& 124.7 & $-1.70$ 	&	\nodata	& $116.0$  & $0.68$			& \nodata & 5 \\
6 & Eridanus   	      	&   66.1853 &   0.10 &  $-$21.1876 &   1.5 & 90.1 &	95.4 	& $-1.43$ 		&	\nodata 	& $-23.6$   & $0.9$			& \nodata 	& 5 \\
7 & Palomar~2 		&   71.5248 &   0.00 &   $+$31.3817 &   0.0 & 27.2 &	35.5 & $-1.42$ 		&	\nodata 	& $-133.0$ & $8.39$			& \nodata & 5 \\
8 & Carina	  	&  100.4065 &   0.58 &  $-$50.9593 &   4.4 & 105.0 	& 106.7 & $-1.72$ 	&	$0.33$ 	& $222.9$  & $6.6\pm1.2$ 	& $34$ & 2, 6 \\
9 & NGC2419 		&  114.5354 &   0.07 &   $+$38.8819 &   0.9 & 82.6 & 90.4	& $-2.10$		&	$0.032$	& $-20.2$ 	 &  $4.4\pm0.5$	& $2.05$ & 7, 8 \\
10 & Koposov~2 	&  119.5715 &   0.26 &   $+$26.2574 &   5.5 &  34.7 	& 42.3 & \nodata 	&	 \nodata	& \nodata 	 & \nodata			& \nodata & 	\\
11 & UMa~II        	&  132.8726 &   3.66 &   $+$63.1335 &   9.2 & 32.0 & 38.5 & $-2.18$		&	$0.66$ 	& $-116.5$ & $6.7\pm1.4$	& $1910$ & 1, 9 \\
12 & Pyxis   		&  136.9869 &   0.15 &  $-$37.2266 &   2.1 & 39.4 &	 41.5 & \nodata 	&	 \nodata	& $35.9$ 	 & $2.5$			& \nodata	 10 \\
13 & Leo~T    		&  143.7292 &   0.52 &   $+$17.0482 &   5.6 & 417.0 & 422.1 & $-1.74$ 	&      $0.54$	& $35$ 	 & $7.5\pm1.6$		& \nodata & 1, 9 \\  
14 & Palomar~3  	&  151.3823 &   0.09 &   $+$0.0718 &   1.4 & 92.5 & 96.0 & $-1.63$ 		&	\nodata	& $83.4$ 	 & $1.17$			& \nodata 	& 5 \\
15 & Segue~1     	&  151.7504 &   2.84 &   $+$16.0756 &  27.4 & 23.0 	& 28.1 & $-2.74$ 	&      $0.75$ 	& $208.5$  & $3.9\pm0.8$ 	& $1530$ & 11 \\
16 & Leo~I     		&  152.1146 &   0.12 &   $+$12.3059 &   1.2 & 254.0  & 256.0 & $-1.45$ 	&	$0.32$ 	& $282.5$  & $9.2\pm1.4$ 	& $4.4$ & 1 \\
17 & Sextans      	&  153.2628 &   0.61 &   $-$1.6133 &   8.1 & 86.0 & 89.2 & $-1.94$		&	$0.47$ 	& $224.3$  & $7.9\pm1.3$		& $110$ & 1, 2 \\
18 & UMa~I         	&  158.7706 &   2.98 &   $+$51.9479 &  15.3 & 97.0 	& 101.9 & $-2.10$	&	$0.65$ 	& $-55.3$ 	 & $7.6\pm1.0$		& $1620$ & 1, 9 \\
19 & Willman~1    	&  162.3436 &   1.04 &   $+$51.0501 &   4.8 & 38.0 & 43.0 & $-2.11$ 	&	$0.557$	& $-12.3$  &  $4.3^{+2.3}_{-1.3}$ 	& $520$ & 12 \\
20 & Leo~II     		&  168.3627 &   0.15 &   $+$22.1529 &   2.0 & 233.0  & 235.7 & $-1.63$ 	&	$0.40$ 	& $79.1$ 	 & $6.6\pm0.7$		& $13$ & 1, 13 \\
21 & Palomar~4  	&  172.3179 &   0.08 &   $+$28.9732 &   1.2 & 108.7 & 111.5 & $-1.41$ 	&	\nodata 	& $74.5$ 	 & $1.06$ 			& \nodata 	& 5 \\
22 & Leo~V     		&  172.7857 &   1.07 &    $+$2.2194 &   8.6 & 178.0  & 178.8 & $-2.00$ 	&	\nodata	& $173.3$  & $3.7^{+2.3}_{-1.4}$	& $215$ & 14 \\
23 & Leo~IV     		&  173.2405 &   0.98 &   $-$0.5453 &  17.0 & 154.0  & 154.6 & $-2.45$ 	&	$0.65$ 	& $132.3$  & $3.3\pm1.7$ 	& $145$ & 1, 9 \\
24 & Koposov~1 	&  179.8253 &   0.34 &   $+$12.2615 &  11.1 & 48.3 	& 49.5 & \nodata 	&	\nodata	& \nodata 	 & \nodata			& \nodata 	& \\
25 & ComBer     	&  186.7454 &   1.02 &   $+$23.9069 &  12.7 & 44.0 	& 45.2 & $-2.25$ 	&	$0.43$ 	& $98.1$ 	 & $4.6\pm0.8$		& $500$ & 1, 9 \\
26 & CVn~II        	&  194.2927 &   0.73 &   $+$34.3226 &  13.3 & 160.0 & 160.7 & $-2.12$	&	$0.59$ 	& $-128.9$ & $4.6\pm1.0$		& $230$ & 1, 9 \\
27 & CVn~I        	&  202.0091 &   1.03 &   $+$33.5521 &   6.6 & 218.0  & 217.8 & $-1.91$ 	&	$0.44$ 	& $30.9$ 	 & $7.6\pm0.4$ 	& $160$ & 1, 9 \\
28 & AM~4 		&  209.0883 &   0.31 &  $-$27.1635 &   4.9 & 33.2 	& 28.5 & $-1.30$ 	&	\nodata 	& \nodata 	 &	$0.30$ 		& \nodata 	& 5 \\
29 & Bootes~II    	&  209.5141 &   1.86 &   $+$12.8553 &  21.3 & 42.0 	& 39.8 & $-2.72$  	&	$0.30$  	& $-117.0$ & $10.5\pm7.4$ 	& $6400$ & 15, 16 \\
30 & Bootes~I    	&  210.0200 &   0.83 &   $+$14.5135 &  15.8 & 66.0 	 & 63.5 & $-2.59$  	&     $0.43$ 	& $99.0$    &	$2.4^{+0.9}_{-0.5}$ 	& $60$ & 17, 18 \\
31 & NGC5694  	&  219.9019 &   0.11 &  $-$26.5390 &   1.3 & 35.0 & 29.1 & $-1.98$		&	\nodata 	& $-140.3$ & $5.8\pm0.8$  	& $1.5$ & 19 \\
32 & Mu\~noz~1	&  225.4490 &   1.00 &   $+$66.9682 &   9.3 &  45.0 	& 47.3 & $-1.46$	&	\nodata 	& $-137.0$ & $0.25\pm0.05$ 	& \nodata  & 20 \\
33 & NGC5824		&  225.9943 &   0.00 &  $-$33.0685 &   0.0 &  32.1 & 25.6 & $-1.91$		&	\nodata 	& $-27.5$ 	  & $11.6\pm0.5$ 	& \nodata  & 19 \\
34 & UMi			&  227.2420 &   1.07 &   $+$67.2221 &   5.7 & 76.0 & 78.0 & $-2.13$ 	&	$0.34$ 	& $-246.9$ &  $9.5\pm1.2$ 	& $70$ & 1, 21 \\
35 & Palomar~14 	&  242.7544 &   0.14 &   $+$14.9584 &   2.2 & 76.5 	& 71.3 & $-1.62$	&	\nodata 	& $72.3$    &  $0.66$ 		& \nodata 	& 5 \\
36 & Hercules     	&  247.7722 &   1.54 &   $+$12.7852 &   8.7 &  132.0 &  126.2 & $-2.39$   &	$0.51$  	& $45.2$    &  $3.7\pm0.9$ 	& $140$ & 1, 22 \\
37 & NGC6229  	&  251.7454 &   0.00 &   $+$47.5276 &   0.0 & 30.5 	& 29.9 & $-1.47$ 	&	\nodata 	& $-154.2$ & $6.07$ 		& \nodata 	& 5 \\
38 & Palomar~15     &  254.9626 &   0.11 &   $-$0.5390 &   1.6 & 45.1 	 & 38.0 & $-2.07$ 	&	\nodata 	& $68.9$ 	  & $0.79$ 		& \nodata 	& 5 \\
39 & Draco       		&  260.0684 &   0.62 &   $+$57.9185 &   3.2 & 76.0 	& 76.0 & $-1.98$ 	&	$0.42$  	& $-291.0$ & $9.1\pm1.2$ 	& $80$ & 1, 23 \\
40 & NGC7006   	&  315.3721 &   0.00 &   $+$16.1871 &   0.0 &  41.2 	& 38.4 & $-1.52$	&	\nodata	& $-384.1$ & $4.37$ 		& \nodata 	& 5 \\
41 & Segue~3   	&  320.3795 &   0.37 &   $+$19.1178 &   4.4 &  27.0 	 & 25.5 & $-1.30$	&	\nodata	& $-167$ 	  & $1.2^{+2.6}_{-1.2}$ 	& \nodata 	& 24 \\
42 & Pisces~II		&  344.6345 &   0.61 &    $+$5.9526 &   5.7 &  182.0 & 181.1 & $-1.9$ 	&	\nodata 	& \nodata    &  \nodata 		& \nodata & \\
43 & Palomar~13 	&  346.6858 &   0.12 &   $+$12.7712 &   2.5 &  26.0 	& 27.1 & $-1.60$ 	&	\nodata	& $25.2$    &  $0.4^{+0.4}_{-0.3}$ 	& \nodata 	& 25 \\
44 & NGC7492   	&  347.1102 &   0.08 &  $-$15.6108 &   1.0 &  26.3 	& 25.4 & $-1.78$ 	&	\nodata 	& $-177.5$ & $1.2\pm1.0$	 	& \nodata 	 & 19 \\
&&&&&&&&&&&&&\\
\multicolumn{14}{c}{{Secondary Sample}} \\
&&&&&&&&&&&&&\\
1 & Triangulum~II 	&   33.3252 &   0.97 &   $+$36.1702 &  19.0 &  30.0 & 36.5 & $-2.24$ & $0.53$ &  $-381.7$ & <4.2 & \nodata & 26 \\
2 & Eridanus~III 	&   35.6952 &   1.47 &  $-$52.2838 &   7.1 &  87.0 	& 87.0	& \nodata 		&	\nodata 	& \nodata 	& \nodata	& \nodata 	 & \\
3 & Horologium~I 	&   43.8813 &   1.71 &  $-$54.1160 &  20.2 &  79.0 	& 79.3	& $-2.76$	 	&	$0.17$ 	& $112.8$ & $4.9$	 &\nodata 	& 27 \\
4 & Horologium~II 	&   49.1077 &  14.15 &  $-$50.0486 &  40.4 &  78.0 &	 79.1 & $-2.10$	 	&	\nodata	& \nodata 	& \nodata  & \nodata & 28 \\
5 & Reticulum~II 	&   53.9203 &   1.63 &  $-$54.0513 &   7.9 &  30.0 	& 31.5	& $-2.46$	 	&	$\sim0.3$ 	& $64.7 $ 	& $3.22$	& \nodata 	& 27  \\
6 & Eridanus~II 	&   56.0925 &   0.84 &  $-$43.5329 &   5.7 &  380.0 	& 381.9 	& $-2.38$		&	$0.47$ 	& $75.6$ 	& $6.9^{+1.2}_{-0.9}$	& \nodata 	& 29 \\
7 & Pictoris~I 	 	&   70.9490 &   1.92 &  $-$50.2854 &   8.6 &  114.0 & 115.7 & \nodata 		&	\nodata 	& \nodata 	& \nodata	& \nodata  &	 \\
8 & Laevens~1 		&  174.0668 &   0.28 &  $-$10.8772 &   2.9 &  145.0 & 144.8 & $-1.65$	 	&	\nodata	& $148.2$ & $2.04$	& \nodata 	& 30 \\
9 & Hydra~II	      	&  185.4251 &   0.91 &  $-$31.9860 &  13.7 &  134.0 &	131.1 & $-2.02$	 &	\nodata 	& $303.1$ & $3.6$	& \nodata 	& 30 \\
10 & Kim~2 	      	&  317.2020 &   1.21 &  $-$51.1671 &  38.3 &  100.0 &	94.0 & \nodata 		&	\nodata 	& \nodata 	& \nodata	& \nodata 	 \\
11 & Balbinot~1 	&  332.6791 &   0.25 &   $+$14.9403 &   4.6 &  31.9 	& 31.2 & $-1.58$	 	&	\nodata 	& \nodata 	& \nodata	& \nodata 	 & 31 \\
 & Kim~1\tablenotemark{$\dagger$} 	&  $+$332.9214 &   0.90 &    7.0271 &  12.9 &  19.8 & 19.2  & $-1.70$ & \nodata & \nodata & \nodata & \nodata & 32 \\
12 & Grus~I 	      	&  344.1797 &   3.58 &  $-$50.1800 &  55.2 &  120.0 &	116.1 & \nodata 	&	\nodata 	& \nodata 	& \nodata	& \nodata 	&  \\
13 & Phoenix~II 	&  354.9960 &   0.90 &  $-$54.4115 &  21.0 &  83.0 	 &	79.9 & \nodata 		&	\nodata 	& \nodata 	& \nodata	& \nodata 	 & 
\enddata
\smallskip
\label{t:cat1}
\tablenotetext{$\dagger$}{Inner halo member}
\end{deluxetable}
{\scriptsize
{\bf Notes for Primary Sample:} 
(1) -- \citet{kirby13b}; 
(2) -- \citet{walker09b}; 
(3) -- \citet{carraro07a}; 
(4) -- \citet[][95\% confidence limit for kinematic data]{kirby13a}; 
(5) -- \citet{webbink85a};
(6) -- \citet{koch06a}; 
(7) -- \citet{willman12a}; 
(8) -- \citet{baumgardt09a}.
(9) -- \citet{simon07a};
(10) -- \citet{palma00a};
(11) -- \citet{simon11a};
(12) -- \citet{willman11a};
(13) -- \citet{koch07a};
(14) -- \citet{walker09c};
(15) -- \citet{belokurov08a};
(16) -- \citet{koch14a};
(17) -- \citet{koposov11a};
(18) -- \citet{lai11a};
(19) -- \citet{harris96a};
(20) -- \citet{munoz12b};
(21) -- \citet{wilkinson04a};
(22) -- \citet{aden09a};
(23) -- \citet{walker07a};
(24) -- \citet{fadely11a};
(25) -- \citet{bradford11a};

{\bf Notes for Secondary Sample:}
(26) -- \citet[][95\% confidence limit for kinematic data]{kirby17a};
(27) -- \citet{koposov15b};
(28) -- \citet{kim15a};
(29) -- \citet{li17a};
(30) -- \citet{kirby15a};
(31) -- \citet{balbinot13a};
(32) -- \citet{kim15c}.
}

\clearpage

\begin{deluxetable}{llrrccccccl}
\tiny
\tablewidth{0pt}
\tablecaption{Satellites Not Included in this Survey: Tertiary Sample}
\tablehead{
\colhead{No.} &
\colhead{Name} &
\colhead{$\alpha_0$\,(J2000)} &
\colhead{$\delta_0$\,(J2000)} &
\colhead{$R_{\odot}$} &
\colhead{$R_{\rm GC}$} &
\colhead{$M_V$} &
\colhead{$r_e$} &
\colhead{$\epsilon$} &
\colhead{$\overline{v}_r$} &
\colhead{Sources} \\
\colhead{} &
\colhead{} &
\colhead{(deg)} &
\colhead{(deg)} &
\colhead{(kpc)} &
\colhead{(kpc)} &
\colhead{(mag)} &
\colhead{(pc)} &
\colhead{} &
\colhead{(km~s$^{-1}$)}
}
& & & & & & & & & & \\
\multicolumn{11}{c}{{Inner Halo ($R_{\rm GC} <$ 25 kpc)}} \\
& & & & & & & & & & \\
1  & Gaia~2 		& ~28.12 		& $+$53.04 		& 5.5 & 12.8 & $-$2.0 & 3 	& $0.18^{+0.20}_{-0.12}$ & \nodata & 1 \\
2  & Gaia~1 		&  101.47 		& $-$16.75 		& 4.6 & 12.1 & $-$5.1 & 9 & \nodata & $57.6$ & 1, 2 \\
3  & Draco~II       	&   238.20 	&  $+$64.57 		& $20\pm3$  &  $22\pm3$  &           $-2.9\pm0.80$ & $19^{+8}_{-6}$   &  $0.24\pm0.25$ & $+347.6\pm1.8$  	& 3, 4 \\
4  & Sagittarius    	&   283.83 	& $-$30.55 		& $26\pm2$  &  $18\pm2$  &        $-13.5\pm0.3$ & $2587\pm219$   &  $0.64\pm0.02$  & $+140.0\pm2.0$  	& 5 \\
5  & Tucana~III     	&   359.15 	& $-$59.60 		&  $25\pm2$  &  $23\pm2$  &           $-2.4\pm0.42$ & $44\pm6$      &  \nodata  & $-102.3\pm0.4$        		& 6, 7 \\
& & & & & & & & & & \\
\multicolumn{11}{c}{{Outer Halo ($R_{\rm GC} \ge$ 25 kpc)}} \\
& & & & & & & & & & \\
 1 & Tucana~IV      	& \phantom{00}0.73 & $-$60.85  	&  $48\pm4$  &  $46\pm4$  &           $-3.5\pm0.28$ & $127\pm24$     & $0.40\pm0.10$  & \nodata        		& 6 \\

 2 & DESJ0034-4902 & 8.45 & $-$40.04 			  	& 87 & 85 & $-3.00^{+0.66}_{-0.41}$ & $9.88\pm7.09$ & $0.69\pm0.24$ & \nodata 						& 8 \\
 3 & SMC            	& \phantom{0}13.19 & $-$72.83 	&  $64\pm4$  &  $61\pm4$  &         $-16.8\pm0.1$ & $1106\pm77$     &  $0.41\pm0.05$  & $+145.6\pm0.6$  	& 5 \\
 4 & DESJ0111-1341 & 17.79 & $-$13.68 				& $26.5\pm1.3$ & $29.4\pm1.3$ & $+0.3^{+0.9}_{-0.6}$ & $4.55^{+1.33}_{-0.95}$ & $0.27^{+0.20}_{-0.17}$ & \nodata & 9 \\
 5 & Cetus~II       	& \phantom{0}19.47 & $-$17.42 	&  $30\pm3$  &  $32\pm3$  & \phantom{+}$0.0\pm0.68$ & $17\pm7$    &  $\le0.4$  & \nodata        			& 6 \\
 6 & DESJ0225+0304 & 36.43 & $+$3.07 				& $23.8^{+0.7}_{-0.5}$ & $29.6^{+0.7}_{-0.5}$ & $-1.1^{+0.5}_{-0.3}$ & $18.6^{+9.2}_{-4.9}$ & $0.61^{+0.14}_{-0.23}$ & \nodata & 9 \\
 7 & Reticulum~III  	& \phantom{0}56.36 & $-$60.45 	&  $92\pm13$ &  $92\pm13$ &           $-3.3\pm0.29$ & $64\pm24$     &  $\le0.4$  & \nodata        			& 6 \\
 8 & LMC            	& \phantom{0}80.89 & $-$69.76 	&  $51\pm2$  &  $50\pm2$  &         $-18.1\pm0.1$ & $2697\pm115$    &  $0.15\pm0.08$ & $+262.2\pm3.4$  	& 5 \\
 9 & Columba~I      	& \phantom{0}82.86 & $-$28.03 	& $182\pm18$ & $186\pm18$ &           $-4.5\pm0.17$ & $103\pm25$    &  $\le0.2$  & \nodata        			& 6 \\
 10 & Crater~II       	&           177.31 & $-$18.41 		& $117\pm2$  & $116\pm2$ & $-7.8\pm0.10$ & $1066\pm86$ & $\le0.1$ & \nodata 						& 10 \\
 11 & Virgo~I & 180.04 & $-$0.68 					& $87^{+13}_{-8}$ & $87^{+13}_{-8}$ & $-0.8\pm0.9$ & $38^{+12}_{-11}$ & $0.44^{+0.14}_{-0.17}$& \nodata 	& 11 \\
 12 & Sagittarius~II 	&           298.17 & $-$22.07 		&  $67\pm5$  &  $60\pm5$  &           $-5.2\pm0.40$ & $38^{+8}_{-7}$  &  $0.23\pm0.20$  & \nodata        		& 3 \\
 13 & Indus~II       	&           309.72 & $-$46.16 		& $214\pm16$ & $208\pm16$ &           $-4.3\pm0.19$ & $181\pm67$    &  $\le0.4$   & \nodata        			& 6 \\
 14 & Laevens~3    	&           316.73 &   $+$14.98 		&  $67\pm3$  &  $64\pm3$  &           $-4.4\pm0.30$ & $7\pm2$      &  $0.21\pm0.21$  & $-140.5\pm2.0$ 		& 3 \\
 15 & Grus~II        	&           331.02 & $-$46.44 		&  $53\pm5$  &  $49\pm5$  &           $-3.9\pm0.22$ & 93$\pm$14     &  $\le0.21$  & \nodata        			&  6 \\
 16 & Pegasus~III   	&           336.10 &   $+$05.41 		&  $205\pm20$ & $203\pm20$ &           $-4.1\pm0.50$ &  $78^{+30}_{-24}$  &  $0.46\pm0.23$  & \nodata        	& 12, 13 \\
 17 & Aquarius~II	& 	     338.48 & $-$9.33 			& $108\pm3$& $105\pm3$& $4.36\pm0.14$& $159\pm24$ & $0.39\pm0.09$ & $71.1\pm2.5$ 				& 14 \\
 18 & Tucana~II      	&           342.98 & $-$58.57 		&  $57\pm5$  &  $53\pm5$  &           $-3.8\pm0.10$ & $165^{+28}_{-19}$  & $0.39\pm0.15$  & $-129.1\pm3.5$ &  15, 16, 17 \\
 19 & Tucana~V       	&           354.35 & $-$63.27 		&  $55\pm9$  &  $52\pm9$  &           $-1.6\pm0.49$ & $17\pm6$     &  $0.70\pm0.15$   & \nodata        		&  6 \\
\enddata
\label{t:cat2}
\end{deluxetable}
{\scriptsize
{\bf Notes for Tertiary Sample:} 
(1) -- \citet{koposov17a}; 
(2) -- \citet{mucciarelli17a}; 
(3) -- \citet{laevens15b}
(4) -- \citet{martin16a}; 
(5) -- \citet{mcconnachie12a}; 
(6) -- \citet{drlica15a};
(7) -- \citet{simon17a};
(8) -- \citet{luque16a};
(9) -- \citet{luque17a};
(10) -- \citet{torrealba16a};
(11) -- \citet{homma16a};
(12) -- \citet{kim15d};
(13) -- \citet{kim16a};
(14) -- \citet{torrealba16b};
(15) -- \citet{koposov15a};
(16) -- \citet{bechtol15a};
(17) -- \citet{walker16a}.
}

\clearpage

\begin{deluxetable}{rlcccccccc}
\scriptsize
\tablewidth{0pt}
\tablecaption{Measured Structural Parameters: Exponential and Plummer Models}
\tablehead{
\colhead{No.} &
\colhead{Object} &
\colhead{$\theta_{\rm exp}$} &
\colhead{$\epsilon_{\rm exp}$} &
\colhead{$r_{\rm {h,exp}}$} &
\colhead{$r_{\rm {h,exp}}$} &
\colhead{$\theta_{\rm p}$} &
\colhead{$\epsilon_{\rm p}$} &
\colhead{$r_{\rm {h,p}}$} &
\colhead{$r_{\rm {h,p}}$} \\
\colhead{} &
\colhead{} &
\colhead{(deg)} &
\colhead{} &
\colhead{(arcmin)} &
\colhead{(pc)} &
\colhead{(deg)} &
\colhead{} &
\colhead{(arcmin)} &
\colhead{(pc)}}
\startdata
&&&&&&&&&\\
\multicolumn{10}{c}{{Primary Sample}} \\
&&&&&&&&&\\
1 & Sculptor	  	& $92\pm1$ 	& $0.36\pm0.01$ & $12.43\pm0.18$ & $311\pm 5$ &$92\pm 1$ & $0.33\pm0.01$ & $11.17\pm0.05$ & $280\pm1$\\ 
2 & Whiting~1  	& $53\pm17$ 	& $0.22\pm0.08$	& $0.70\pm0.08$	& $6.1\pm0.7$ 		&$48\pm16$  		& $0.23\pm0.07$	& $0.64\pm0.08$	& $5.6\pm0.7$\\
3 & Segue~2     	& $166\pm16$ 	& $0.21\pm0.07$	& $3.64\pm0.29$	& $37.1\pm2.9$ 	&$164\pm14$ 		& $0.22\pm0.07$	& $3.76\pm0.28$	& $38.3\pm2.8$\\
4 & Fornax	  	& $45\pm1$ 	& $0.28\pm0.01$	& $18.5\pm0.2$\tablenotemark{ a} 	& $791\pm 9$ 	&$45\pm1$ 		& $0.29\pm0.01$	& $19.6\pm0.08$\tablenotemark{$\dagger$} 	& $838\pm3$\\ 
5 & AM~1	  	& $41\pm15$ 	& $0.17\pm0.06$	& $0.48\pm0.03$	& $17.2\pm 1.1$ 	&$42\pm23$ 		& $0.07\pm0.05$	& $0.45\pm0.03$	& $16.1\pm1.1$\\ 
6 & Eridanus   	& $35\pm29$ 	& $0.09\pm0.04$	& $0.65\pm0.03$	& $17.0\pm0.8$ 	&$32\pm24$ 		& $0.09\pm0.04$	& $0.64\pm0.04$	& $16.8\pm1.0$\\
7 & Palomar~2 	& $72\pm13$ 	& $0.06\pm0.02$	& \nodata	& \nodata 		&$72\pm14$ 		& $0.05\pm0.02$	& $1.12\pm0.08$	& $8.9\pm0.2$\\
8 & Carina	  	& $60\pm1$ 	& $0.37\pm0.01$	& $10.2\pm0.1$	& $311\pm 3$	 	&$60\pm1$ 		& $0.36\pm0.01$	& $10.1\pm0.10$	& $308\pm3$\\ 
9 & NGC2419 	& $103\pm6$ 	& $0.04\pm0.01$	& \nodata		& \nodata 		&$103\pm6$ 		& $0.05\pm0.01$	& $0.85\pm0.01$			& $20.4\pm0.2$ \\
10 & Koposov~2 	& $-36\pm20$ 	& $0.45\pm0.15$	& $0.42\pm0.08$	& $4.2\pm0.8$ 		&$-35\pm18$  		& $0.43\pm0.14$	& $0.44\pm0.07$	& $4.4\pm0.7$\\
11 & UMa~II       	& $-76\pm2$ 	& $0.55\pm0.03$	& $13.9\pm0.4$ 	& $129\pm4$		&$-76\pm2$ 		& $0.56\pm0.03$	& $13.8\pm0.50$	& $128\pm5$\\
12 & Pyxis	  	& $-8\pm28$ 	& $0.04\pm0.02$	& $1.59\pm0.03$	& $18.2\pm 0.3$ 	&$-12\pm31$  		& $0.04\pm0.01$	& $1.62\pm0.03$	& $18.6\pm0.3$\\ 
13 & Leo~T	        & $-104\pm20$ 	& $0.24\pm0.09$	& $1.27\pm0.13$	& $154\pm 16$ 	&$-104\pm18$  	& $0.23\pm0.09$	& $1.26\pm0.13$	& $153\pm16$\\ 
14 & Palomar~3  	& $11\pm30$ 	& $0.06\pm0.03$	& $0.71\pm0.02$	& $19.1\pm0.5$ 	&$1\pm28$  	 	& $0.03\pm0.02$	& $0.71\pm0.02$	& $19.1\pm0.5$\\
15 & Segue~1     	& $75\pm18$ 	& $0.32\pm0.13$	& $3.93\pm0.42$	& $26.3\pm2.8$ 	&$77\pm15$ 		& $0.33\pm0.10$	& $3.62\pm0.42$	& $24.2\pm2.8$\\
16 & Leo~I		& $78\pm1$ 	& $0.31\pm0.01$	& $3.53\pm0.03$	& $261\pm 2$ 		&$78\pm1$ 		& $0.30\pm0.01$	& $3.65\pm0.03$	& $270\pm2$\\ 
17 & Sextans      	& $57\pm1$ 	& $0.29\pm0.01$	& $16.9\pm0.1$ 	& $423\pm3$		&$57\pm1$ 		& $0.30\pm0.01$	& $16.5\pm0.10$	& $413\pm3$\\
18 & UMa~I         	& $67\pm2$ 	& $0.59\pm0.03$	& $8.13\pm0.31$ 	& $229\pm9$		&$67\pm2$ 		& $0.59\pm0.03$	& $8.31\pm0.35$	& $234\pm10$\\
19 & Willman~1    	& $74\pm4$ 	& $0.47\pm0.06$	& $2.52\pm0.21$ 	& $27.9\pm2.3$ 	&$73\pm4$ 		& $0.47\pm0.06$	& $2.51\pm0.22$	& $27.7\pm2.4$\\
20 & Leo~II	  	& $40\pm9$ 	& $0.07\pm0.02$	& $2.46\pm0.03$	& $167\pm 2$	 	&$38\pm8$ 	 	& $0.07\pm0.01$	& $2.52\pm0.03$	& $171\pm2$\\ 
21 & Palomar~4  	& $94\pm31$ 	& $0.02\pm0.01$	& $0.64\pm0.02$ 	& $20.2\pm0.6$ 	&$87\pm25$ 		& $0.03\pm0.01$	& $0.64\pm0.02$	& $20.2\pm0.6$\\
22 & Leo~V	  	& $-64\pm33$ 	& $0.45\pm0.18$	& $1.05\pm0.39$	& $54.4\pm20.2$ 	&$-71\pm26$ 		& $0.43\pm0.22$	& $1.00\pm0.32$	& $51.8\pm16.6$\\
23 & Leo~IV	  	& $-28\pm30$ 	& $0.19\pm0.09$	& $2.61\pm0.32$	& $117\pm14$  		&$-28\pm38$ 		& $0.17\pm0.09$	& $2.54\pm0.27$	& $114\pm12$\\ 
24 & Koposov~1	& $2\pm19$ 	& $0.54\pm0.16$	& $0.68\pm0.18$	& $9.6\pm 2.5$ 	&$7\pm21$ 		& $0.45\pm0.15$	& $0.62\pm0.18$	& $8.7\pm2.5$\\ 
25 & ComBer     	& $-58\pm4$ 	& $0.37\pm0.05$	& $5.67\pm0.32$ 	& $72.6\pm4.0$ 	&$-57\pm4$ 		& $0.37\pm0.05$	& $5.64\pm0.30$	& $72.1\pm3.8$\\
26 & CVn~II       	& $9\pm13$ 	& $0.41\pm0.13$	& $1.43\pm0.24$ 	& $66.6\pm11.1$	&$9\pm15$ 		& $0.40\pm0.13$	& $1.52\pm0.24$	& $70.7\pm11.2$\\
27 & CVn~I        	& $80\pm2$ 	& $0.45\pm0.02$	& $7.48\pm0.20$ 	&$474\pm13$		&$80\pm2$ 		& $0.44\pm0.03$	& $7.12\pm0.21$	& $452\pm13$\\
28 & AM~4  		& $33\pm23$ 	& $0.27\pm0.15$	& $0.74\pm0.18$	& $7.1\pm1.7$ 		&$32\pm24$ 		& $0.08\pm0.16$	& $0.68\pm0.15$	& $6.6\pm1.4$\\ 
29 & Bootes~II    	& $-71\pm33$ 	& $0.23\pm0.11$	& $3.07\pm0.44$ 	& $37.5\pm5.4$ 	&$-68\pm27$ 	 	& $0.25\pm0.11$	& $3.17\pm0.42$	& $38.7\pm5.1$\\
30 & Bootes~I    	& $6\pm3$ 	& $0.26\pm0.02$	& $10.5\pm0.2$ 	& $202\pm4$		&$6\pm3$ 		& $0.30\pm0.03$	& $9.97\pm0.27$	& $191\pm5$\\
31 & NGC5694  	& $65\pm13$ 	& $0.06\pm0.02$	& \nodata		& \nodata 		&$68\pm11$ 		& $0.06\pm0.02$	& \nodata			& \nodata \\
32 & Mu\~noz~1	& $136\pm50$ 	& $0.35\pm0.17$	& $0.49\pm0.19$	& $6.4\pm 2.5$   	&$139\pm46$ 		& $0.34\pm0.17$	& $0.49\pm0.15$	& $6.4\pm2.0$\\ 
33 & NGC5824	& $40\pm6$ 	& $0.04\pm0.01$	& \nodata		& \nodata 		&$40\pm7$ 		& $0.03\pm0.01$	& \nodata			& \nodata \\
34 & UMi	 		& $50\pm1$ 	& $0.55\pm0.01$	& $18.2\pm0.1$ 	& $404\pm2$		&$50\pm1$ 		& $0.55\pm0.01$	& $18.3\pm0.11$	& $407\pm2$\\
35 & Palomar~14 	& $81\pm18$ 	& $0.10\pm0.06$	& $1.42\pm0.08$ 	& $31.6\pm1.8$ 	&$86\pm17$ 		& $0.09\pm0.05$	& $1.36\pm0.06$	& $30.3\pm1.3$\\
36 & Hercules     	& $-74\pm2$ 	& $0.70\pm0.03$	& $5.83\pm0.65$ 	& $224\pm25$		&$-73\pm2$ 		& $0.69\pm0.03$	& $5.63\pm0.46$	& $216\pm17$\\
37 & NGC6229  	& $-53\pm23$ 	& $0.03\pm0.01$	&  \nodata  		& \nodata 		&$102\pm76$ 		& $0.03\pm0.01$	& $0.41\pm0.01$			& $3.63\pm0.09$ \\
38 & Palomar~15 	& $93\pm18$ 	& $0.04\pm0.02$	& $1.45\pm0.04$ 	& $19.0\pm0.5$ 	&$91\pm18$ 		& $0.04\pm0.02$	& $1.49\pm0.03$	& $19.5\pm0.4$\\
39 & Draco     		& $87\pm1$ 	& $0.30\pm0.01$	& $9.61\pm0.10$ 	& $212\pm2$		&$87\pm1$	 	& $0.29\pm0.01$	& $9.67\pm0.09$	& $214\pm2$\\
40 & NGC7006   	& $111\pm6$ 	& $0.06\pm0.02$	&  \nodata  		& \nodata 		& $108\pm6$ 		& $0.05\pm0.01$	& \nodata			& \nodata \\
41 & Segue~3	  	& $55\pm29$ 	& $0.25\pm0.13$	& $0.54\pm0.11$	& $4.2\pm 0.9$ 	& $51\pm38$ 	 	& $0.23\pm0.11$	& $0.49\pm0.08$	& $3.8\pm0.6$\\ 
42 & Pisces~II  		& $98\pm13$ 	& $0.39\pm0.10$	& $1.18\pm0.20$	& $62.5\pm 10.6$ 	& $78\pm20$  	 	& $0.34\pm0.10$	& $1.12\pm0.16$	& $59.3\pm8.5$\\ 
43 & Palomar~13 	& $6\pm37$ 	& $0.04\pm0.04$	& $1.26\pm0.10$ 	& $9.5\pm0.8$ 		& $11\pm32$ 		& $0.05\pm0.05$	& $1.14\pm0.10$	& $8.6\pm0.8$\\
44 & NGC7492   	& $125\pm21$ 	& $0.03\pm0.01$	&    \nodata 		& \nodata 		& $-51\pm33$ 		& $0.03\pm0.02$	& $1.28\pm0.02$			& $9.79\pm0.16$ \\
&&&&&&&&&\\
\multicolumn{10}{c}{{Secondary Sample}} \\
&&&&&&&&&\\
1 & Triangulum~II	& $28\pm19$	& $0.48\pm0.17$	& $2.34\pm0.58$	& $20.4\pm5.1$	& $44\pm18$		& $0.46\pm0.16$ 	& $1.99\pm0.49$ 	& $17.4\pm4.3$\\
2 & Eridanus~III 	& $73\pm28$	& $0.57\pm0.20$	& $0.34\pm0.23$	& $8.6\pm5.8$		& $60\pm28$		& $0.58\pm0.25$	& $0.30\pm0.24$	& $7.6\pm6.1$	\\
3 & Horologium~I	& $53\pm27$	& $0.32\pm0.13$	& $1.71\pm0.37$	& $39.3\pm8.5$	& $57\pm25$		& $0.27\pm0.13$	& $1.59\pm0.31$	& $36.5\pm7.1$\\
4 & Horologium~II & $137\pm12$	& $0.71\pm0.17$	& $2.17\pm0.59$	& $49.2\pm13.3$	& $140\pm12$		& $0.72\pm0.16$	& $1.94\pm0.61$	& $44.0\pm13.8$\\
5 & Reticulum~II	& $69\pm2$	& $ 0.56\pm0.03$	& $5.41\pm0.18$	& $47.2\pm1.6$	& $70\pm2$		& $0.58\pm0.02$	& $5.52\pm0.19$	& $48.2\pm1.7$\\
6 & Eridanus~II	& $82\pm8$	& $0.38\pm0.07$	& $1.80\pm0.16$	& $199\pm17$		& $82\pm8$		& $0.35\pm0.06$	& $1.77\pm0.17$	& $196\pm18.8$\\
7 & Pictoris~I		& $69\pm21$	& $0.57\pm0.19$	& $0.89\pm0.36$	& $29.5\pm11.9$	& $72\pm21$		& $0.63\pm0.21$	&  $0.88\pm0.38$	& $29.2\pm12.6$\\
8 & Laevens~1	& $102\pm20$	& $0.20\pm0.11$	& $0.50\pm0.06$	& $21.1\pm2.5$	& $111\pm16$		& $0.17\pm0.12$	& $0.51\pm0.16$	& $21.5\pm6.7$\\
9 & Hydra~II		& $13\pm28$	& $0.25\pm0.16$	& $1.65\pm0.39$	& $64.3\pm15.2$	& $16\pm25$		& $0.24\pm0.16$	& $1.52\pm0.28$	& $59.2\pm10.9$\\
10 & Kim~2		& $3\pm26$	& $0.72\pm0.30$	& $0.70\pm0.46$	& $20.4\pm13.3$	& $8\pm20$		& $0.32\pm0.28$	& $0.48\pm0.41$	& $14.0\pm11.9$\\
11 & Balbinot~1	& $156\pm13$	& $0.33\pm0.12$	& $0.86\pm0.20$	& $8.0\pm1.8$		& $154\pm10$		& $0.37\pm0.15$	& $0.87\pm0.20$	& $8.1\pm1.9$	\\
    & Kim~1		     & $120\pm26$	& $0.64\pm0.19$	& $0.93\pm0.25$	& $5.4\pm1.4$		& $120\pm20$		& $0.59\pm0.22$	& $1.09\pm0.25$	& $6.3\pm1.4$	\\
12 & Grus~I		      & $6\pm33$	& $0.55\pm0.25$	 	& $1.50\pm0.68$	& $52.4\pm23.8$	& $23\pm18$		& $0.45\pm0.30$	& $0.81\pm0.66$	& $28.3\pm23.0$\\
13 & Phoenix~II	& $-19\pm15$	& $0.62\pm0.19$	& $1.60\pm0.33$	& $38.6\pm8.0$	& $-20\pm18$		& $0.67\pm0.22$	& $1.49\pm0.53$	& $36.0\pm12.8$
\enddata
\smallskip
\tablenotetext{$\dagger$}{\citet{battaglia06a}}
\label{t:structural1}
\end{deluxetable}

\clearpage

\begin{deluxetable}{rlcccccccc}
\scriptsize
\tablewidth{0pt}
\tablecaption{Measured Structural Parameters: King Models}
\tablehead{
\colhead{No.} &
\colhead{Object} &
\colhead{$\theta_{\rm k}$} &
\colhead{$\epsilon_{\rm k}$} &
\colhead{$r_{\rm {c,k}}$} &
\colhead{$r_{\rm {t,k}}$} \\
\colhead{} &
\colhead{} &
\colhead{(deg)} &
\colhead{} &
\colhead{(arcmin)} &
\colhead{(arcmin)}}
\startdata
&&&&&\\
\multicolumn{6}{c}{{Primary Sample}} \\
&&&&&\\
1 & Sculptor  		& $91\pm1$ 	& $0.37\pm0.02$ & $7.03\pm0.05$ & $74.1\pm0.4$  \\ 
2 & Whiting~1  	& $46\pm19$ 	& $0.18\pm0.07$	& $0.20\pm0.03$	& $8.41\pm1.69$ 		 \\ 
3 & Segue~2     	& $167\pm17$ 	& $0.22\pm0.07$	& $2.93\pm0.82$	& $16.8\pm3.8$ 		 \\
4 & Fornax	  	& $46\pm1$ 	& $0.28\pm0.01$	& $17.6\pm0.2$	& $69.1\pm 0.4$\tablenotemark{$\dagger$} 		 \\ 
5 & AM~1	  	& $42\pm15$ 	& $0.16\pm0.06$	& $0.28\pm0.04$	& $2.63\pm 0.36$ 	 \\ 
6 & Eridanus   	& $37\pm29$ 	& $0.10\pm0.05$	& $0.36\pm0.05$	& $4.05\pm0.65$ 		\\
7 & Palomar~2 	& $68\pm21$ 	& $0.04\pm0.02$	& $0.33\pm0.01$	& $8.91\pm0.26$ 		\\
8 & Carina	  	& $60\pm1$ 	& $0.38\pm0.01$	& $7.97\pm0.16$ 	& $58.4\pm0.98$ 	  \\ 
9 & NGC2419 	& $106\pm7$ 	& $0.04\pm0.01$	& $0.27\pm0.01$	& $10.97\pm0.07$ 		 \\
10 & Koposov~2 	& $-36\pm16$ 	& $0.41\pm0.15$	& $0.23\pm0.14$	& $2.93\pm1.89$ 		  \\
11 & UMa~II       	& $-77\pm2$ 	& $0.56\pm0.03$	& $11.7\pm1.2$ 	& $59.8\pm3.1$	 \\
12 & Pyxis	  	& $-11\pm21$ 	& $0.04\pm0.02$	& $1.10\pm0.06$	& $8.17\pm 0.49$ 	  \\ 
13 & Leo~T	        & $-110\pm14$ 	& $0.24\pm0.10$	& $0.86\pm0.57$	& $6.25\pm 1.10$ 	  \\ 
14 & Palomar~3  	& $7\pm21$ 	& $0.05\pm0.02$	& $0.54\pm0.03$	& $3.38\pm0.16$ 		  \\
15 & Segue~1     	& $74\pm16$ 	& $0.34\pm0.10$	& $3.24\pm1.56$	& $16.4\pm2.6$ 		 \\
16 & Leo~I		& $78\pm1$ 	& $0.31\pm0.01$	& $3.60\pm0.10$	& $13.5\pm 0.3$ 		 \\ 
17 & Sextans      	& $58\pm1$ 	& $0.30\pm0.01$	& $20.1\pm0.5$ 	& $60.5\pm0.6$	 \\
18 & UMa~I         	& $67\pm3$ 	& $0.57\pm0.03$	& $13.3\pm2.9$ 	& $24.0\pm1.9$		 \\
19 & Willman~1    	& $74\pm4$ 	& $0.47\pm0.06$	& $1.29\pm0.26$ 	& $16.5\pm3.5$ 	 	 \\
20 & Leo~II	  	& $43\pm8$ 	& $0.07\pm0.02$	& $2.25\pm0.10$	& $9.82\pm 0.41$ 		  \\ 
21 & Palomar~4  	& $84\pm43$ 	& $0.03\pm0.01$	& $0.38\pm0.03$ 	& $3.61\pm0.36$ 		 \\
22 & Leo~V	  	& $-66\pm21$ 	& $0.46\pm0.21$	& $0.44\pm0.19$	& $9.27\pm5.85$ 	 \\
23 & Leo~IV	  	& $-30\pm26$ 	& $0.20\pm0.09$	& $2.14\pm0.82$	& $11.9\pm3.1$  	 \\ 
24 & Koposov~1	& $7\pm13$ 	& $0.46\pm0.19$	& $0.29\pm0.12$	& $4.3\pm 2.1$ 		 \\ 
25 & ComBer     	& $-58\pm4$ 	& $0.38\pm0.05$	& $4.25\pm0.73$ 	& $26.1\pm3.9$ 		 \\
26 & CVn~II       	& $10\pm19$ 	& $0.38\pm0.15$	& $0.89\pm0.65$ 	& $5.37\pm4.00$	 	 \\
27 & CVn~I        	& $80\pm2$ 	& $0.47\pm0.02$	& $6.70\pm0.50$ 	&$ 30.9\pm1.0$	 	 \\
28 & AM~4  		& $34\pm10$ 	& $0.30\pm0.14$	& $0.36\pm0.11$	& $5.99\pm1.85$ 		 \\ 
29 & Bootes~II    	& $-73\pm30$ 	& $0.25\pm0.12$	& $2.50\pm1.47$ 	& $12.9\pm8.1$ 		  \\
30 & Bootes~I    	& $6\pm3$ 	& $0.26\pm0.02$	& $11.9\pm1.0$ 	& $37.5\pm0.9$	 \\
31 & NGC5694  	& $66\pm10$ 	& $0.06\pm0.02$	& $0.05\pm0.01$		& $8.64\pm0.11$	 	 \\
32 & Mu\~noz~1	& $148\pm39$ 	& $0.34\pm0.17$	& $0.24\pm0.23$	& $4.43\pm 1.93$ 	 	 \\ 
33 & NGC5824	& $45\pm1$ 	& $0.04\pm0.01$	& $0.05\pm0.01$		& $13.21\pm0.21$ 	 \\ 
34 & UMi	 		& $50\pm1$ 	& $0.55\pm0.01$	& $13.53\pm0.3$ 	& $77.3\pm0.7$	\\
35 & Palomar~14 	& $89\pm15$ 	& $0.09\pm0.05$	& $0.64\pm0.04$ 	& $11.0\pm0.8$ 		 \\
36 & Hercules     	& $-73\pm2$ 	& $0.69\pm0.03$	& $3.29\pm0.54$ 	& $39.9\pm4.7$		 \\
37 & NGC6229  	& $-42\pm20$ 	& $0.04\pm0.01$	& $0.09\pm0.01$		& $5.25\pm0.06$		 \\
38 & Palomar~15 	& $93\pm23$ 	& $0.04\pm0.02$	& $0.91\pm0.06$ 	& $7.92\pm0.76$ 		 \\
39 & Draco     		& $87\pm1$ 	& $0.30\pm0.01$	& $6.62\pm0.15$ 	& $48.1\pm1.3$	 	 \\
40 & NGC7006   	& $110\pm6$ 	& $0.07\pm0.02$	& $0.12\pm0.01$		& $6.35\pm0.03$	     	 \\
41 & Segue~3	  	& $59\pm16$ 	& $0.21\pm0.13$	& $0.27\pm0.12$	& $3.46\pm 0.86$ 		  \\ 
42 & Pisces~II  		& $99\pm11$ 	& $0.37\pm0.10$	& $0.72\pm0.58$	& $7.65\pm 2.69$ 	   \\ 
43 & Palomar~13 	& $5\pm32$ 	& $0.04\pm0.04$	& $0.31\pm0.05$ 	& $15.9\pm1.5$ 		 \\
44 & NGC7492   	& $62\pm49$ 	& $0.03\pm0.02$	& $0.89\pm0.02$		& $6.36\pm0.07$ 	 	 \\
&&&&&\\
\multicolumn{6}{c}{{Secondary Sample}} \\
&&&&&\\
1 & Laevens~2	& $36\pm16$	& $0.41\pm0.16$	&	$1.39\pm0.60$	& $12.9\pm3.81$	 	\\
2 & Eridanus~III 	& $80\pm19$	& $0.63\pm0.24$	&	$0.32\pm0.21$	& $1.45\pm1.0$	\\
3 & Horologium~I	& $29\pm27$	& $0.28\pm0.16$	&	$1.33\pm1.13$ 	& $6.61\pm3.32$	\\
4 & Horologium~II& $135\pm10$	& $0.70\pm0.10$	& 	$1.67\pm1.23$	& $8.07\pm2.76$	\\
5 & Reticulum~II	& $69\pm2$	& $0.56\pm0.03$	&	$6.35\pm0.75$	& $19.2\pm0.9$ \\
6 & Eridanus~II	& $82\pm6$	& $0.37\pm0.07$	&	$1.84\pm0.9$	& $6.72\pm1.14$	 \\
7 & Pictoris~I		& $72\pm14$	& $0.58\pm0.20$	&	$0.64\pm0.44$	& $5.55\pm2.22$	\\
9 & Laevens~1	& $111\pm15$	& $0.14\pm0.10$	&	$0.45\pm0.14$	& $2.01\pm0.65$	 \\
9 & Hydra~II		& $14\pm20$	& $0.23\pm0.13$	&	$1.23\pm0.57$	& $8.22\pm2.21$	\\
10 & Indus~I		&	\nodata	&	\nodata	&	\nodata	&		\nodata		\\
11 & Balbinot~1	& $154\pm7$	& $0.32\pm0.13$	& $0.38\pm0.13$	& $6.02\pm1.00$	\\
 & Kim~1		     & $119\pm11$	& $0.60\pm0.19$	& $1.15\pm0.25$	& $3.07\pm0.93$	\\
12 & Grus~I		     & $9\pm16$	& $0.57\pm0.23$	& $1.26\pm0.47$	& $5.43\pm3.73$	\\
13 & Phoenix~II	& $-19\pm11$	& $0.76\pm0.14$	& $0.91\pm0.72$	& $8.14\pm4.46$	
\enddata
\smallskip
\tablenotetext{$\dagger$}{\citet{battaglia06a}}
\label{t:structural2}
\end{deluxetable}

\clearpage

\begin{deluxetable}{rlccccc}
\scriptsize
\tablewidth{0pt}
\tablecaption{Measured Structural Parameters: S\'ersic Models}
\tablehead{
\colhead{No.} &
\colhead{Object} &
\colhead{$\theta_{\rm s}$} &
\colhead{$\epsilon_{\rm s}$} &
\colhead{$n_s$} &
\colhead{$r_{\rm {e,s}}$} &
\colhead{$r_{\rm {e,s}}$} \\
\colhead{} &
\colhead{} &
\colhead{(deg)} &
\colhead{} &
\colhead{} &
\colhead{(arcmin)} &
\colhead{(pc)}}
\startdata
&&&&&&\\
\multicolumn{7}{c}{{Primary Sample}} \\
&&&&&&\\
1 & Sculptor  		& $94\pm1$ 	& $0.37\pm0.01$ 	& $1.16\pm0.01$ & $12.33\pm0.05$ & $308\pm1$ \\ 
2 & Whiting~1  		& $55\pm13$ 	& $0.24\pm0.05$	& $2.19\pm0.26$ & $0.73\pm0.07$	& $6.39\pm0.61$ \\ 
3 & Segue~2     	& $166\pm15$ 	& $0.21\pm0.07$	& $0.82\pm0.16$ & $3.64\pm0.29$	& $37.1\pm2.9$ \\ 
4 & Fornax	  	& $46\pm1$ 	& $0.28\pm0.01$	& $0.71\pm0.01$ & $18.4\pm0.2$	& $787\pm9$ \\ 
5 & AM~1	  	     	& $43\pm12$ 	& $0.16\pm0.06$	& $1.08\pm0.13$ & $0.46\pm0.03$	& $16.5\pm1.1$ \\  
6 & Eridanus   		& $35\pm25$ 	& $0.09\pm0.04$	& $1.18\pm0.14$ & $0.64\pm0.04$	& $16.8\pm1.1$ \\ 
7 & Palomar~2 		& $71\pm12$ 	& $0.05\pm0.02$	& $1.69\pm0.04$ & $0.99\pm0.02$	& $7.83\pm0.16$ \\ 
8 & Carina	  	& $60\pm1$ 	& $0.37\pm0.01$	& $0.94\pm0.01$ & $11.43\pm0.12$ & $349\pm4$ \\ 
9 & NGC2419 		& $104\pm5$ 	& $0.05\pm0.01$	& $1.71\pm0.02$ & $1.07\pm0.01$	& $25.7\pm0.2$ \\ 
10 & Koposov~2 	& $-36\pm25$ 	& $0.48\pm0.12$	& $1.35\pm0.70$ & $0.43\pm0.09$	& $4.34\pm0.91$ \\ 
11 & UMa~II       	& $-77\pm2$ 	& $0.56\pm0.03$	& $0.89\pm0.10$ & $13.95\pm0.46$ & $130\pm4$ \\ 
12 & Pyxis	  	& $-10\pm14$ 	& $0.04\pm0.02$	& $0.99\pm0.05$ & $1.62\pm0.04$	& $18.6\pm0.5$ \\  
13 & Leo~T	  	& $-107\pm16$ & $0.23\pm0.09$	& $1.03\pm0.26$ & $1.25\pm0.14$	& $152\pm17$ \\ 
14 & Palomar~3  	& $23\pm16$ 	& $0.07\pm0.03$	& $0.87\pm0.05$ & $0.72\pm0.02$	& $19.4\pm0.5$ \\ 
15 & Segue~1     	& $75\pm16$ 	& $0.34\pm0.11$	& $0.85\pm0.28$ & $3.95\pm0.48$	& $26.4\pm3.2$ \\ 
16 & Leo~I		& $78\pm1$ 	& $0.30\pm0.01$	& $0.77\pm0.02$ & $3.30\pm0.03$	& $244\pm2$ \\  
17 & Sextans      	& $58\pm1$ 	& $0.30\pm0.01$	& $0.60\pm0.01$ & $17.67\pm0.17$ & $442\pm4$ \\ 
18 & UMa~I         	& $67\pm2$ 	& $0.57\pm0.03$	& $0.47\pm0.08$ & $8.34\pm0.34$	& $235\pm10$ \\ 
19 & Willman~1    	& $74\pm4$ 	& $0.47\pm0.06$	& $1.34\pm0.20$ & $2.53\pm0.22$	& $28.0\pm2.4$ \\ 
20 & Leo~II	  	& $43\pm8$ 	& $0.07\pm0.02$	& $0.71\pm0.02$ & $2.48\pm0.03$	& $168\pm2$ \\ 
21 & Palomar~4  	& $80\pm15$ 	& $0.03\pm0.02$	& $1.12\pm0.08$ & $0.64\pm0.02$	& $20.2\pm0.6$ \\ 
22 & Leo~V	  	& $-65\pm21$ 	& $0.35\pm0.07$	& $1.70\pm0.36$ & $1.00\pm0.22$	& $51.8\pm11.4$ \\ 
23 & Leo~IV	  	& $-29\pm27$ 	& $0.19\pm0.09$	& $0.86\pm0.26$ & $2.61\pm0.31$  & $117\pm14$ \\ 
24 & Koposov~1	& $1\pm17$ 	& $0.55\pm0.15$	& $1.27\pm0.56$ & $0.72\pm0.18$  & $10.1\pm2.5$ \\ 
25 & ComBer     	& $-58\pm4$ 	& $0.37\pm0.05$	& $0.93\pm0.12$ & $5.63\pm0.30$  & $72.1\pm3.8$ \\
26 & CVn~II       	& $10\pm11$ 	& $0.46\pm0.11$	& $0.59\pm0.49$ & $1.51\pm0.23$	& $70.3\pm10.7$	 \\ 
27 & CVn~I        	& $80\pm2$ 	& $0.46\pm0.02$	& $0.78\pm0.04$ & $7.67\pm0.23$	& $486\pm14$ \\ 
28 & AM~4  		& $34\pm15$ 	& $0.29\pm0.14$	& $1.44\pm0.33$ & $0.76\pm0.14$	& $7.34\pm1.35$	 \\  
29 & Bootes~II    	& $-70\pm27$ 	& $0.24\pm0.12$	& $0.71\pm0.43$ & $3.05\pm0.45$	& $37.3\pm5.5$ \\ 
30 & Bootes~I    	& $7\pm3$ 	& $0.25\pm0.02$	& $0.64\pm0.03$ & $11.26\pm0.27$ & $216\pm5$ \\ 
31 & NGC5694  	& $67\pm9$ 	& $0.06\pm0.02$ 	& $3.20\pm0.08$ & $0.42\pm0.01$ & $4.28\pm0.10$ \\ 
32 & Mu\~noz~1	& $188\pm15$ 	& $0.50\pm0.05$	& $1.89\pm0.31$ & $1.70\pm0.32$ & $22.2\pm4.2$ \\ 
33 & NGC5824		& $47\pm4$ 	& $0.04\pm0.01$	& $3.82\pm0.05$ & $0.53\pm0.01$	& $4.95\pm0.09$ \\ 
34 & UMi	 		& $50\pm1$ 	& $0.55\pm0.01$	& $0.82\pm0.01$ & $17.32\pm0.11$ & $383\pm2$ \\ 
35 & Palomar~14 	& $90\pm10$ 	& $0.11\pm0.04$	& $1.49\pm0.08$ & $1.44\pm0.06$ & $32.0\pm1.3$ \\ 
36 & Hercules     	& $-73\pm2$ 	& $0.69\pm0.04$	& $1.19\pm0.17$ & $5.99\pm0.58$ & $230\pm22$ \\ 
37 & NGC6229  	& $-23\pm12$ 	& $0.02\pm0.01$	& $2.62\pm0.08$ & $0.36\pm0.01$ & $3.19\pm0.09$ \\ 
38 & Palomar~15 	& $93\pm11$ 	& $0.05\pm0.02$	& $1.04\pm0.06$ & $1.45\pm0.03$ & $19.0\pm0.4$ \\ 
39 & Draco      		& $87\pm1$ 	& $0.30\pm0.01$	& $0.96\pm0.02$ & $9.93\pm0.09$	& $219\pm2$ \\ 
40 & NGC7006   	& $110\pm5$ 	& $0.07\pm0.01$	& $2.55\pm0.07$ & $0.51\pm0.01$ & $6.11\pm0.12$	 \\ 
41 & Segue~3	  	& $65\pm26$ 	& $0.22\pm0.09$	& $1.30\pm0.30$ & $0.52\pm0.09$	& $4.08\pm0.71$ \\  
42 & Pisces~II  		& $99\pm13$ 	& $0.40\pm0.10$	& $1.12\pm0.34$ & $1.22\pm0.20$	& $64.6\pm10.6$ \\  
43 & Palomar~13 	& $3\pm19$ 	& $0.10\pm0.06$	& $2.22\pm0.19$ & $1.26\pm0.09$ & $9.53\pm0.68$ \\ 
44 & NGC7492   	& $-5\pm18$ 	& $0.02\pm0.02$	& $1.00\pm0.02$ & $1.25\pm0.01$	& $9.56\pm0.01$ \\ 
&&&&&&\\
\multicolumn{7}{c}{{Secondary Sample}} \\
&&&&&&\\
1 & Laevens~2		& $36\pm16$	& $0.39\pm0.11$	& $1.45\pm0.45$ & $2.00\pm0.40$	& $17.4\pm3.5$ \\
2 & Eridanus~III 	& $62\pm11$	& $0.32\pm0.13$	& $1.64\pm0.27$ & $0.29\pm0.23$	& $7.34\pm5.82$	 \\
3 & Horologium~I	& $50\pm26$	& $0.31\pm0.16$	& $0.98\pm0.47$ & $1.54\pm0.34$	& $35.4\pm7.8$ \\
4 & Horologium~II 	& $130\pm16$	& $0.86\pm0.19$	& $1.09\pm0.37$ & $2.83\pm1.31$	& $64.2\pm29.7$ \\
5 & Reticulum~II	& $69\pm2$ 	& $0.56\pm0.03$	& $0.60\pm0.05$ & $5.59\pm0.21$	& $48.8\pm1.8$ \\
6 & Eridanus~II		& $82\pm7$ 	& $0.37\pm0.06$	& $0.77\pm0.19$ & $1.81\pm0.17$	& $200\pm19$ \\
7 & Pictoris~I		& $72\pm10$ 	& $0.24\pm0.19$	& $1.51\pm0.31$ & $0.66\pm0.32$	& $21.9\pm10.6$ \\
8 & Laevens~1		& $109\pm25$ 	& $0.11\pm0.10$	& $0.77\pm0.36$ & $0.49\pm0.07$	& $20.7\pm2.9$ \\
9 & Hydra~II		& $29\pm25$ 	& $0.17\pm0.13$	& $1.20\pm0.46$ & $1.50\pm0.32$	& $58.5\pm12.5$	 \\
10 & Indus~I		& $5\pm20$ 	& $0.72\pm0.29$	& $1.22\pm0.44$ & $0.87\pm0.45$	& $25.3\pm13.0$ \\
11 & Balbinot~1	& $157\pm10$ 	& $0.35\pm0.10$	& $1.48\pm0.23$ & $0.84\pm0.11$	& $7.79\pm1.02$ \\
 & Kim~1	      		& $127\pm24$ 	& $0.67\pm0.22$	& $1.24\pm0.55$ & $0.93\pm0.22$	& $5.36\pm1.27$ \\
12 & Grus~I	       	& $11\pm32$ 	& $0.54\pm0.26$	& $1.33\pm0.31$ & $2.08\pm0.87$	& $72.6\pm30.4$ \\
13 & Phoenix~II	& $-19\pm14$ 	& $0.61\pm0.15$	& $1.14\pm0.27$ & $1.61\pm0.27$	& $38.9\pm6.5$	  
\enddata
\smallskip
\label{t:structural3}
\end{deluxetable}

\clearpage

\begin{deluxetable}{rlrrrcll}
\scriptsize
\tablewidth{0pt}
\tablecaption{Derived Parameters: Magnitude, Luminosity and Surface Brightness Measurements}
\tablehead{
\colhead{No.} &
\colhead{Object} &
\colhead{$M_{\rm g}$} &
\colhead{$M_{\rm r}$} &
\colhead{$M_{\rm V}$} &
\colhead{$\log{L_{V}/{L_{V,\odot}}}$} &
\colhead{$\mu_{V,0}$} &
\colhead{$\mu_{V,e}$} \\
\colhead{} &
\colhead{} &
\colhead{(mag)} &
\colhead{(mag)} &
\colhead{(mag)} &
\colhead{} &
\colhead{(mag arcsec$^{-2}$)} &
\colhead{(mag arcsec$^{-2}$)}}
\startdata
&&&&&&\\
 & \multicolumn{7}{c}{{Primary Sample}} \\
&&&&&&\\
1 & Sculptor  		& $-10.57\pm 0.10$ 	& $-10.98\pm 0.10$ &$-10.82\pm0.14$	&  	 $6.262\pm0.056$  & $23.29\pm0.15$  	&  $25.46\pm0.15$ \\ 
2 & Whiting~1  		& $-2.26\pm0.23$	& $-2.73\pm0.38$ 	&$-2.55\pm0.44$	&	 $2.951\pm0.176$  & $21.43^{+0.64}_{-0.66}$  	&  $25.83\pm0.65$ \\
3 & Segue~2     	& $-1.52\pm0.41$	& $-2.08\pm0.78$ 	&$-1.86\pm0.88$	&	 $2.676\pm0.352$  & $28.48\pm1.06$  	&  $29.91\pm1.06$ \\
4 & Fornax	  	& $-13.22\pm 0.10$	& $-13.61\pm 0.10$ 	&$-13.46\pm0.14$	&	 $7.317\pm0.056 $  & $23.59\pm0.16$  	&  $24.77\pm0.16$ \\ 
5 & AM~1	  	     	& $-4.65\pm0.16$	& $-5.28\pm0.20$ 	&$-5.03\pm0.26$	&	 $3.944\pm0.104 $  & $23.18^{+0.40}_{-0.41}$  	&  $25.17^{+0.40}_{-0.38}$ \\
6 & Eridanus   		& $-4.59\pm0.16$	& $-5.15\pm0.21$ 	&$-4.93\pm0.26$	&	 $3.904\pm0.104 $  & $23.23\pm0.40$  	&  $25.44\pm0.40$ \\
7 & Palomar~2 		& $-8.76\pm0.05$	& $-9.26\pm0.05$ 	&$-9.07\pm0.07$	&	 $5.558\pm0.028 $  & $16.55\pm0.11$  	&  $19.87\pm0.11$ \\
8 & Carina	  	& $-9.10\pm0.04$	& $-9.65\pm0.03$ 	&$-9.43\pm0.05$	&	 $5.706\pm0.020 $  & $25.35\pm0.07$  	&  $27.02\pm0.07$ \\
9 & NGC2419 		& $-9.16\pm0.02$	& $-9.45\pm0.02$ 	&$-9.35\pm0.03$	&	 $5.670\pm0.012 $  & $18.82\pm0.05$  	&  $22.17\pm0.05$ \\
10 & Koposov~2 	& $-0.59\pm0.52$	& $-1.14\pm0.62$ 	&$-0.92\pm0.81$	&	 $2.302\pm0.324 $  & $23.39^{+1.32}_{-1.22}$  	&  $25.96^{+1.32}_{-1.22}$ \\
11 & UMa~II       	& $-3.86\pm0.16$	& $-4.50\pm0.20$ 	&$-4.25\pm0.26$	&	 $3.630\pm0.104 $  & $28.07\pm0.33$  	&  $29.64\pm0.33$ \\ 
12 & Pyxis	  	& $-5.43\pm0.13$	& $-5.88\pm0.14$ 	&$-5.71\pm0.19$	&	 $4.215\pm0.076 $  & $23.06\pm0.24$  	&  $24.86\pm0.24$ \\
13 & Leo~T	  	& $-7.40\pm0.10$	& $-7.72\pm0.10$ 	&$-7.60\pm0.14$	&	 $4.973\pm0.056 $  & $25.42^{+0.40}_{-0.37}$  	&  $27.30^{+0.40}_{-0.37}$ \\
14 & Palomar~3  	& $-5.08\pm0.13$	& $-5.76\pm0.17$ 	&$-5.49\pm0.21$	&	 $4.127\pm0.084 $  & $23.54\pm0.27$  	&  $25.07\pm0.27$ \\
15 & Segue~1     	& $-1.08\pm0.46$	& $-1.43\pm0.57$ 	&$-1.30\pm0.73$	&	 $2.452\pm0.292 $  & $28.06^{+1.01}_{-0.98}$  	&  $29.55^{+1.01}_{-0.98}$ \\ 
16 & Leo~I		& $-11.43\pm0.20$	& $-12.00\pm0.20$ 	&$-11.78\pm0.28$	&	 $6.642\pm0.112 $  & $22.61\pm0.30$  	&  $23.92\pm0.30$ \\ 
17 & Sextans       	& $-8.36\pm0.04$	& $-8.95\pm0.04$ 	&$-8.72\pm0.06$	&	 $5.419\pm0.024 $  & $27.22\pm0.08$  	&  $28.17\pm0.08$ \\ 
18 & UMa~I        	& $-4.51\pm0.16$	& $-5.55\pm0.35$ 	&$-5.12\pm0.38$	&	 $3.981\pm0.152 $  & $29.11\pm0.47$  	&  $29.77\pm0.47$ \\
19 & Willman~1    	& $-2.20\pm0.37$	& $-2.74\pm0.64$ 	&$-2.53\pm0.74$	&	 $2.943\pm0.296 $  & $25.87\pm0.94$  	&  $28.42\pm0.92$ \\
20 & Leo~II	  	& $-9.32\pm0.02$	& $-10.02\pm0.03$ 	&$-9.74\pm0.04$	&	 $5.828\pm0.016 $  & $24.24\pm0.07$  	&  $25.42\pm0.07$ \\
21 & Palomar~4  	& $-5.58\pm0.10$	& $-6.31\pm0.12$ 	&$-6.02\pm0.16$	&	 $4.339\pm0.064 $  & $22.73\pm0.23$  	&  $24.80\pm0.23$ \\
22 & Leo~V	  	& $-4.06\pm0.22$	& $-4.62\pm0.28$ 	&$-4.40\pm0.36$	&	 $3.692\pm0.144 $  & $24.89^{+0.90}_{-0.79}$  	&  $28.23^{+0.90}_{-0.79}$ \\
23 & Leo~IV	  	& $-4.70\pm0.16$	& $-5.18\pm0.21$ 	&$-4.99\pm0.26$	&	 $3.930\pm0.104 $  & $27.80^{+0.53}_{-0.50}$  	&  $29.31^{+0.53}_{-0.50}$ \\ 
24 & Koposov~1	& $-0.78\pm0.42$	& $-1.20\pm0.55$ 	&$-1.04\pm0.69$	&	 $2.348\pm0.276 $  & $25.10^{+1.31}_{-1.17}$  	&  $27.50^{+1.31}_{-1.17}$ \\
25 & ComBer     	& $-4.03\pm0.16$	& $-4.60\pm0.19$ 	&$-4.38\pm0.25$	&	 $3.682\pm0.100 $  & $26.98\pm0.37$  	&  $28.65\pm0.37$ \\
26 & CVn~II       	& $-4.80\pm0.25$	& $-5.42\pm0.20$ 	&$-5.17\pm0.32$	&	 $4.002\pm0.128 $  & $26.50^{+0.68}_{-0.63}$  	&  $27.43^{+0.68}_{-0.63}$ \\
27 & CVn~I        	& $-8.43\pm0.05$	& $-9.04\pm0.04$ 	&$-8.80\pm0.06$	&	 $5.451\pm0.024 $  & $26.78\pm0.13$  	&  $28.12\pm0.14$ \\
28 & AM~4  		& $-0.67\pm0.51$	& $-1.04\pm0.63$ 	&$-0.90\pm0.81$	&	 $2.293\pm0.324 $  & $24.72^{+1.25}_{-1.18}$  	&  $27.50^{+1.25}_{-1.18}$ \\
29 & Bootes~II    	& $-2.55\pm0.31$	& $-3.19\pm0.67$ 	&$-2.94\pm0.74$ 	&	 $3.106\pm0.296 $  & $27.55^{+1.09}_{-1.04}$  	&  $28.74^{+1.09}_{-1.04}$ \\
30 & Bootes~I     	& $-5.71\pm0.11$	& $-6.21\pm0.23$ 	&$-6.02\pm0.25$ 	&	 $4.338\pm0.100 $  & $28.38\pm0.30$  	&  $29.42\pm0.30$ \\ 
31 & NGC5694  	& $-7.61\pm0.07$	& $-8.15\pm0.06$ 	&$-7.94\pm0.09$	&	 $5.107\pm0.036 $  & $13.41\pm0.14$  	&  $20.00\pm0.14$ \\
32 & Mu\~noz~1	& $-0.20\pm0.62$ 	& $-0.67\pm0.74$ 	&$-0.49\pm0.97$	&	 $2.127\pm0.388 $  & $26.32^{+1.42}_{-1.34}$  	&  $30.07^{+1.42}_{-1.34}$ \\ 
33 & NGC5824		&  $-9.07\pm0.03$	& $-9.42\pm0.03$ 	&$-9.29\pm0.04$ 	&	 $5.648\pm0.016 $  & $11.14\pm0.08$  	&  $19.08\pm0.08$ \\
34 & UMi	  		& $-8.70\pm0.03$	& $-9.25\pm0.04$ 	&$-9.03\pm0.05$ 	&	 $5.546\pm0.020 $  & $25.77\pm0.08$  	&  $27.19\pm0.06$ \\
35 & Palomar~14	& $-4.95\pm0.16$ 	& $-5.71\pm0.18$ 	&$-5.40\pm0.24$	& 	 $4.093\pm0.096 $  & $23.58\pm0.33$  	&  $26.46\pm0.33$ \\
36 & Hercules     	& $-5.46\pm0.11$	& $-6.08\pm0.13$ 	&$-5.83\pm0.17$	& 	 $4.266\pm0.068 $  & $26.82\pm0.39$  	&  $29.05\pm0.38$ \\
37 & NGC6229  	& $-7.74\pm0.11$	& $-8.24\pm0.12$ 	&$-8.05\pm0.16$ 	&	 $5.150\pm0.064 $  & $13.86\pm0.22$  	&  $19.19\pm0.22$ \\
38 & Palomar~15 	& $-5.24\pm0.13$ 	& $-5.95\pm0.14$ 	&$-5.66\pm0.19$	& 	 $4.198\pm0.076 $  & $23.06\pm0.23$  	&  $24.96\pm0.23$ \\ 
39 & Draco     		& $-8.35\pm0.03$	& $-8.95\pm0.04$ 	&$-8.71\pm0.05$ 	&	 $5.417\pm0.020 $  & $25.12\pm0.07$  	&  $26.85\pm0.07$ \\
40 & NGC7006   	& $-7.10\pm0.06$	& $-7.63\pm0.06$	&$-7.42\pm0.08$	& 	 $4.901\pm0.032 $  & $15.98\pm0.12$  	&  $21.16\pm0.12$ \\ 
41 & Segue~3	  	& $-0.74\pm0.37$	& $-0.93\pm0.56$ 	&$-0.87\pm0.67$	& 	 $2.280\pm0.268 $  & $23.84^{+1.09}_{-1.02}$  	&  $26.31^{+1.09}_{-1.02}$ \\ 
42 & Pisces~II		& $-3.87\pm0.24$	& $-4.45\pm0.30$ 	&$-4.22\pm0.38$	&	 $3.620\pm0.152 $  & $26.52^{+0.77}_{-0.71}$  	&  $28.60^{+0.77}_{-0.71}$ \\ 
43 & Palomar~13 	& $-2.49\pm0.29$	& $-3.06\pm0.47$ 	&$-2.84\pm0.55$	& 	 $3.066\pm0.220 $  & $22.14\pm0.70$  	&  $26.60\pm0.70$ \\
44 & NGC7492   	& $-5.75\pm0.03$ 	& $-6.34\pm0.03$ 	&$-6.11\pm0.04$	& 	 $4.375\pm0.016 $  & $21.22\pm0.06$  	&  $23.04\pm0.06$ \\ 
&&&&&&\\
 & \multicolumn{7}{c}{{Secondary Sample}} \\
&&&&&&\\
1 & Laevens~2		& $-1.46\pm0.42$	& $-1.67\pm0.63$ 	&$-1.60\pm0.76$	&	 $2.572\pm0.304 $  & $25.72^{+1.24}_{-1.16}$  	&  $28.52^{+1.24}_{-1.16}$ \\
2 & Eridanus~III 	& $-2.01\pm0.60$	& $-2.61\pm0.61$ 	&$-2.37\pm0.86$	&	 $2.881\pm0.344 $  & $22.84^{+4.28}_{-2.13}$  	&  $26.04^{+4.28}_{-2.12}$ \\
3 & Horologium~I 	& $-3.31\pm0.37$	& $-3.69\pm0.42$ 	&$-3.55\pm0.56$	&	 $3.351\pm0.224 $  & $26.28^{+1.10}_{-0.99}$  	&  $28.05^{+1.10}_{-0.99}$ \\
4 & Horologium~II 	& $-1.28\pm0.69$	& $-1.73\pm0.75$ 	&$-1.56\pm1.02$	&	 $2.555\pm0.408 $  & $27.64^{+2.37}_{-1.85}$  	&  $29.65^{+2.37}_{-1.85}$ \\
5 & Reticulum~II	& $-3.65\pm0.24$	& $-4.01\pm0.29$ 	&$-3.88\pm0.38$	&	 $3.482\pm0.152 $  & $26.77\pm0.46$  	&  $27.72\pm0.46$ \\
6 & Eridanus~II		& $-6.89\pm0.06$	& $-7.41\pm0.07$ 	&$-7.21\pm0.09$	&	 $4.815\pm0.036 $  & $26.63\pm0.30$  	&  $27.95\pm0.30$ \\
7 & Pictoris~I		& $-3.16\pm0.41$	& $-3.63\pm0.44$ 	&$-3.45\pm0.60$	&	 $3.311\pm0.240 $  & $24.50^{+2.04}_{-1.46}$  	&  $27.42^{+2.04}_{-1.46}$ \\
8 & Laevens~1		& $-4.62\pm0.22$	& $-4.90\pm0.25$ 	&$-4.80\pm0.33$	&	 $3.852\pm0.132 $  & $24.48^{+0.66}_{-0.62}$  	&  $25.80^{+0.66}_{-0.62}$ \\
9 & Hydra~II		& $-4.32\pm0.25$	& $-4.77\pm0.27$ 	&$-4.60\pm0.37$	&	 $3.771\pm0.148 $  & $26.14^{+0.89}_{-0.79}$  	&  $28.39^{+0.89}_{-0.79}$ \\
10 & Indus~I		& $-3.01\pm0.43$	& $-3.52\pm0.45$ 	&$-3.32\pm0.62$	&	 $3.260\pm0.248 $  & $24.38^{+2.20}_{-1.53}$  	&  $26.67^{+2.20}_{-1.53}$ \\
11 & Balbinot~1	& $-0.94\pm0.60$	& $-1.40\pm0.66$ 	&$-1.22\pm0.89$	&	 $2.421\pm0.356 $  & $24.36^{+1.19}_{-1.16}$  	&  $27.22^{+1.19}_{-1.16}$ \\
 & Kim~1		     	& $+0.98\pm0.71$	& $+0.58\pm0.78$ 	&$+0.73\pm1.05$	&	 $1.639\pm0.420 $  & $25.21^{+1.64}_{-1.51}$  	&  $27.54^{+1.64}_{-1.51}$ \\
12 & Grus~I  		& $-3.27\pm0.39$	& $-3.59\pm0.44$ 	&$-3.47\pm0.59$	&	 $3.321\pm0.236 $  & $26.86^{+1.77}_{-1.35}$  	&  $29.39^{+1.77}_{-1.35}$ \\
13 & Phoenix~II	& $-3.09\pm0.42$	& $-3.42\pm0.47$ 	&$-3.30\pm0.63$	&	 $3.252\pm0.252 $  & $25.85^{+1.03}_{-0.97}$  	&  $27.96^{+1.03}_{-0.97}$ \\
\enddata
\smallskip
\label{t:mags}
\end{deluxetable}

\clearpage

\clearpage

\clearpage

\clearpage

\clearpage

\clearpage

\clearpage

\clearpage

\clearpage

\clearpage

\clearpage

\clearpage

\clearpage

\clearpage

\clearpage

\clearpage

\clearpage

\clearpage

\clearpage

\clearpage

\clearpage

\clearpage

\clearpage

\clearpage

\clearpage

\clearpage

\clearpage

\clearpage

\clearpage

\clearpage

\end{document}